\documentclass[preprint,authoryear,12pt]{elsarticle}

\usepackage{amssymb}
\usepackage[normal]{subfigure}
 \usepackage{lineno}

\pdfoutput=1
\journal{Planetary and Space Science}

\begin{document}

\begin{frontmatter}



\title{Nonequilibrium thermodynamics of circulation regimes in optically-thin, dry atmospheres}


\author[label1]{Salvatore Pascale}
\author[label1]{Francesco Ragone}
\author[label1,label2]{Valerio Lucarini}
\author[label3]{Yixiong Wang}
\author[label1]{Robert Boschi}

\address[label1]{KlimaCampus, Meteorologisches Institut, Universit\"at  Hamburg, Hamburg, Germany}
\address[label2]{Department of Mathematics and Statistics, University of Reading, Reading, U}
\address[label3]{Department of Physics, University of Oxford, Clarendon Laboratory, Parks Road, Oxford, UK}

\begin{abstract}
An extensive analysis of an optically-thin, dry  atmosphere at different values of the  thermal Rossby number $\mathcal{R}o$  and of the Taylor number $\mathcal{F}_f$ is performed with a general circulation model by varying  the rotation rate $\Omega$ and the surface drag $\tau$  in a wide parametric range.
 By using   nonequilibrium thermodynamics   diagnostics   such  as material entropy production, efficiency, 
    meridional heat transport and kinetic energy dissipation we characterize in a new way the different circulation regimes. Baroclinic circulations  feature high mechanical dissipation, meridional heat transport, material entropy production  and are fairly efficient in converting heat into mechanical work.
  The thermal dissipation associated with the  sensible heat flux is found to depend mainly on the surface properties, almost independent  from the rotation rate and very low for quasi-barotropic circulations and regimes approaching equatorial super-rotation.   Slowly rotating, axisymmetric circulations   have the highest  meridional heat transport.
At high rotation rates  and intermediate-high drag,  atmospheric circulations are zonostrohic   with very low mechanical dissipation,  meridional heat transport and efficiency. When $\tau$ is interpreted as a tunable parameter associated with the turbulent boundary layer transfer of momentum and sensible heat,    our results   confirm the possibility of using  the Maximum Entropy Production Principle as a tuning guideline in the range of values of $\Omega$.
 This study suggests the effectiveness of using fundamental nonequilibrium thermodynamics 
 for investigating the properties of planetary atmospheres and extends our knowledge of the thermodynamics of the atmospheric circulation regimes.
\end{abstract}
\begin{keyword}
 Circulation regimes  \sep nonequilbrium thermodynamics \sep  terrestrial planetary atmospheres \sep baroclinic instabilty \sep entropy production

\MSC[2010] 85A20 \sep 80A17 \sep  86A10 \sep 76U05 \sep 37G35

\end{keyword}

\end{frontmatter}

\linenumbers


\section{Introduction}
\label{intro}

In the last two decades, more than 700  planets outside the solar system (exoplanets) have been
discovered \citep{Udry},  and the Kepler Space Telescope has recently  located over 2,000 exoplanet candidates \citep{Borucki}. 
The study of exoplanets and their climates   is in its early stage and it is quickly developing \citep{SeagerDeming}. Observational  data are still poor and difficult to obtain, particularly for those planets -- super-Earths \citep{Char} -- that might be  capable of sustaining liquid water and thus potentially suitable for life.    Nevertheless,  the discovery of exoplanets is extending the scope
   of planetary sciences towards the study of  the so-called  ``exoclimates''\citep{Heng, Burrows, Heng2, Showman2, Joshi, Merlis, Lewis, Pierre, Thrasta, Rauscher, Dobbs}.
  Exoplanets    and their atmospheres are  in general   capable     of supporting a broad set  of circulation regimes since they are         characterized by  a range of physical  (atmospheric composition, rotation rate, dimension, surface)  and orbital (obliquity, eccentricity, distance from the parental star, spectral type of the parental star, presence or not of phase locking)   parameters even  wider than that of   Solar System  planets \citep{WilliamsPollard}. 
  Planetary science aims at  predicting and classifying  in a concise but comprehensive way exoclimates once the main orbital and physical parameters are known.  
  
  Recently  \cite{Read} noted that  the large variety of circulation regimes  may be better understood 
 by adopting the fluid-dynamical  method of similarity, i.e. by defining a set of dimensionless  numbers that fully 
characterise the planetary circulations. Two climate states that share the same set of dimensionless  numbers  are   dynamically
 equivalent and so  the statistical properties of one can be   mapped onto those of the other.  Obviously the set 
 of parameters is fairly large,  and one of the main objectives of planetary science is  to understand what is the  minimal number of dimensionless  parameters  needed to define virtually equivalent circulations    \citep{Yixiong, Showman}.
  In this study we  focus on  the impact of two parameters, the rotation rate  $\Omega$ and on the surface turbulent  exchange rate $\tau$, on the atmospheric circulation of an Earth-like dry atmosphere. The choice of such parameters naturally leads to the definition of two dimensionless numbers, the thermal Rossby number $\mathcal{R}o$ and the Taylor (frictional) number $\mathcal{F}_f$ \citep{Read}.

   Over the last three decades, the effect of the planetary rotation on atmospheric circulation has been investigated in some details with the aid  of general circulation models \citep{Hunt_rot, Williams_A, Williams_B,Navarra, DelGenio, Geisler, Read, VallisMeridional}.
 Variations in the value of  $\Omega$ impacts directly  the size of the baroclinic waves and the extent of the Hadley cell, which are the main features  of the large-scale Earth atmospheric circulation.
The size of the baroclinic disturbances, being proportional to  the Rossby
 deformation radius \citep{Eady}, scales  as $1/\Omega$. The latitudinal extent of
  the Hadley cell   also scales as $ 1/\Omega$ \citep{HeldHou}.   
 Numerical simulations of slowly rotating  Earth-like  planets and of Solar System planets like Venus and Titan \citep{Clancy, Hourdin} have shown the presence of one  poleward-extended Hadley cell in each
hemisphere and the weakening or complete disappearing of the midlatitude  baroclinic disturbances.
  On the other hand, at fast rotation rates  the emergence of multiple cells in the meridional circulation and multiple  
jets in the zonal circulation has been observed both in numerical simulations   \citep{Williams_A, Williams_jup} and observations (e.g. Jupiter).

   The dynamical effects  of the  solid lower boundary of terrestrial planets on the atmospheric circulation  
   is also quite important in order to understand planetary circulations  and has not been  fully addressed yet  \citep{Showman}.
   The characteristics of the surface have  been recognised as  
 a key factor in shaping Earth's atmospheric  circulation \citep{Ja, JamesGray}, although  
 this topic has received less attention than that related to  $\Omega$. The surface of a terrestrial planet, due to its roughness, 
  affects the turbulent flow within  the planetary boundary and thus the exchange of momentum 
 and energy  between the surface and the  atmosphere \citep{Arya}.   It has been shown  \citep{JamesGray, James2, Klei03}
  that the reduction of the surface drag leads to
  strong horizontal barotropic  shears in the zonal mean flow. By using 
  a two-level quasi-geostrophic model, \cite{James2}   showed that    the growth rate  of the most 
  unstable baroclinic modes is reduced  considerably by the   strong horizontal wind shears.
  This is related to the general fact that the linearised baroclinic instability  equations
  obey the Squire's theorem \citep{Kundu}.
   The role of drag  has received some attention in the exoplanets context  \citep{Rauscher} but,  to the authors' knowledge,  has not been systematically investigated so far for rotation rates which are different
  from the  Earth's.   In this study we investigate the combined effect
  of   rotation speed and surface roughness  on the dynamics, linking it to the nonequilibrium thermodynamics of the system. 
 
  Thermodynamics provide a way  for characterizing concisely  
a complex physical system, bringing together comprehensive but minimal physical information. 
%This issue is closely related to the definition of robust observables \citep{LucaComplex}.    
The  atmosphere of a planet is an example of a nonequilibrium system  \citep{Gallavotti,Maz, Klei09}, and its general circulation  redistributes energy in order to  compensate for  the radiative differential heating between hot and cold regions. The atmospheric  circulation therefore is fuelled  by the conversion of available  potential energy due to large temperature  gradients  into
  kinetic energy.   The atmosphere, in other terms, produces mechanical work,   acting   as a heat engine \citep{Lor67,Peix,JonRan,Lucarini}. It seems therefore natural to adopt nonequilibrium thermodynamics as 
  a general framework for studying exoclimates. Such an approach has 
been, for example,    applied    in   \cite{LucFr} and \cite{Boschi} for studying
the bistability of an Earth-like  planet.
Furthermore,  thermodynamical  disequilibrium  drives a variety of irreversible processes, from  frictional dissipation to chemical reactions.  The  irreversibility of climatic processes 
 is quantified by  the material entropy production \citep{Goody00, MEP, Klei09}.  The interest in studying climate 
 material entropy production  largely stemmed from the proposal of the maximum entropy production principle (MEPP) 
 by Paltridge \citep{Pal, Pal78, Pal2}, who suggested that the climate  adjusts in such a way as to maximize the 
 material entropy production. In its weak form, the MEPP suggests to use the entropy production as a target function to be maximized when tuning an empirical or uncertain parameter of a model \citep{Klei03,Kunz}. Whereas the theoretical foundations of MEPP
 are still unclear \citep{De2,Grgr, Goody07}, such a conjecture has also been proposed as a  way to estimate the meridional
 heat transport of other planets, such as  Mars and Titan \citep{Lorenz, Jupp} and potentially to exoplanets too,  and has stimulated the re-examination of  climatic dissipative processes \citep{Peix,Goody00,Paul,Paul2,MEP,Frae,Pascale}. 

%  Recently, a link has been found among the Carnot efficiency, the intensity of the Lorenz energy cycle,  the material entropy production, and the degree of irreversibility of the climate system \citep{Lucarini}  -- namely, the  efficiency of the equivalent thermal machine also  sets  the proportionality between the internal  entropy fluctuation of the system and the lower bound to entropy production by the fluid compatible  with the second law of thermodynamics. Such a bound is basically given by the entropy produced  by the dissipation of the mechanical energy, whereas the excess of entropy production is due to the  turbulent transport of heat down the gradient of  the temperature field.

In this study we   perform a large ensemble of  numerical simulations with an Earth-like general circulation model for many different values of $\Omega$ and $\tau$ in order to  compute  the dissipative  properties $\zeta$ (where $\zeta$ is any dissipative function, e.g. material entropy production) of circulations of dry atmospheres at different thermal Rossby and Taylor numbers,  $\zeta(\mathcal{R}o, \mathcal{F}_f)$. We  relate, for the first time,  the properties of  $\zeta(\mathcal{R}o, \mathcal{F}_f)$  to  the different  circulation regimes and extend our knowledge on the global thermodynamic properties of rotating fluids. We anticipate that particular regimes (e.g. baroclinic, zonostrophic, super-rotation) are effectively characterized in terms of their  thermodynamic properties. We conclude with a brief analysis of how effectively the MEPP can be used to infer the optimal value for an uncertain or empirical parameter, in this case exactly the time scale controlling the exchange of momentum and energy between free atmosphere and the surface. 

The paper is organized as follows, In Section \ref{adim}  we will shortly  discuss the dimensionless  parameters relvant for this study. In Section    \ref{setup} the model and the experimental setup are presented.   The characterization  of different dynamical regimes  is the subject of    Section \ref{regimes} whereas    in Section \ref{ther_an}  the thermodynamical properties of the circulation regimes are analysed. In Section \ref{discussione}  the main conclusions are summarized.

\section{Parametric range of general circulations and dimensionless numbers}
\label{adim}

The role of the rotation rate in planetary circulations has been first investigated   in laboratory 
experiments  with a thermally driven rotating annulus \citep{Hide, Hide2, HideMason, Read_an,Read_01, Wordsworth08,Hide3}. 
   The system consists of a fluid confined between coaxial cylinders maintained at two different temperatures and rotating at an angular velocity $\Omega$.  When the basic parameters $\Omega$ and $\Delta T$ (temperature difference between 
 the inner and outer cylinder) are varied,  a wide variety of flow patterns is observed. Different dynamical  regimes can be identified  if results are grouped with respect to two dimensionless parameters, the   \emph{thermal Rossby number}:
\begin{equation}
\mathcal{R}o=\frac{g \alpha D \Delta T}{\Omega^2 L^2},
\label{thermal_rossby}
\end{equation}
and the \emph{Taylor number}:  
\begin{equation}
\mathcal{T}a=\frac{4\Omega^2 L^5}{\nu^2 D},
\label{taylor}
\end{equation}
in which $L$ is the channel width, $D$ its depth, $\nu$ the kinematic viscosity of
 the fluid, $\alpha$  its volumetric expansion coefficient,  and $g$ the gravitational acceleration.
% The parameter   $\mathcal{T}a$ measures the strength of the viscous dissipation with respect to the Coriolis force and turns out to be \citep{Read}:
%\begin{equation}
%\mathcal{T}a\propto4\Omega^2 \tau_{\nu}^2,
%\label{taylor_2}
%\end{equation}
%since the viscous timescale is $\tau_{\nu}=L^2/\nu$ and $L/D\sim 1$. On the other hand,  $\mathcal{R}o$ is regarded as 
%a measure of the strength of buoyancy relative to Coriolis forces.  

\cite{Read} has extended the definition of the thermal Rossby number  and of the  Taylor number     to the case of atmospheric circulations. 
 The analogous of the thermal Rossby number  is defined as:
\begin{equation}
\mathcal{R}o=\frac{R \Delta \theta_h}{\Omega^2 a^2},
\label{thermal_rossby2}
\end{equation}
where  $a$ is the planet's radius, $R$ the specific gas constant
 and $\Delta\theta_h$ the horizontal (potential) temperature contrast between equator and poles.
  A  difference between the definitions in eq. (\ref{thermal_rossby}) and eq.
  (\ref{thermal_rossby2})  is that $\Delta\theta_h$ is not fixed externally
  but rather determined by the circulation itself. 
  %  Attempts have been made to link $\Delta\theta_h$ to the radiative-convective temperature contrast    $\Delta\theta_{hE}$ \cite{Stone,Held} (which in turn is determined  only by applied solar heating) but this problem is still poorly understood. 
  In the following we will take $\Delta\theta_{h}=\Delta\theta_{hE}$, as done for example in \cite{Mitch}, where $\theta_{hE}$ is the radiative-convective equilibrium potential temperature,   since this is  
  externally determined by the incoming stellar radiative energy  and thus a  more objective quantity to
  describe the horizontal differential driver for the circulation.
  A   Taylor  number can be  defined  analogously to the case of the rotating annulus as:
\begin{equation}
\mathcal{F}_f=4\Omega^2\tau^2_f
\label{friction}
\end{equation}
in which $\tau_f$ is the typical timescale for kinetic energy dissipation. We note that $\mathcal{F}_f\propto (\tau_f/\tau_{rot})^2$ , where $\tau_{rot}=2\pi/\Omega$, i.e. $\mathcal{F}_f$ is proportional to the ratio of (the squares of) the typical timescales associated with turbulent dissipation of kinetic energy and rotation.  For planets with a solid core,  $\tau_f$ is the surface drag timescale and is in general determined by the characteristics of the surface. 
The use of (\ref{thermal_rossby2}) and (\ref{friction}) has been proved to be very useful in classifying atmospheric circulation \citep{Yixiong}.

\section{Model and experimental setup}
\label{setup}

\subsection{The Planet Simulator}
\label{PlaSim}

Numerical simulations  have been performed with the Planet Simulator (PlaSim),  a general circulation
 model of intermediate complexity \citep{PLASIM}. The model is freely available at www.mi.uni-hamburg.de/plasim.  
 PlaSim is a fast running model and it is therefore suitable for large-ensemble numerical experiments. Moreover,  a full set of  thermodynamic  diagnostics is available, thus making it well suited for this work \citep{Frae, LucFr}.
 
 The atmospheric dynamic core uses the  primitive equations, which are solved using a spectral transform method \citep{Eliasen, Orszag}.  Interaction between radiation and atmosphere is dealt with using   simple but realistic   longwave \citep{Sasamori} and shortwave \citep{Lacis} radiative schemes. In particular  the incoming  solar flux $F_{SW}^{toa}$ at the top of the atmosphere (TOA)  is
\begin{equation}
F_{SW}^{toa}=S_0 \cos Z
\label{solar_toa}
\end{equation}
where $S_0$ is the solar constant ($1365$ W\,m$^{-2}$) and $Z$ the zenith angle, which is in general a function on the latitude, time of the year and time of the day, and it is computed following \cite{Berger}. All  simulations have been performed with orbital parameter -- obliquity, eccentricity, distance from the Sun, typical of Earth.   Other sub-grid scale parametrisations include  interactive clouds \citep{Stephens, Stephens2, SlingoSlingo}, moist \citep{Kuo, Kuo2} and dry convection, large scale precipitation, boundary layer  fluxes  and vertical and horizontal diffusion \citep{Louis, Louis2, Laursen}.  More information  can be found in PlaSim reference manual, freely  available at www.mi.uni-hamburg.de/Downloads-un.245.0.html. 

 In all simulations  the lower boundary is a  flat surface with prescribed  albedo and heat capacity (see Table \ref{tabella1_bis}).  This is implemented with a shallow  energy-conserving slab-ocean model with  an areal heat capacity ($C_{slab}=10^7$ J\,K$^{-1}$\,m$^{-2}$) comparable to that chosen in \cite{Friers} and \cite{Heng3}.  
In this way we   avoid fixed surface temperature and have a simple but energetically consistent climate model. The surface temperature evolves in time according to $C_{slab}\dot{T}_s=F_{SW}^{surf}+F_{LW}^{-}=\sigma T_s^4-F_T$ ($F_{SW}^{surf}$ net solar radiation at the surface, $F_{LW}^{-}$ downward longwave radiation  at the surface, $F_T$ surface sensible heat flux).  We set the depth of the mixed layer to $5$ m  in order to have an areal heat capacity ($C_{slab}=10^7$ J\,K$^{-1}$\,m$^{-2}$) comparable to that chosen in \cite{Friers} and \cite{Heng3}. % A surface albedo  equal to $0.2$ is chosen  as this is true for most of Solar System planets. 
 We have checked our result at $C_{slab}=10^8$ J\,K$^{-1}$\,m$^{-2}$ too, finding little effects on the circulations and on the global thermodynamical properties.
%Recently, the possibility of running the model in a fully coupled atmosphere-ocean setting has been implemented \citep{Eileen}.  
  Simulations are performed   at T42  spectral resolution  ($2.8^{\circ}\times 2.8^{\circ}$)  with ten  levels (T42/10LEV in the following).
 
 %, one at T21 spectral resolution (triangular truncation at wave number $21$, $ 5.6^{\circ} \times 5.6^{\circ}$ or about $500$ Km) with five vertical levels (T21/5LEV in the following) and one at T42  spectral resolution  ($2.8^{\circ}\times 2.8^{\circ}$ or about $250$ Km)  with ten vertical levels (T42/10LEV in the following). 
  %The resolution  T21  inevitably leads to an underestimation of the strength of the Lorenz energy cycle and of the material entropy production  \citep[e.g. of a factor $0.7$ in ][]{Klei03}, as the smallest  baroclinic  eddies are not fully resolved, and  it  has been chosen in order to have  a fast running model.
 
  In this study we consider  dry atmospheres. Dry atmospheres are relevant for planetary (e.g. Mars) and paleoclimatological (e.g. Snowball Earth) studies and, moreover, allow us to avoid the role of phase transitions associated with condensing substances,   simplifying the problem and making neater the connection between dynamics and thermodynamics of the system.   Such configuration  is obtained   by switching off the surface evaporation module and starting from a dry atmospheric condition.   Water vapour is consequently  not inserted within the atmosphere, which  remains dry for all  timesteps.
  %In PlaSim  the dry configuration is obtained  by simply switching off the subroutine dealing with the  surface fluxes of latent heat and starting from a dry condition. Consequently  water vapour is not inserted within the atmosphere which remains dry for all  timestep.
  
 % Such experimental setup is therefore representative of a dry rocky planet with atmospheric composition similar to the one of Earth. 
 %In  a subsequent  study we shall also consider  the impact of the hydrological cycle and its feedbacks.

%In order to assess the effect of the climate's heat capacity, we perform two sets of experiments where we consider a high and a low value for the heat capacity  of the surface. In the first case (HIGHHC), the  model has a heat capacity $C \approx 2\times 10^{8}$ J\,m$^{-2}$\,K$^{-1}$, which is a  typical value for a 50-meter deep ocean mixed layer. In the second set of experiments (LOWHC) the heat capacity is $C \approx 2.5\times 10^{7}$ J\,m$^{-2}$\,K$^{-1}$,  which is typical  for a rocky planet. For both set of runs we take a surface albedo  equal to $0.2$, also typical for terrestrial planets \citep{Wordsworth}. }

 \subsection{The strength of the turbulent surface exchanges}
\label{simplification}

In order to have a  wide and controlled variation in $\mathcal{F}_f$  (Eq. \ref{friction}),
 we simplify the representation of the surface fluxes. In PlaSim the temperature tendency 
 of the first atmospheric layer (of thickness $dz$) due to the turbulent sensible
  heat flux, $(\partial T/\partial t)_{shf}$,  is computed as:
\begin{equation}
\left(\frac{\partial T}{\partial t}\right)_{shf}=-\frac{F_T}{\rho c_p dz}=     \frac{\gamma_h|\mathbf{u}|}{dz}(T_s-\xi T)=       \frac{T_s-\xi T}{\tau_h(\mathbf x,t)},
\label{def_tau}
\end{equation}
in which  $F_T= \gamma_h|\mathbf{u}|(T_s-\xi T)$ is the surface sensible heat flux,     $\gamma_h=(k/\ln(z/z_0))^2 f(Ri,z_0)$ is the heat transfer coefficients ($z$ is height from the surface, $k$ is the von-Karman parameter, $z_0$ is the surface roughness, and $f$ is an empirical function dependent on stability (as expressed by the Richardson number $Ri$) and surface roughness), $\xi$ is the Exner factor  \citep[for more details see ][]{Louis, PlasimManual}.
  The parameter  $\tau_h$ has time dimension and in a standard    run  is a function of space  and time, $\tau_h(x,,y,z,t)=dz/(\gamma_h(x,y,t) |\mathbf{u}(x,y,t)|)$ but remains of the same order of magnitude. 
   Since we are interested in variations of orders of magnitude in $\tau_h$, we substitute the locally 
    computed $\tau_h$ with a fixed (in space and time) time scale $\tau_h$ as:
\begin{equation}
\left(\frac{\partial T}{\partial t}\right)_{shf}=-\frac{\xi T-T_s}{\tau_h}.
\label{tau_sens}
\end{equation}
Similarly to eq. (\ref{def_tau}),  for the wind tendency due to the surface stress, $(\partial \mathbf{u}/\partial t)_{stress}$,  we have:
\begin{equation}
\left(\frac{\partial \mathbf{u}}{\partial t}\right)_{stress}=-\frac{\mathbf u}{\tau_m(\mathbf x,t)}.
\end{equation}
with   $\tau_m(x,y,z,t)=dz/(\gamma_m(x,y,t) |\mathbf{u}(x,y,t)|)$ and the drag coefficient $\gamma_D$ defined similarly to $\gamma_h$.  Again we
substitute the locally compute $\tau_m(\mathbf{x},t)$ with a fixed (in space and time) drag timescale $\tau_m$ (Rayleigh friction timescale).
Generally the  drag and heat transfer coefficients $\gamma_D$ and $\gamma_h$ -- and therefore 
the time constants $\tau_m$ and $\tau_h$ -- have  similar magnitude. This is  
particularly true in the case of neutral flows,  for which $\gamma_D=\gamma_h$ is indeed 
a very good approximation \citep{ Arya, Louis}. For non-neutral flows,  $\gamma_h$ and
$\gamma_D$ are different but still of the same order of magnitude, as can be seen in Fig. 11.6 of \cite{Arya}. 
On the base of this and since in this study 
we are going to explore a wide parametric range, we assume for the sake of simplicity:
\begin{equation}
\tau_m=\tau_h=\tau.
\end{equation}
Experiments are performed for  $\Omega^*=\Omega/\Omega_E=1/10$, $1/5$, $1/2$, $1$, $2$, $4$, $8$, where
 $\Omega_E$ is the  Earth rotation rate. For each value of  $\Omega^*$ we
 run the model with $\tau=2700$, $3600$, $10800$, $21600$, $43200$, $86400$, $(86400\times3)$ , $(86400\times 10)$, $(86400\times 30)$, $(864000\times 100)$,  $(864000\times 500)$ seconds,  
 that is from $45$ minutes (model timestep for $\Omega/\Omega_E\le1$ )  to $500$ days. 
 Simulations with very large  $\tau$ are representative of an atmosphere with
 no solid lower boundary \citep{Ja, Menou, Heng3}.
 %For $\Omega^*=2$ a timestep of $30$ min is used, for $\Omega^*=4$ of $10$ min and higher values of $\Omega^*$, $5$ min.

Let us note that as  $\Omega$ increases, the typical size of the baroclinic
 disturbances $L_c$ decreases as \citep{Eady}
\begin{equation}
 L_c=2.4\pi L_R,
 \label{eady}
 \end{equation}
 with the Rossby deformation radius $L_R=NH/f$ \citep{Ja,Williams_A}, $N$ the buoyancy
 frequency, $H$ the height scale and $f=2\Omega\sin\varphi$ the Coriolis parameter.  
 For our dry-atmosphere simulations  an order-of-magnitude estimate  at the midlatitudes for $\Omega^*=8$ leads to $\Delta \theta\approx 110$ K, $\overline{\theta}\approx 240$ K  (see, e.g., Fig.\ref{circ2_t42_h}), $\Delta z=9$ km,  $N\approx(g/\overline{\theta}(\Delta\theta/\Delta z))^{1/2}\approx 2\times 10^{-2}$ s$^{-1}$ and therefore to $L_R\sim 200$ Km.  This implies that  T42 simulations (spatial resolution about $250$ Km) should be able to capture at least the largest eddies at $\Omega^*=8$ and more than adequate for $\Omega^*\le 4$. 
% For a moist atmosphere   $L_c\sim 6000$ km for $\Omega^*=1$ and  $L_c\sim 1500$ km for $\Omega^*=4$  \citep{Williams_A},  thus indicating that  $L_c<750$ km   for $\Omega^* > 8$.  At T21 resolution ($\sim 500$ km) therefore the   model  does not resolve anymore  baroclinic eddies for  $\Omega^* \ge 8$.  Because of this we have not performed  simulations at $\Omega^* > 8$.

\section{Circulation regimes at different  $\mathcal{R}o$ and $\mathcal{F}_f$}
\label{regimes}

%In the estimates of $\mathcal{R}o$ we have taken $\Delta\theta_h\sim 60$ K which corresponds approximately to the potential temperature difference  near  the surface of the radiative  solution for such a dry atmosphere. The range of variation  of the dimensionless    parameters   $\mathcal{R}o$ and $\mathcal{F}_f$ is shown in Tab.~\ref{tabella1}. We see that   $\mathcal{R}o$ varies approximately over four orders of magnitude,  whereas $\mathcal{F}_f$  varies over thirteen  orders of magnitude, thus extending considerably the parametric  range for $\mathcal{F}_f$  considered by  \cite{Williams_A, Williams_B} and \cite{Read} ($4$ orders   of magnitude for both parameters).  Typical values for  Earth, Mars and Titan  of $\mathcal{R}o$ and $\mathcal{F}_f$  are  $6\times 10^{-2}$ , $2\times 10^{-1}$,   $18$  and $1.6\times 10^{4}$, $44$, $7.5\times 10^4$ respectively  \citep{Read}.
  
%In Fig. (\ref{diag2}) we show the portion of the  space of the physical parameters ($\Omega^*$, $\tau$) relevant for the  simulations  which have been performed in this study. With the grey bold lines we have marked the boundaries between the main circulation regimes and thermodynamics properties as we will discuss in the following in Section \ref{risultati}. In order to help to connect ($\Omega^*$, $\tau$) with ($\mathcal{F}_f$,$\mathcal{R}o$), the dimensionless numbers are over-plotted ($\mathcal{F}_f$ dotted, $\mathcal{F}_f$ dotted-dashed).

 The diagram in  Fig. \ref{diag1} shows  the  dimensionless space ($\mathcal{F}_f$,$\mathcal{R}o$). The over-plotted bullet points represent numerical experiments performed at $\Omega^*=0.1$ (circles, denoted as ``slow rotation''), $\Omega^*=1$ (squares, ``intermediate rotation'') and $\Omega^*=8$ (triangles, ``fast rotation'') for strong, intermediate and weak drag condition ($\tau$ equal to $45$ minutes, $1$ day and $500$ days respectively) whose mean  meridional and zonal circulations are shown in Fig.~\ref{circ1_t42} and Fig.~\ref{circ2_t42}   and delimit the portion of  the ($\mathcal{F}_f$,$\mathcal{R}o$) space     covered by the numerical simulation performed in this study.
 We have over-plotted the corresponding values of $\Omega^*$ (horizontal dot-dashed lines) and $\tau$ (dotted lines) in order to highlight the connection between the dimensionless numbers and the physical parameters $\Omega^*$ and $\tau$.  Note that $\Omega^*$ and $\mathcal{R}o$ as well as $\tau$ and $\mathcal{F}_f$ point in opposite directions.  
 In order to help  to set the stage for the reader to understand the results in the following and make it easier  to interpret the montage of figures (\ref{circ2_t42}) and (\ref{circ1_t42}),  we anticipate the main characteristics of the simulated circulations:
   
\begin{itemize}

\item [1.] At high thermal Rossby number ($\mathcal{R}o\geq8$), the decrease of the surface drag controls the transition from counter- to super-rotating (SR in Fig.\ref{diag2}) equatorial flow. Super-rotation is approached for $\mathcal{F}_f\ge 10^4$;
 \item[2.] At intermediate rotation speed     ($1\le \mathcal{R}o \le 0.01$), strong drag ($\mathcal{F}_f \le 10$)  is associated with axisymmetric circulations (AR in Fig. \ref{diag2}).     The decrease of $\tau$  leads to the appearance of the  indirect Ferrel cell   for   $10 \le \mathcal{F}_f \le 10^5$  characterized by  baroclinic activity (BC in Fig.\ref{diag2}); further decrease of the surface drag ($\mathcal{F}_f\ge 10^5$) leads to the emergence of a barotropic flow (BT in Fig.\ref{diag2})  characterised by   a large reduction  in the  vertical  shears of the zonal wind  and the the complete disappearing of the Ferrel cell; 
\item[3.]   For fast rotations ($\mathcal{R}o  \le 10^{-3}$) the increase the of Taylor frictional number ($\mathcal{F}_f > 10^4$)  leads to the appearance of a  multi-jet, zonostrophic flow  (ZN in Fig. \ref{diag2}) for $\mathcal{F}_f>10^3$ ($\tau>6$ hours)
\end{itemize}
Boundaries between the different regimes are schematically sketched in Fig. \ref{diag2}. In the following we give a detailed description of the different regimes.

\subsection{Slow rotation ($\mathcal{R}o=8$)}
\label{slow}

Fig. \ref{circ1_t42_a}, \ref{circ1_t42_b}, \ref{circ1_t42_c} 
 and Fig. \ref{circ2_t42_a}, \ref{circ2_t42_b}, \ref{circ2_t42_c}    show the slow rotation rate ($\mathcal{R}o=8$). Such circulations are dominated  by one Hadley cell in each hemisphere  which extends northward up to the poles (this regime is denoted AS in the Fig.\ref{diag2}). 
 This is a general consequence of the
 conservation of angular momentum and  in agreement with the  theory
 of the Hadley circulation of \cite{HeldHou}. The temperature features almost no latitudinal dependence, especially in the middle atmosphere. This is    typical of slowly rotating planets \citep{Williams_A, Navarra}, and is due to
  the strong  Hadley cell  circulation.
  It is interesting to note the effect of the surface drag on shaping the Hadley circulation. 
   By comparing Fig. \ref{circ1_t42_e} to Fig. \ref{circ1_t42_b} ($\mathcal{R}o,  10^{-1}\rightarrow 8$; $\mathcal{F}_f$, $10^{3}\rightarrow 10$)  and Fig.\ref{circ1_t42_c} to Fig.\ref{circ1_t42_f} ($\mathcal{R}o, 10^{-1}\rightarrow 8$;  $\mathcal{F}_f$, $10^{7}\rightarrow 10^{5}$)  we note a  decrease of the counter-rotating westward upper-level equatorial jet    approaching the  beginning of the equatorial  super-rotation \citep[for example compare Fig \ref{circ1_t42_c} to Fig. 13 of][]{HengVogt}. Equatorial super-rotation is indeed expected to take place when $\mathcal{R}o\gg 1$ \citep{Mitch}. Therefore simulations with $\Omega^*< 1/10$ and moderate or high drag are needed in order to obtain fully super-rotating atmospheric circulations (as is the case, for example, for Venus to Titan). %If we look at   Fig.~\ref{circ1_t42_a}-\ref{circ1_t42_d}, no evidence of the onset of super-rotation  is instead observed even though $\mathcal{R}o$  increases from $0.1$ to $8$. This is because    $\mathcal{F}_f$ is too small ($\mathcal{F}_f$, $10^{-1}\rightarrow 10^{-3}$) and therefore  the flow remains in the  axisymmetric region within the $(\mathcal{R}o, \mathcal{F}_f)$ space  \citep[Figure 1 of][]{Yixiong}.  We have marked the edges of the super-rotating regimes as SR in Fig.~\ref{diag2}.

%  For  strong  surface drag ($\tau=45 $ min),  there is a strong kinetic energy dissipation at the surface, where we observe very weak winds (Fig. \ref{circ1_a}, \ref{circ1_b}). The intensity of the winds    increases,  and the position of the jet moves further northward  as $\tau$ increases.   Horizontal shear increases too, and for  low surface exchange rate ($\tau=500$ days) an equatorial westward jet appears.  The atmosphere tends to cool down since the sensible    heat flux from the surface decreases, and therefore  the atmosphere is more and more   decoupled from the surface -- which in turn tends to warm up --,  and the meridional temperature structure is more and more flattened. 
   %The case with $\mathcal{R}=\approx 10$, $\mathcal{F}_f\approx 4\times 10^4$ is indeed 
 %  not too far from Titan ($\mathcal{R}o=\approx 18$, $\mathcal{F}_f\approx  7.5\times 10^4$)
%  Comparison of Fig. \ref{circ2_a}, \ref{circ2_b}, \ref{circ2_c} and  Fig. \ref{circ2_bis_a} , \ref{circ2_bis_b},\ref{circ2_bis_c} reveals that the most vigorous    meridional circulation is associated with the intermediate case ($\tau=3$ days).
%Interestingly, for the LOWHC simulations a secondary circulation develops  within the Hadley cell (Fig. \ref{circ2_bis_a}).

\subsection{Intermediate rotation ($\mathcal{R}o=0.08$)}
\label{medium}

In the medium rotation case ($\mathcal{R}o=0.08$),   we have atmospheric circulations characterized by strong eastward zonal jets 
 at about $50-60^{\circ}$ and by a thermally direct (Hadley)  and indirect (Ferrel) meridional  cell  (Fig. \ref{circ1_t42} (d,e,f) and Fig. \ref{circ2_t42} (d,e,f)).  The general circulation is considerably affected by the different surface properties. In particular
 we note that at large $\mathcal{F}_f$, the flow develops
 strong   barotropic horizontal shears,  as first discussed by \cite{JamesGray}.
Note that, as we are considering a dry optically-thin atmosphere,  none of the three circulations shown in Fig. (\ref{circ1_t42_d}-\ref{circ1_t42_f}) is close
to the one we observe on Earth (e.g. \cite{Peix2}) but rather  similar   to that  of Mars \citep{Lewis}.

% This is not surprising however,  since in this simulation setup we have removed completely the water vapour, which indeed has a large effect on the atmospheric circulation as also shown by these numerical simulations. However the climate obtained for $\mathcal{R}o\approx0.1$ and $\mathcal{F}_f\approx 10^3$ shows  a good similarity with the mean state of Mars (e.g. see figure 2 of \cite{Read}), which in fact has a dry atmosphere  and relatively similar  dimensionless numbers ($\mathcal{R}o\approx0.2$, $\mathcal{F}_f\approx 50$).
%In particular,  when we consider the circulation obtained for $\mathcal{F}_f\approx 50$ (Fig. \ref{mars}), the resulting patterns are even more similar to those of Mars.

 The effect of the surface drag is particularly evident in the meridional circulation, which
  is largely  modified by the surface properties. A clear thermally direct-indirect
  cell structure emerges in the intermediate cases  $\mathcal{F}_f\sim 10^2$ ($\tau\sim 1$ day), with the boundaries 
  of the Hadley cell at about $40^{\circ}$. The intensity and the extent of the indirect  cell is greatly reduced
  in the high drag  ($\mathcal{F}_f\le 10^{-1} $)  case, when the baroclinic waves are largely suppressed and the flow tend to become axisymmetric. The Ferrel cell is instead  completely suppressed in the low drag ($\mathcal{F}_f \ge 10^{5}$) case, where the flow   becomes   barotropic.   The large impact of the surface properties      on the meridional circulation is related to their impact on the baroclinic disturbances \citep{JamesGray}, which normally develop at the edge of the thermally direct (Hadley) and indirect (Ferrel) cells.  The Ferrel  cell is related to the presence of eddy momentum convergence, a key ingredient of baroclinic disturbances \citep{Holton}, and its disappearance    points out   the suppression or weakening of the midlatitude disturbances.  In the presence of weak surface drag,  zonal winds tend to have high values at the surface  which remain fairly constant with height but change sign at the midlatitudes  from westward to eastward going from the equator to the poles (e.g. Fig.~\ref{circ1_t42_f})  thus generating a strong horizontal shear.   Such strong horizontal shears inhibit the growth of baroclinic waves, as demonstrated  in  \citep{James2}.  On the other hand,  with a surface characterized by  a high drag, baroclinicity is   suppressed too, because the system frictional dissipation is too high and  kinetic energy is rapidly extracted not giving   eddies the chance to grow and develop \citep{Klei03}. 
  
  Let us also note in Fig. \ref{circ2_t42}  the presence of shallow cells embedded close to the surface embedded in a larger one. This is a characteristic of optically-thin atmospheres of rocky planets  in which the solid lower boundary with low thermal inertia respond very quickly to diurnal and seasonal solar heating \citep{Caballero}. Similar features are indeed observed in Mars circulation \citep[see e.g. figure 2 of][]{Lewis}.    Such shallow cells disappears in fact in the additional runs we have performs at $C_{slab}=10^{8}$  J\,K$^{-1}$\,m$^{-2}$ (not shown) and have very little effect on the thermodynamic properties we are going to discuss in the following sections.

\subsection{Fast rotation ($\mathcal{R}o=10^{-3}$)}
\label{fast}

Finally,  in the fast rotation runs ($\mathcal{R}o=10^{-3}$)   we  observe multiple jets (Fig. \ref{circ1_t42}(g)-(i)) and 
 multiple meridional cells   (Fig. \ref{circ2_t42}(g)-(i)) in agreement with previous studies \citep{Hunt_rot, Williams_A} and
 with the scaling of the Rossby deformation radius (eq. \ref{eady}). The decrease of  $L_R$ 
 with the rotation rate makes baroclinic waves less and less efficient in the poleward heat transporting process and reduction of the meridional temperature contrast. The temperature field in fact shows larger contrast in the meridional and
vertical profile,  and
 the thermal structures    tend to be in   radiative-convective  equilibrium. The effect of $\tau$ is mainly observed in the zonal
 wind profiles (Fig. \ref{circ1_t42} (g,h,i)) and in the   meridional stream function (Fig. \ref{circ2_t42}(g,h,i)). 
 Multi-jet, zonostrophic flow \citep{Yixiong}   emerges as the surface drag decreases for $\mathcal{F}_f>10^3$, as can be seen
 in Fig. \ref{circ2_t42_i}. 
 
 %It is evident  that the decrease of surface drag tends to make the  multi-jet structure emerge. 

\section{Thermodynamic analysis}
\label{ther_an}

\subsection{Thermodynamic diagnostics}
\label{thermodynamics}

The general circulation is the result  of the conversion of the available  potential energy  generated
 by radiative differential heating into mechanical work (winds),  as first shown by \cite{Lor55,EdLor60, Lor67}. 
 For an atmosphere in a statistical steady state,  the rate of generation 
  of available potential energy,  $G$,  the rate of conversion into kinetic energy, $W$,  and the rate
  of dissipation of kinetic energy  through the turbulent cascade (and ultimately  via viscous
  dissipation), $D$,  have to be equal when averaged over long time periods (e.g. a year or longer), 
  $\overline{G}=\overline{W}=\overline{D}$ ($\overline{(\cdot)}$ denotes the time mean). They   are
  therefore  equivalent ways of measuring  the strength of the  Lorenz energy cycle \citep{Lor55}. 
  
  The energy cycle introduced by Lorenz has been set onto a thermodynamic 
  framework through the consideration of the effective Carnot engine 
  describing the ability of the atmosphere to perform work \citep{JonRan, Renn, Lucarini}. 
   The atmosphere is seen as a heat engine which generates
   mechanical work at average rate  $\overline{W}$ from the differential 
   heating due to radiative and material (e.g. latent heat release) diabatic processes.
   If   $\dot{Q}^{+}$ and $\dot{Q}^{-}$ are 
  the local positive and negative diabatic heating rate (i.e. $\dot{Q}^{+}=\dot{Q} $
  where $\dot{Q} > 0$ and 
  $\dot{Q}^{+}=0$ where $\dot{Q}<0$ and similarly
   for $\dot{Q}^{-}$) with
   \begin{equation}
   \Phi^{\pm}=\int  \dot{Q}^{\pm} \rho dV,
   \label{phi}
   \end{equation}
      we have that  $\overline{\Phi^+} +\overline{\Phi^-}=\overline{W}\ge0$. Moreover,
      one can define     an efficiency $\eta$:
\begin{equation}
\eta=\frac{\overline{\Phi^+} + \overline{\Phi^-}}{\overline{\Phi^+}}
\label{carnot}
\end{equation}
which gives us an indication about the capability of the general circulation
 of generating kinetic energy given the net heating input $\Phi^+$.  From Eq. (\ref{carnot})
 it follows that
 \begin{equation}
\overline{W}= \eta\overline{\Phi^+}
 \label{carnot2}
 \end{equation}
in full analogy with the definition of efficiency  of a heat engine \citep{Fermi}.
Such a quantity has been proved to be particularly relevant in marking the climatic
 shifts between the present day climates and  the Snowball Earth \citep{ LucFr, Boschi} 

Dissipation, and therefore irreversibility, is ubiquitous  in planetary atmospheres and, more generally,
 in nonequilibrium  systems. The kinetic energy of the atmospheric flow is ultimately
  transferred through a turbulent cascade  to smaller scales where it is then dissipated into heat by 
   friction due to viscosity.  Thermal dissipation due to  sensible heat fluxes  
   between the surface and lower atmosphere   is another  irreversible process  which may take place in planetary atmospheres.
  Planets  whose atmospheres  allow phase transitions of one or more of their chemical
  substances (e.g. water on Earth or methane on Titan) also experience  further irreversible processes as
  evaporation/condensation and diffusion \citep{Goody00, Paul2}. Irreversible processes  are associated with a positive-defined 
material entropy production \citep{Peix,Maz,Pri,Frae,Klei09}. 
%The material entropy production is a fundamental quantity in nonequilibrium Thermodynamics \citep{Maz,Pri} and gives information about the irreversible processes taking place within the system.
 General  discussions   about the entropy budget of the climate system and about how to estimate it
    from climate models can be found in 
    \cite{Peix}, \cite{Goody00}, \cite{MEP}, \cite{Klei09}, \cite{Pascale}, \cite{Pascale2}, \cite{Luc10}.  
For a climate with a dry atmosphere the material entropy production  is due to two kinds
 of processes: dissipation of kinetic energy and sensible heat fluxes.  If $\epsilon^2$ is the local 
  rate of kinetic energy dissipation such that $D = \int \epsilon^2 \rho dV$,
   the entropy production associated with it reads:
\begin{equation}
\dot{S}_{kediss}=  \int \frac{\epsilon^2}{T} \rho dV.
\label{matke}
\end{equation}
 In PlaSim the dissipation of
 kinetic energy is due to: (i) turbulent stresses in the surface boundary layer
 (which accounts for more than $50\%$ of the overall dissipation) and,  gravity 
 wave drag, implemented as a Rayleigh friction at the highest level with a
 timescale of $50$ days, which we define as $D_{phys}$. Such contribution to the total mechanical dissipation is diagnosed in the model as $1/2 \int \rho dz (\mathbf{v}^2_a-\mathbf{v}^2_b)$ where $\mathbf{v}_b$ and $\mathbf{v}_a$ is the velocity   before and after the application of the boundary layer scheme and Rayleigh friction; (ii) numerical dissipation due to numerical  diffusion 
 of momentum \citep{Johnson}, which we call $D_{num}$.  More precisely, in PlaSim horizontal diffusion is implemented by  a  $8$th order  hyperdiffusion term applied to the vertical component of the relative vorticity $\zeta=\mathbf{k}\cdot(\nabla\times \mathbf{v})$ and horizontal wind divergence $\delta=\nabla_h\cdot\mathbf v$, $\kappa\nabla^8(\zeta,\delta)$, where $\kappa$ is a coefficient of numerical diffusion -- the prognostic equations for the horizontal velocity are transformed into equations for $\zeta$ and $\delta$, for more details on PlaSim dynamical core  see \cite{PlasimManual}   --.  Although it is hard to interpret $D_{num}$
 as representative of  small scale dissipative processes \citep{Jabl}  --  the
 hyperdiffusion schemes do not usually  match the symmetry requirements of
 the stress tensor needed to ensure the conservation of the angular
 momentum \citep{Becker}  --   these contributions are produced by the model and  will  be taken into account 
 in order to be consistent with  the model  itself \citep{Johnson, Egger, Wool}.
  The total  dissipation of kinetic energy of the model is therefore $D=D_{phys}+D_{num}$.

Sensible heat  in PlaSim is associated with  turbulent surface fluxes $F_T$  driven by the temperature difference existing between  the lowermost part of the atmosphere and the surface and with numerical vertical and horizontal diffusion (of the same kind of that used for momentum) and dry convection.
    The material entropy production associated with  $F_T$ is:
\begin{equation}\dot{S}_{F}=\int F_T \left(\frac{1}{T_a}-\frac{1}{T_S}\right) dA,
\label{matsens}
\end{equation}
where $T_a$ is the temperature of the first atmospheric level
 (where $F_T$ is absorbed thus heating it) and $T_S$ the surface temperature. The material entropy production associated therefore to sensible heat  is the sum of the material entropy production due to surface turbulent fluxes, $\dot{S}_{sens}$ and to  the other sources of sensible heat (diffusion and dry convection), $\dot{S}_{sens}$, and it reads
 \begin{equation}
 \dot{S}_{sens}=\dot{S}_{F}+\dot{S}_{diff}.
 \label{matsens2}
 \end{equation}
 The total material entropy production of the system is therefore:
\begin{equation}
\dot{S}_{mat}=\dot{S}_{kediss}+\dot{S}_{sens}.
\label{mat}
\end{equation}
The ratio 
\begin{equation}
 \alpha=\dot{S}_{sens}/\dot{S}_{kediss}
\label{beja}
\end{equation}
  is a measure of the degree of irreversibility  of the system, which is zero 
if all the production of entropy is due to the Ð unavoidable Ð  dissipation of the mechanical
energy      \citep{LucFr}.   The parameter $\alpha$ introduced above is related to the Bejan number $\mathcal{B}e$ as $\mathcal{B}e=\alpha+1$ \citep{Paoletti}. Systems with large $\alpha$ are instead characterized by
high thermal dissipation  relatively to the mechanical viscous dissipation and therefore
by a  higher degree of irreversibility.
% A relation exists which links all the previous quantities \citep{Lucarini}:
%\begin{equation}
%\overline{S}_{mat}\approx\eta\overline{\Sigma^+}(1+\alpha)
%\label{link}
%\end{equation}
%in which $\Sigma^\pm=\int dV \rho Q^\pm/T$ and for a steady state $\overline{\Sigma^+} +\overline{\Sigma^-}=0 $.

\subsection{Dissipative properties of circulation regimes}
\label{diss}

In this section we analyse the dissipative properties of the different circulations described
 in Sec. \ref{regimes} as the parameters $\Omega$ and $\tau$, and consequently 
 $\mathcal{R}o$ and $\mathcal{F}_f$,  are varied.  Sensitivity studies of dissipative properties  have
  been proposed first by \cite{Kunz}  and then used extensively in \cite{Pascale2} and  \cite{Boschi}
  as an insightful way to assess the models'  tuning and their thermodynamical properties.
  In the following,  we plot quantities in the  ($\Omega^*$, $\tau$) plane for practical purposes,  
  and we overplot  the values of  $\log_{10} \mathcal{R}o$ and $\log_{10} \mathcal{F}_f$ (Fig.~\ref{kediss} to Fig.~\ref{Bej}).
  % Lines of constant $\mathcal{R}o$ are parallel to lines of constant $\Omega^*$, since, given the definition (\ref{thermal_rossby2}) and the fact that we take $\Delta\theta_h=\Delta\theta_{hE}$, $\mathcal{R}o$ only depends on $\Omega$.  Since $\mathcal{F}_f$ is instead proportional to $1/\Omega$, its isolines   have negative slope  (e.g. Fig~\ref{kediss_oc}).
  
% Before starting our analysis,  let us note that all steady states are at a global mean surface temperature of about $275$ K (\textbf{check this with T42}) since in a dry atmosphere and in a climate without sea ice there are no  strong feedback mechanisms  capable of altering the spatially integrated energy balance at the surface and at the top of the atmosphere.  

 \paragraph{Kinetic energy dissipation and meridional heat transport}
 \label{kediss}
 
 %On the base of the discussion in Section~\ref{circ}, we can anticipate that at low values of $\Omega$, the baroclinic eddies become larger than the size of the exoplanet (see equation (\ref{eady}) and related discussion) and thus do not develop; at high values of $\Omega$ they become too small and dissipate fast. Furthermore, strong drag leads to kinetic energy  extraction before the baroclinic eddies can grow, whereas weak drag leads to the build-up of barotropic shear. Therefore, the optimal situation  is for intermediate values of $\Omega$ and surface drag. In thus study we have tried to quantify this through simulations and   using thermodynamical diagnostics.  
 %Furthermore, strong drag leads to KE extraction before the baroclinic eddies can grow, whereas weak drag leads to the build-up of barotropic shear. Thus, intermediate values of drag is optimal for the baroclinic eddies to develop and participate in KE dissipation
 
  In Fig. \ref{kediss},  the results of the numerical simulations show  that   for  $10^{-2} < \mathcal{R}o  < 1$  and $1<\mathcal{F}_f<10^3$   there is the highest   total   dissipation of kinetic energy, $D$. We observe a non-trivial dependence on $\Omega$ 
  and $\tau$. The most intense dissipation  is  centered around $\mathcal{R}o\approx 0.1$ and $\mathcal{F}_f \approx 10^2$ ($\tau=12$ hours and $\Omega^*=1$), with  $D\approx 0.45$ W\,m$^{-2}$.   This is mainly associated with the dissipation of kinetic energy in the boundary layer, as can be seen in Fig. \ref{kediss_bl} where $D_{phys}$ is shown. 
  On the base of the discussion in Section~\ref{regimes}, we can speculate that at low values of $\Omega$, the baroclinic eddies become larger than the size of the exoplanet (see equation (\ref{eady}) and related discussion) and thus do not develop; at high values of $\Omega$ they become too small, convert inefficiently available potential energy into kinetic energy \citep{Hunt_rot}, and dissipate quickly. Furthermore, the  surface properties  have a dramatic impact on the circulation, as shown also by \cite{JamesGray}, because
  the growth rate of the most unstable baroclinic waves  is strongly  inhibited  by
  horizontal shears \citep{James2} observed, for example, in Fig. \ref{circ1_t42_e}. This explains the drop of $D$ at high $\mathcal{F}_f$ and intermediate $\mathcal{R}o$. On the other hand, strong drag leads to kinetic energy  extraction   early in the development of baroclinic eddies. Therefore, the optimal situation  is expected  for intermediate values of $\Omega$ and surface drag.  Our results     are in agreement with those of \cite{Klei03, Klei06}, who  considered the case $\Omega^*=1$ only.

Moving on  to fastly rotating planets,  there is a significant decrease of $D$ at  low    thermal Rossby number ($\mathcal{R}o <10^{-2}$) for any value of $\mathcal{F}_f$ (zonostrophic flow, ZN). The strength of the
      Lorenz energy cycle therefore tends to become more  insensitive to the surface properties.  
Interestingly, also circulations  of  slowly rotating planets with low drag  ($\mathcal{R}o > 1$, $\mathcal{F}_f > 10^4$, corresponding with the super-rotation regime, see Fig.\ref{diag1})  have very weak kinetic energy dissipation.  The dissipation rate   remains high for  slow rotation  and    for strong drag  ($\mathcal{F}_f \le 0.1$, $\mathcal{R}o\ge 10$, AS circulations, Fig.\ref{diag2}). This is consistent with the fact that in the low rotation, axisymmetric  circulations,  baroclinicity is mostly absent, and  the dissipation of kinetic
    energy is simply related to the strength of the surface drag, which extracts kinetic
     energy from the mean flow, thus  causing very weak winds near the surface.  

The meridional heat transport \citep{Peix} is in general a very important quantity in planetary atmospheres \citep{Lorenz} and it is associated  with the  radiative imbalance between high     and low temperature 
  regions. The zonal mean of the meridional heat transport $T(\vartheta)$ is worked out  at each latitude $\vartheta$ by integrating  the longitudinally averaged top-of-the-atmosphere (TOA)  radiation budget \citep{LucRag}.
   %$ASR-OLR$ \citep[e.g.][]{Battisti}:
   %\begin{equation}
   %Tr(\vartheta)=2\pi a^2\int_{\pi/2}^{\vartheta} (ASR(\vartheta^\prime)-OLR(\vartheta^\prime)) d\vartheta^\prime
   %\label{toarad}
   %\end{equation}
    %where $ASR$ is the absorbed stellar radiation and $OLR$ the outgoing long wave radiation at TOA. 
%    For planets  with obliquity less that $54^{\circ}$ N \citep{WilliamsPollard}, $T>0$ for $0<\vartheta<\pi/2$ and $Tr<0$ for $-\pi/2<\vartheta<0$, that is the equatorial regions are warmer than the polar regions and the atmospheric circulation moves heat from the equator towards the poles.  
 A scalar index, $MHT$, of the meridional heat  transport is then defined as 
     half of the difference of the  values of the poleward heat transport in the two hemispheres at $30^{\circ}$ latitude,
  \begin{equation}
  MHT=1/2(Tr(\pi/3)-Tr(-\pi/3)).
  \label{mer}
  \end{equation}
  $MHT$ thus represents the net heat flowing out of the equatorial region through  zonal walls  placed at $30^{\circ}$. 

 Overall we observe that the meridional heat transport  increases with $\mathcal{R}o$,  in agreement with the results   found in \cite{VallisMeridional}.  This general feature   is due to  the inefficiency of the too small  baroclinic eddies at high $\Omega$ in transporting heat (eq. \ref{eady}).   
 
 Furthermore,    it evident that  for intermediate rotation rates ($1/5\le\Omega^*\le 2$)  $MHT$ peaks at $\tau\approx1$ day ($\approx 1$ PW), that is in the region of baroclinic circulations (Fig. \ref{diag1} and \ref{diag2}),  coinciding with the maximum in dissipation (Fig.~\ref{kediss}). It is well known in fact that  midlatitude  eddies constitute a very important  mechanism  of meridional heat transport\citep{Lor67, Ja}. This is  also clear  from   the zonal mean of the transient eddy  flux  $\overline{v^{\prime}T^{\prime}}$ (not shown), 
  which reaches  the highest values $\approx 8 $ K\,ms$^{-1}$ at $900$ hPa and $50$ N/S for the values of $\tau$ maximizing $D$, compared   to $0.5$ K\,ms$^{-1}$ for $\tau=45$ min (at $700$ hPa and $60$ N/S)  and $4$ K\,ms$^{-1}$ for $\tau=500$ days
  (at $1000$ hPa and $50$ N/S).  Just for the sake of comparison, let us note that  for earth's 
  circulation $\overline{v^{\prime}T^{\prime}} |_{max} \approx $ 15   K\,m\,s$^{-1}$ at $850$ hPa and $50$ N/S \citep[e.g. ][]{Ja}.
  In the slow rotation region ($\mathcal{R}o\approx 10$) we   have the largest  heat transport ($\approx 1.5$ PW)  at high drag ($\tau$ of few hours), which may be explained by lower wind velocities in the lower branch of the    Hadley cell (equatorwards motion).

\paragraph{Efficiency and material entropy production}
\label{alltherest}

The efficiency  diagram (Fig. \ref{carnot})  shows that the highest value of $\eta$  lay in the intermediate rotation range with values of $\approx 3\%$ in correspondence of  the baroclinic and axisymmetric circulations.  At low rotations, the high-drag circulations ($\mathcal{F}_f <1$) are the most efficient. Interestingly, we note that circulations tending toward equatorial super-rotation have a quite substantial drop in efficiency which reduces to $\approx 1\%$. At low $\mathcal{R}o$ the thermodynamic efficiency drops below $1\%$ because of  the drastic drop in $D$ associated with the weakening of the Lorenz energy cycle, therefore zonostrophic flows are very inefficient circulation regimes in terms of converting heat into mechanical work. Let us note that although we are dealing with a dry atmosphere, and therefore  very different from a moist one (in which the magnitude of the heat losses and gain is much higher, for example the latent heat gives a positive heating contribution of $\sim 80$ W\,m$^{-2}$), $\eta$ has comparable values \cite[see e.g. ][]{LucFr} and does not generally exceeds $3\%$.

The material entropy production terms (eq. (\ref{matke}, \ref{matsens2} and \ref{mat})) are shown  in Fig. \ref{entropybudget}-\ref{entropybudget3}.   Fig. \ref{entropybudget}    shows the contribution  due to thermal dissipation $\dot{S}_{sens}$ (\ref{matsens}). This is dominated by $\dot{S}_{F}$, which accounts for almost $2/3$ of $\dot{S}_{sens}$ and is almost independent from $\mathcal{R}o$, having its highest values    for $\tau\sim 3$ days. Such a pattern  is explained by a trade-off  mechanism  between the sensible heat     flux,  which decreases with $\tau$ independently at any $\mathcal{R}o$ (not shown),  and the temperature difference between the surface and the near-surface atmosphere, which      increases with $\tau$ since, due to eq. (\ref{tau_sens}), surface and atmospheres tend to be more decoupled.     The entropy production  associated with the  dissipation 
    of kinetic energy,  $\dot{S}_{kediss}$  (Fig. \ref{entropybudget2}) closely follows  the    pattern of $D$ (Fig. \ref{kediss}) as evident from its own definition (eq. (\ref{matke})). 
 
 The total material entropy production (\ref{mat}) is the sum of the two, so  its properties  are determined mainly
   by $\dot{S}_{sens}$ which is generally larger than $\dot{S}_{kediss}$ ($\sim 1-2$ times in the  at low-intermediate rotation rates, as can    be seen in Fig.~\ref{Bej}   where the irreversibility  parameter $\alpha$ is shown, and up to $10$ times for fast rotating planets). 
    The    region of highest material entropy production  ($\approx 3.5 $ mW\,m$^{-2}$\,K$^{-1}$) is
   observed for $0.1\le \mathcal{R}o\le 0.01$ and $10^2\le \mathcal{F}_f \le 10^3$, and generally  the whole region of the diagram in Fig. \ref{diag1} with $0.5\, \textrm{day}\le\tau\le 5$ days have large material entropy production.
  Overall, the material entropy production tends to be fairly  low  ($\approx 1.5$ mW\,m$^{-2}$\,K$^{-1}$) for fast rotation speeds (e.g. $\mathcal{R}o\sim 10^{-3}$) where we have very low values of $\dot{S}_{sense}$ and lower values of $\dot{S}_{kediss}$. Let us note that the portion of the diagram corresponding to super-rotating fluids (SR in Fig. \ref{diag2})  is characterized by  very low  mechanical and thermal dissipation and therefore very low material entropy production. In this respect super-rotating flows  are quite interesting since such circulations are also characterized by very low efficiency. In other terms they seem to have a behavior close to  inviscid, non-dissipative fluids (for which $D=0$ and $\dot{S}_{sens}=0$ by definition).  \cite{Mitch} also pointed out some peculiar  dynamical properties of super-rotating flows, as for example the fact that the equatorial, strong eastwards jet, once  established, do not need eddy-forcing to be maintained.
Interestingly, these results make clear that there is no obvious correspondence between the presence of large amount of kinetic energy in the atmosphere and the presence of an intense Lorenz energy cycle to support its generation. This matter has been hotly debated in a rather different scientific context, where the possibility of extracting massive amounts of energy from the atmospheric circulation by wind turbines is discussed \citep{Miller}.

  A schematic diagram summarising the main thermodynamical properties discussed so far for the different circulation regimes is shown in Fig. \ref{diag1}:
  \begin{itemize}
  \item [1.]  Baroclinic regime (BC): high $D$, high $\eta$, relatively high $MHT$;
  \item [2.]  Super-rotation (SR): low $D$, low $\eta$, low $\dot{S}_{mat}$;
  \item [3.]  Zonostrophic flow (ZN): low $D$, low $MHT$, low $\eta$;
  \item [4.] Axisymmetric flow (AS): high $MHT$ and $D$ for $\mathcal{R}o> 1$, high $\eta$ for $1<\mathcal{R}o<0.1$, low  $D$, $MHT$ and $\eta$ for $\mathcal{R}o<0.01$.
  \end{itemize}

%    \textbf{maybe some bullets points as in the previous sub-section}
   
  % Finally let us note that the difference in the soil heat capacities do not significantly    alter the patterns of circulation regimes. The most relevant difference  in the circulation    is the   splitting of the Hadley cell for low $\tau$ at low $\Omega^*$ (Fig. \ref{circ2_bis}). Apart from this we observe  an overall weakening of the  general circulation in HIGHHC case. This is particularly evident in  the dissipation of kinetic energy (i.e. Lorenz energy cycle, Fig. \ref{kediss_oc}  and  Fig. \ref{kediss_rock}), that  in HIGHHC runs is about a factor $0.8$ smaller than in LOWHC, in the   meridional heat transport (factor $0.9$, Fig. \ref{meridional_oc} and   Fig. \ref{meridional_rock}) and in the efficiency (factor $0.6$, Fig. \ref{carnot_oc} and Fig. \ref{carnot_rock}).The decrease of the surface heat capacity is reflected in the decrease of the timescale needed to respond to fluctuations of surface  energy fluxes, e.g. due to the seasonal cycle. In the limit of  very small heat capacitythe surface temperature would adapt instantly to  such fluctuations,   whereas for infinite heat capacity,the surface temperature would not respond to it at all. This loss of positive temporal covariance between temperature and diabatic heating is reflected in a loss in the generation of available potential energy -- which in fact, in Lorenz  formulation \citep{EdLor60}, is given by the space and time  covariance between temperature and diabatic heating.

   \subsection{Implications for the Maximum Entropy Production Principle}
   \label{MEPP}

   In this section we briefly describe our results in the context of the Maximum Entropy Production Principle \citep[MEPP,][]{Pal, Pal78,Pal2}, as this conjecture has   gained some momentum also in the planetary science community \citep{Lorenz, TaylorBook}. MEPP has been used  as a closure condition for climatic toy-models \citep{Lorenz} or simple energy balance climate  models \citep[e.g.][]{Pal} in order to determine dynamical quantities as the meridional heat transport. A further, possible application was shown by \cite{Klei03} and \cite{Kunz}, who suggested   to use MEPP  as a guide  for tuning sub-grid motion parameters of PUMA, an atmospheric  general circulation models \citep{PUMA}.  For example, let us consider the Rayleigh drag constant   $\tau$ (eq. \ref{def_tau} and following discussion) depends on the drag coefficient $\gamma_h$ which in turn depends on both surface roughness and dynamical quantities. Therefore different values of $\tau$ can be thought of associated with either different surface properties (as done in the rest of the paper) or to different strengths of the turbulent transfer in the planetary boundary layer. Following the second interpretation,  \cite{ Klei03} showed that the value of  $\tau$ giving the most realistic atmospheric state was that maximizing the entropy production of the system. However, one  major criticism that MEPP has encountered  is  that it does not take into account the effects of the rotation speed \citep{Rodgers, Goody07, Jupp}. This was related to the criticisms on whether one could use MEPP to infer the meridional energy transport.
 % This issue has been considered   by \cite{Jupp} through a two-box model with very simple dynamical assumptions. For sufficiently small $\Omega$,  the revised model by \cite{Jupp}  reproduces to some extent   the  MEPP states obtained without taking  the planetary rotation rate into account. 
  %In this study we show that, within the MEPP conceptual  framework, we have possibility  to extend the results of \cite{Klei03}. 
  In this study we are in a position to have a broader look on  the results of \cite{Klei03}  since a  more detailed diagnostics for the dissipative properties and a larger dynamical range for atmospheric circulations are available. Of course our aim is not, and we do not claim, to prove or disprove MEPP, for which a rigorous demonstration is still missing \citep{De2,Grgr}.

    In order to test MEPP,  we perform control runs  in which the full boundary layer scheme \citep{Louis, Louis2} is  employed without  the simplification of Sect. \ref{simplification} (so $\tau$ is not prescribed but dynamically determined depending on the winds and vertical stability). In the following we shall refer to them and to   quantities evaluated for such simulations with the label ``BLS'' (boundary layer scheme).    In BLS simulations  the drag coefficient is consistently determined at each timestep and each grid-point according to the Monin-Obukhov theory \citep[e.g.][]{Arya} and not prescribed as a constant parameter.  Since this set up employes a more refined and realistic representation of the boundary layer physics, we consider it as our ``reality'' towards which   comparing simulations in which the rougher, tunable $\tau$-scheme is used. Zonal means of the BLS simulations are shown in Fig. \ref{circ1_contr} --  cross sections of temperature and zonal winds -- and in Fig.\ref{circ2_contr} --  meridional  stremfunctions -- for  simulations for $\Omega^*=1/10, 1, 8$ respectively.
 For each $\Omega^*$, we consider $\tau$ as a tunable parameter and select   the value $\tau_{max}(\Omega^*)$  maximising $\dot{S}_{mat}$ (which can be easily visualized in Fig. \ref{entropybudget3}). Furthermore,  we take into account also  $\dot{S}_{kediss}$ (Fig. \ref{entropybudget2}), so that we can be informative also on the maximum dissipation principle \citep{Lor67,Ozawa, Schu, Pascale2}. We denote with  $\tilde{\tau}_{max} (\Omega^*)$  the values of $\tau$ maximising $\dot{S}_{kediss}$. As can be seen in Fig. \ref{entropybudget2}-\ref{entropybudget3}, $\tau_{max}$ and $\tilde{\tau}_{max}$ differ
mostly for $\Omega^* \le 1/2$ (where the maximum dissipation steady states occur for $\tau$ of few hours) whereas they are mostly the  same ($1$ day) for $\Omega^*>2$ days ($\tau\approx 1$ day).

In Fig. \ref{mep1} and \ref{mep2} we compare $\dot{S}_{mat}(\Omega^*; \tau_{max})$ and $\dot{S}_{kediss}(\Omega^*; \tau_{max})$ (dashed line) with $\dot{S}_{mat}^{BLS}(\Omega^*)$ and $\dot{S}^{BLS}_{kediss}(\Omega^*)$ respectively (continuous lines). On the same diagrams we also show the same quantities for $\tau=0.1\,\tau_{max}(\Omega^*)$ (dotted line) and $\tau=10\,\tau_{max}(\Omega^*)$ (dotted-dashed line) in order to provide an indication of the sensitivity of $\dot{S}_{mat}$ and $\dot{S}_{kediss}$ with respect to $\tau_{max}$. The MEPP estimate of $\dot{S}_{mat}$ slightly overestimate the values obtained in controls runs ($\le 5\%$)  but, impressively,  captures fairly well the dependence on $\Omega^*$.   Similarly, the values of $\dot{S}_{kediss}$  obtained for $\tau_{max}$  compare relatively well with the  ones obtained in the controls runs. Circulations corresponding to $\tau_{max}$  are indeed fairly similar to BLS circulations, as can be seen by comparing   Fig. \ref{circ1_contr}(a,b,c)  with Fig. \ref{circ1_t42}(b,e,h)   and Fig. \ref{circ2_contr}(a,b,c)  with Fig. \ref{circ2_t42}(b,e,h).

When the values of $\tilde{\tau}_{max}(\Omega^*)$ associated with  the maximum of $\dot{S}_{kediss}$ is instead taken into account  (Fig. \ref{mep3}-\ref{mep4}),  we observe that   $\dot{S}_{mat}(\Omega^*, \tilde{\tau}_{max})$  provides again a  quite good estimate of   $\dot{S}_{mat}^{BLS}$, with a slight underestimate ($\approx 9\%$) for $\Omega^*<1/2$, due to the fact that  for such values of the rotation rate  $\tilde{\tau}_{max}$ bends towards smaller $\tau$ where $\dot{S}_{mat}$ tends to decrease (Fig. \ref{entropybudget3}). More unsatisfactory is  $\dot{S}_{kediss}(\Omega^*, \tilde{\tau}_{max})$ again for $\Omega^*<1/2$, with a difference of about $16\%$ with respect to $\dot{S}_{kediss}^{BLS}$. 

In the end, both maximum entropy production and maximum dissipation principle provide fairly reasonable estimates of $\dot{S}_{kediss}^{BLS}$ and $\dot{S}_{mat}^{BLS}$, with the maximum entropy production one having better skills  at low $\Omega^*$. The quasi-equivalence of the the two methods is due to the fact that, for such simulations,  both $\dot{S}_{mat}$ and $\dot{S}_{kediss}$  have their maxima in the $(\Omega^*, \tau)$ almost in the same regions. These results seem to confirm, in a relatively large range of dynamical regimes, the possibility  of using MEPP in its weak form, as a a guide for tuning  sub grid parameters associated with turbulent motions, as indicated by \cite{Klei03}.

\section{Conclusions}
\label{discussione}

Stimulated by the   ongoing development of exoplanet sciences,   in this study we have  investigated  the nonequilibrium thermodynamic  properties (kinetic energy dissipation, material entropy production, efficiency, meridional heat transport) of  optically-thin, non-condensing   planetary atmospheres  at different values of the thermal Rossby number $\mathcal{R}o$ and the Taylor number $\mathcal{F}_f$ through a systematic variation of the rotation rate $\Omega$ and surface drag time constant $\tau$.  The most relevant achievement of this study  has been the  characterization of  the nonequilbrium properties of the different  circulation regimes (axisymmetric, super-rotation, baroclinic, barotropic, zonostrophic) obtained  with  numerical simulations with some interesting connection to the Maximum Entropy Production Principle (MEPP).

Slowly rotating planets ($\mathcal{R}o>1$) circulation are mostly   Hadley cell-dominated but tend to equator; super-rotation for $\mathcal{F}_f>10^5$.  For intermediate rotation rates ($1<\mathcal{R}o<0.01$) an axisymmetric ($\mathcal{F}_f<10$), baroclinic ($10<\mathcal{F}_f<10^5$) and barotropic ($\mathcal{F}_f>10^5$) regime are found.
 At high rotation rates ($\mathcal{R}o<0.01$) circulations are characterized by multiple jets (zonostrophic) for $\mathcal{F}_f>10^4$.
 
The   baroclinic regime has high values of  $D$ and  $MHT$ since midlatitude baroclinic waves provide  a very effective way to convert available potential energy into mechanical kinetic energy and transport energy from low to high latitudes. Such mechanism is inhibited by strong barotropic shears characterizing the barotropic regime and therefore both $D$ and $MHT$ experience lower values. 
The  axisymmetric regime has different thermodynamic properties depending on the value of $\mathcal{R}o$ at which it is realised. For  $\mathcal{R}o> 1$,  a very intense Hadley cell develops associated with  high $MHT$ and $D$; for  $1<\mathcal{R}o<0.1$ such quantities are weaker but circulations are more efficient in converting heat into mechanical work (high $\eta$); at faster rotation speeds
  ($\mathcal{R}o<0.01$) a dramatic drop in $D$, $MHT$ and $\eta$ is observed.
   A very interesting case is that of circulation approaching equatorial    super-rotation ($\mathcal{R}o\le 10$, $\mathcal{F}_f>10^5$), for which  low $D$, low $\eta$, low $\dot{S}_{mat}$ occurs, thus showing a  behavior close to  inviscid, non-dissipative fluids (for which $D=0$ and $\dot{S}_{sens}=0$ by definition).
    Zonostrophic flows  low, typical of fast rotating, low surface drag planets, have a very weak atmospheric energy cycle (low $D$), are very  inefficient in converting potential energy into work  and have very low meridional heat transport  $MHT$, therefore showing a temperaure profile  close to the radiative-convective equilibrium (which by definition has $MHT=0$).
%   Axisymmetric flow (AS): high $MHT$ and $D$ for $\mathcal{R}o> 1$, high $\eta$ for $1<\mathcal{R}o<0.1$, low  $D$, $MHT$ and $\eta$ for $\mathcal{R}o<0.01$.

The thermal dissipation $\dot{S}_{sens}$ is instead fairly insensitive to $\mathcal{R}o$ and is determined mainly by 
the timeconstant $\tau$, due to a trade-off mechanism between the temperature difference and
the heat flux. 
%This also determines the properties of the total material entropy production, since 
%the contribution due to the heat flux is always larger at least of a factor $3$.
 
%Let us also note the effect of the two different kinds of surfaces we have considered: HIGHHC, with ocean and albedo heat capacity, and LOWHC, with albedo and heat capacity typical of rock. In spite of similar  dissipative properties, we observe that LOWHC has generally a more intense circulation and larger Carnot  efficiency.  The weakening of the circulation is associated with the  loss of positive temporal covariance between temperature and diabatic heating, which in turn  is reflected into a loss in the generation of available potential energy.  Also a curious feature in the circulation is the   splitting of the Hadley cell for low $\tau$ at low $\Omega^*$ (Fig. \ref{circ2_bis}).

Moreover, we have shown that the possibility of applying MEPP in its weak form, e.g. as a tool for providing guidance in tuning subgrid scale, seems to work relatively well in the range of values of the rotation rate considered in this study, thus extending the results obtained by  \cite{Klei03} when considering the terrestrial rotation rate only. Interestingly, there is broad agreement between what prescribed by applying MEPP and the maximum dissipation principle.

This is a first preliminary study for a special case of dry atmosphere. The presence of the hydrological cycle has a huge effect on the circulation and on the energetics and would be definitely worth investigating.  Another issue is the role of the surface heat capacity, which would also deserve a systematic investigation.   Furthermore,  thermodynamic and dynamical properties of slowly rotating planets, e.g. from $\Omega^*=1/10$ up to phase-locked planets, are still poorly known and would deserve more investigation too. 

\paragraph{Aknowledgments}
The authors thank  S. Ehrenreich,  K. Fraedrich, N. Iro, E. Kirk, J. Lloyd,  F. Lunkeit, R. Plant and P. Read  for their helpful and useful comments.  This work was supported by the EU-FP7 ERC grant NAMASTE. SP, VL,  FR and RB acknowledge the support of CLISAP. We thank the two anonymous referees for their insightful comments  which lead to a significant improvement of the manuscript.

\clearpage

%\begin{table}[ht]
%\centering
%\begin{tabular}{l l ll l  }
%\hline
%\hline
%     $\Omega^*$                & $\mathcal{R}o$      &    $\max(\mathcal{F}_f)$ & $\min(\mathcal{F}_f)$ \\
%                     \hline
%$1/10$       & $7.98$                                                &  $1.5\times 10^{-3} $ &   $3.9\times 10^{5}$\\
%$1/5$         & $1.99$                                                 &   $6.2\times 10^{-3}$ &   $1.5\times 10^{6}$\\
%$1/2$         & $0.31$                                                &  $3.8 \times 10^{-2}$ &   $9.9\times 10^{6}$\\
%$1$            & $0.08$                                                 &  $0.15$                        &   $ 3.9\times 10^{7}  $\\
%$2$            & $1.9\times 10^{-2}$                           &  $0.62$                        &   $1.6\times 10^{8}$\\
%$4$            & $4.9\times 10^{-3}$                          &  $2.5$                          &    $6.3\times 10^{8}$\\             
%$8$            &  $1.2\times 10^{-3}$                         &  $9.9$                          &    $2.5\times 10^{9}$\\
%\hline
%\hline
%\end{tabular}
%\caption{Range of variation for $\mathcal{R}o$ and $\mathcal{F}_f$. For each value of the rotation rate the minimum and maximum value of $\mathcal{F}_f$   (corresponding to the smallest and largest value of $\tau$, see definitions (\ref{thermal_rossby2}) and (\ref{friction})) are shown.   A value of $\Delta\theta_h=\Delta\theta_{hE}\approx 60$ K  has been chosen. \label{tabella1}}
%\end{table}

\begin{table}[ht]
\centering
\caption{Parameters and symbols list \label{tabella1_bis}}
\begin{tabular}{l l ll l  }
\hline
     parameter/symbol                & explanation       &    value \\
                     \hline
         $\Omega_E$            &   Earth's rotation rate  & $7.29\cdot10^{-5} $ rad$^{-1}$\\
  $ c_{d}$    &     specific heat of dry air             &  1004 J\,kg$^{-1}$K$^{-1}$\\
   $ c_{pw}$   &      specific heat of mixed layer model            &   $4180$ J\,kg$^{-1}$K$^{-1}$ \\
   $g$          &  gravitational  acceleration &  $9.81$ m\,s$^{-2}$ \\ 
    $\rho_w$   & ocean water density                   &  $1030$ kg\,m$^{3}$   \\
 $h_{ml}$      &       mixed layer depth           &  $5$ m\\
$C_{slab}$       &      slab-ocean areal heat capacity &     $10^{-7}$ J\,K$^{-1}$ m$^{-2}$\\
$\alpha_s$       &       surface albedo          &  0.2\\
 $S_0$      &       solar constant          &  $1365$ W\,m$^{-2}$\\
 $a$ &  planet's radius & $6300$ km \\
 $\mathcal{R}o$          &   thermal Rossby number        &          \\
 $\mathcal{F}_f$ & ``frictional''  Taylor number& \\
 $ASR$ & absorbed stellar radiation at TOA\\
 $OLR$ & outgoing long wave radiation at TOA\\
$F_T$ & surface sensible heat flux &\\
 $F_{SW}^{toa}$ & & \\
  $F_{SW}^{surf}$ & & \\
  $F_{LW}^{-}$  & & \\
 $\gamma_h$ &   heat transfer coefficient & \\
  $\gamma_D$ &   drag coefficient & \\
 $MHT$    &    meridional heat transport index&\\
 $L_R$ & Rossby deformation radius & \\
 $N$ & buoyancy frequency & \\
 $\alpha$  & irreversibility parameter & \\
 \hline
\hline
\end{tabular}
\end{table}

\clearpage

\section*{Figures' captions}
\begin{figure}
 \centering
   \subfigure[]{      
     \includegraphics[angle=-90, width=0.85\textwidth]{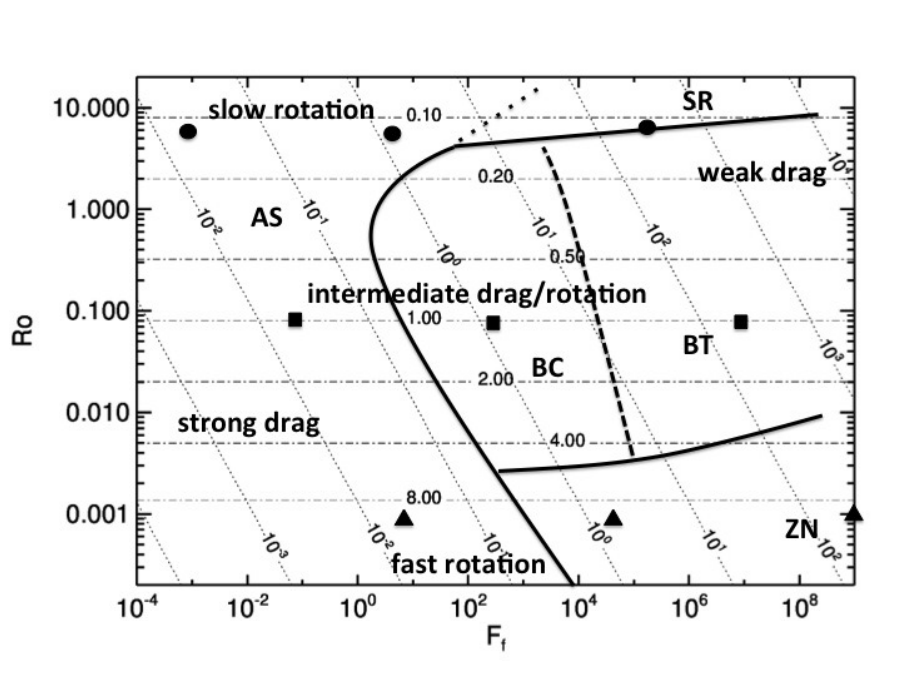}
    \label{diag2}
   }
   \subfigure[]{
    \includegraphics[angle=-90, width=0.85\textwidth]{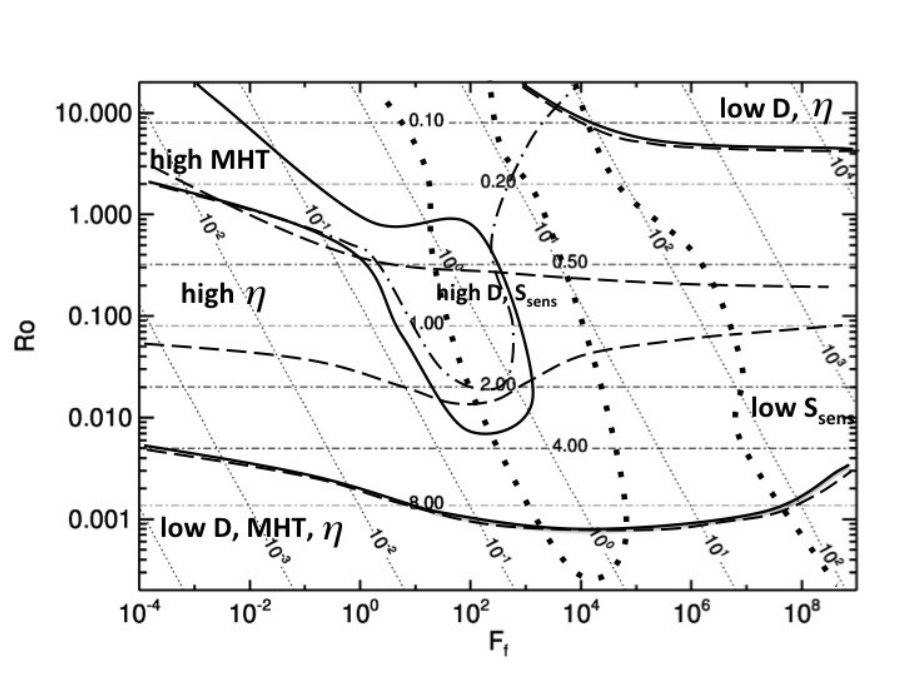}
     \label{diag1}
      } 
\caption{  \label{diags}}
\end{figure}

\begin{figure}
 \centering
\subfigure[$\mathcal{F}_f=1.5\times 10^{-3}$, $ \mathcal{R}o=8 $]{
     \includegraphics[angle=-0, width=0.3\textwidth]{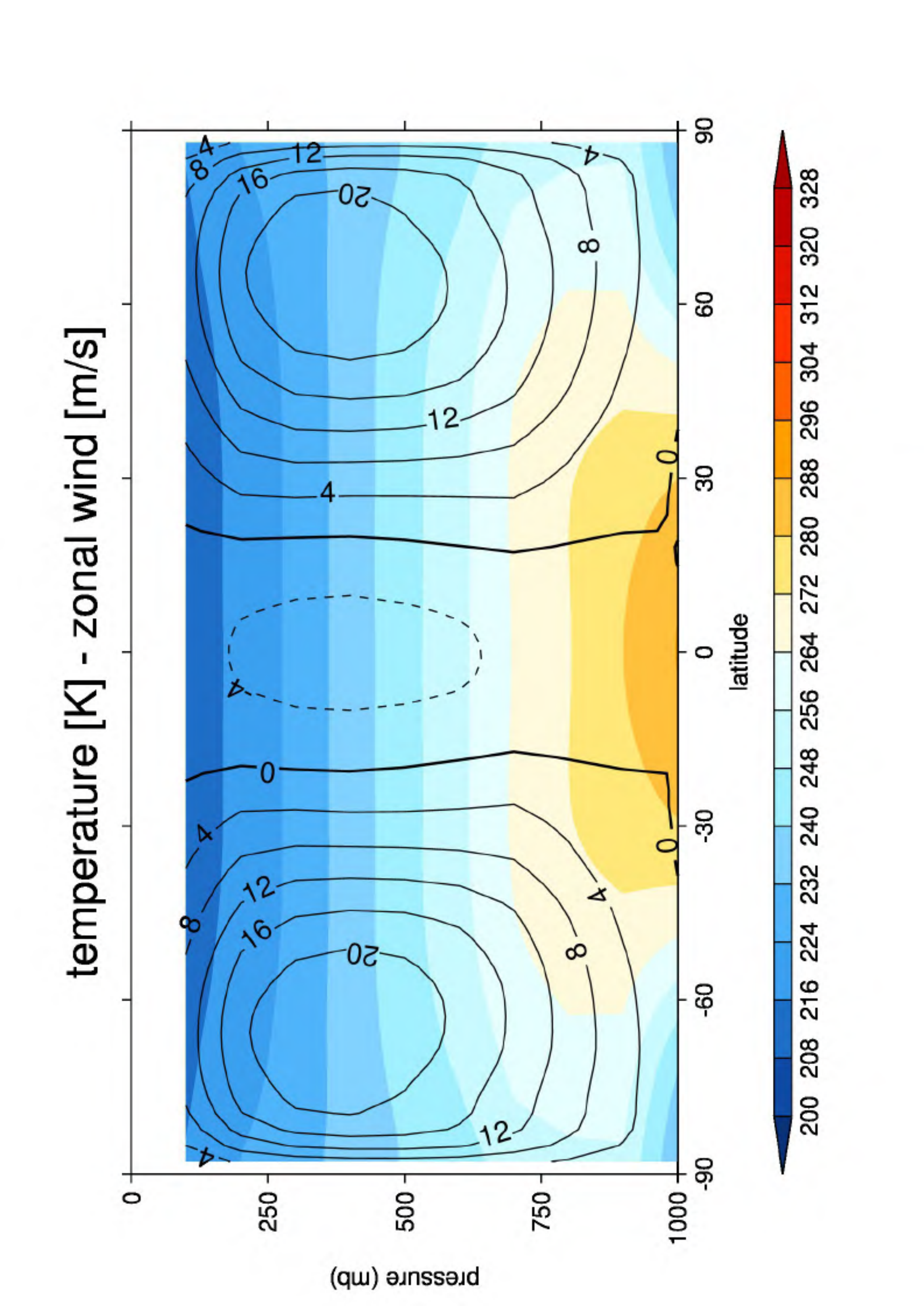}
    \label{circ1_t42_a}
   }
   \subfigure[$\mathcal{F}_f=1$, $\mathcal{R}o=8$]{
    \includegraphics[angle=-0, width=0.3\textwidth]{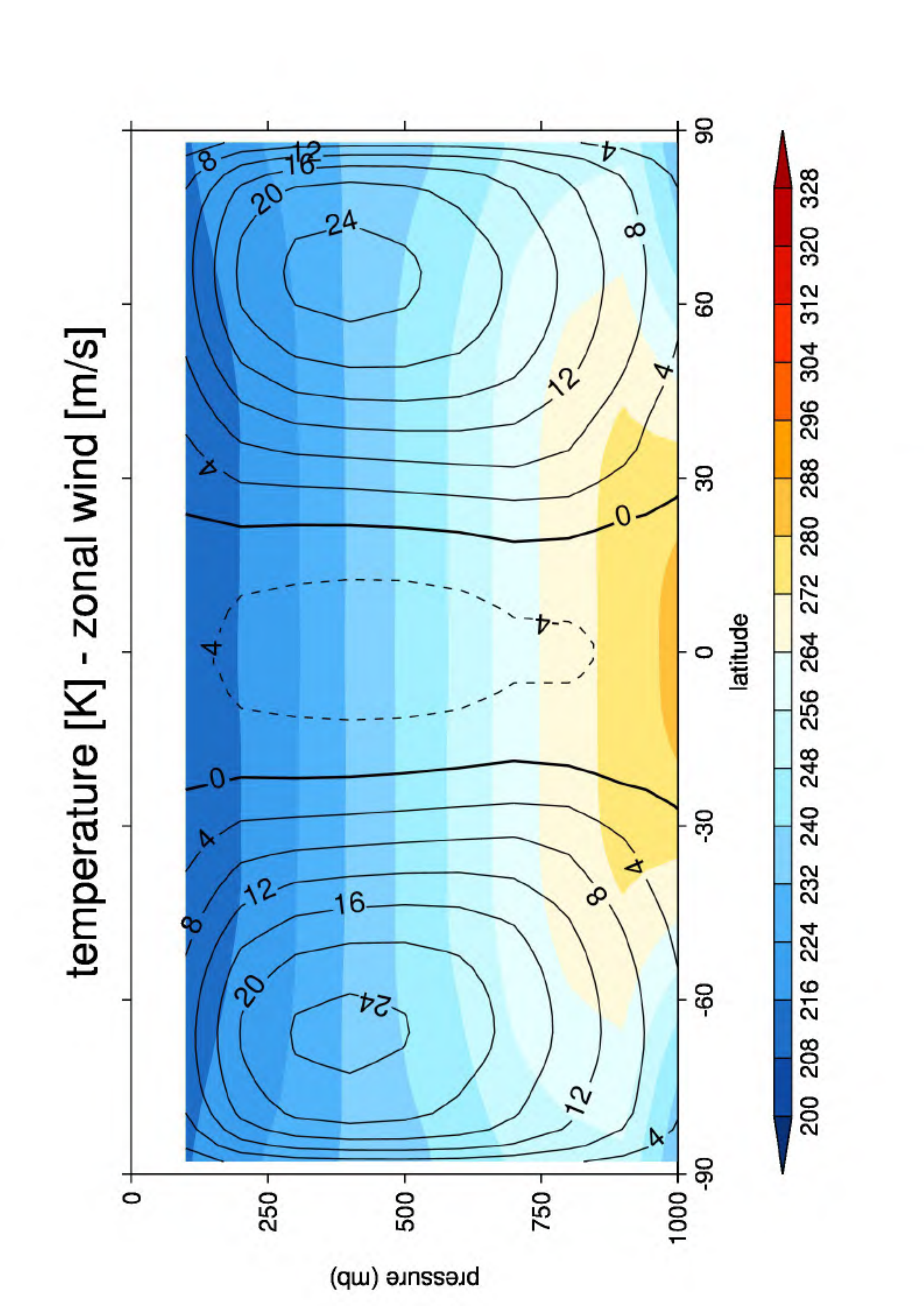}
    \label{circ1_t42_b}
     } 
   \subfigure[$\mathcal{F}_f=4\times 10^{5}$, $\mathcal{R}o=8$]{      
     \includegraphics[angle=-0, width=0.3\textwidth]{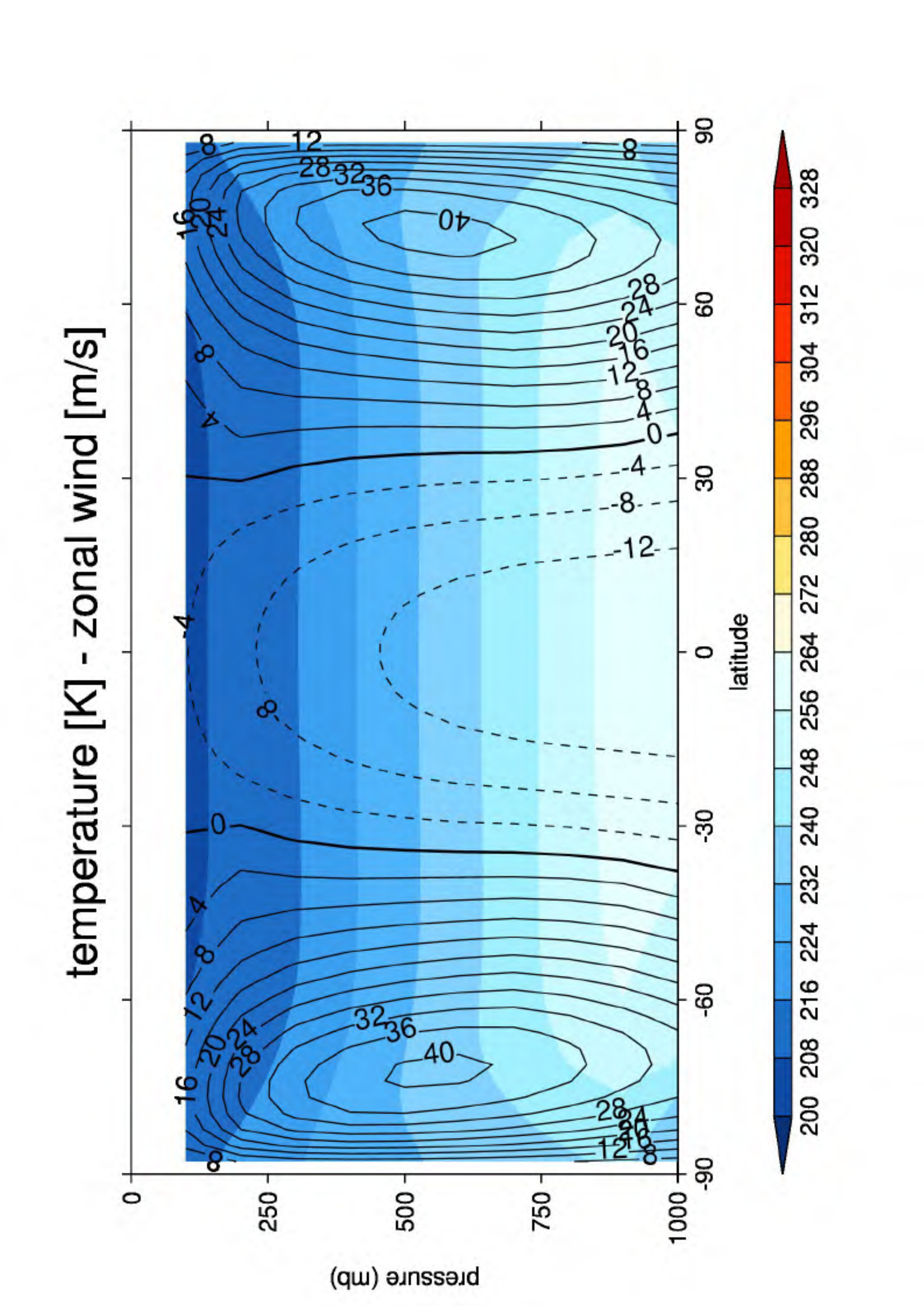}
    \label{circ1_t42_c}
   }
   \subfigure[$\mathcal{F}_f= 10^{-1}$, $\mathcal{R}o=0.08$]{
    \includegraphics[angle=-0, width=0.3\textwidth]{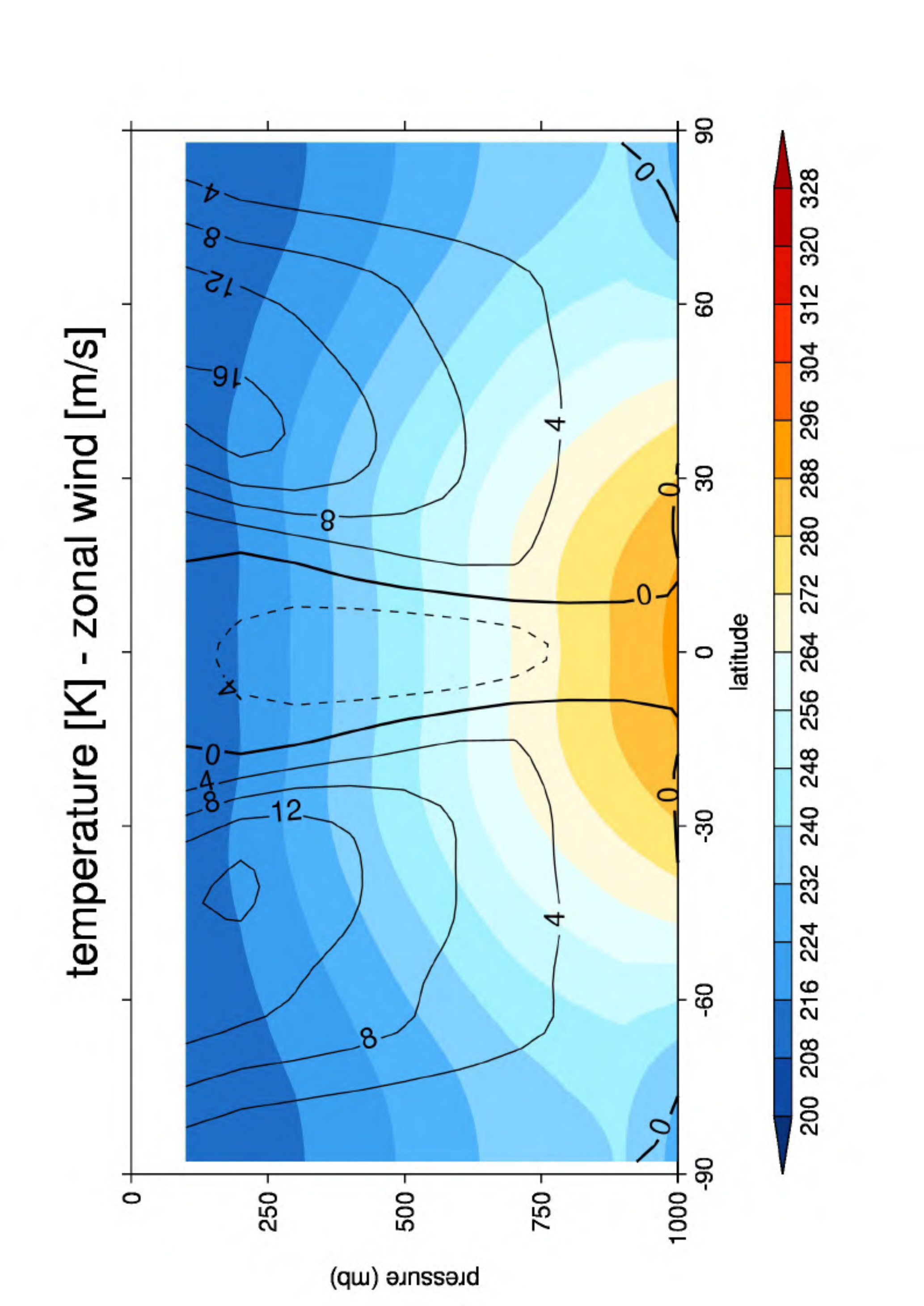}
     \label{circ1_t42_d}
      } 
 \subfigure[$\mathcal{F}_f=10^{2}$, $\mathcal{R}o=0.08$ ]{
    \includegraphics[angle=-0, width=0.3\textwidth]{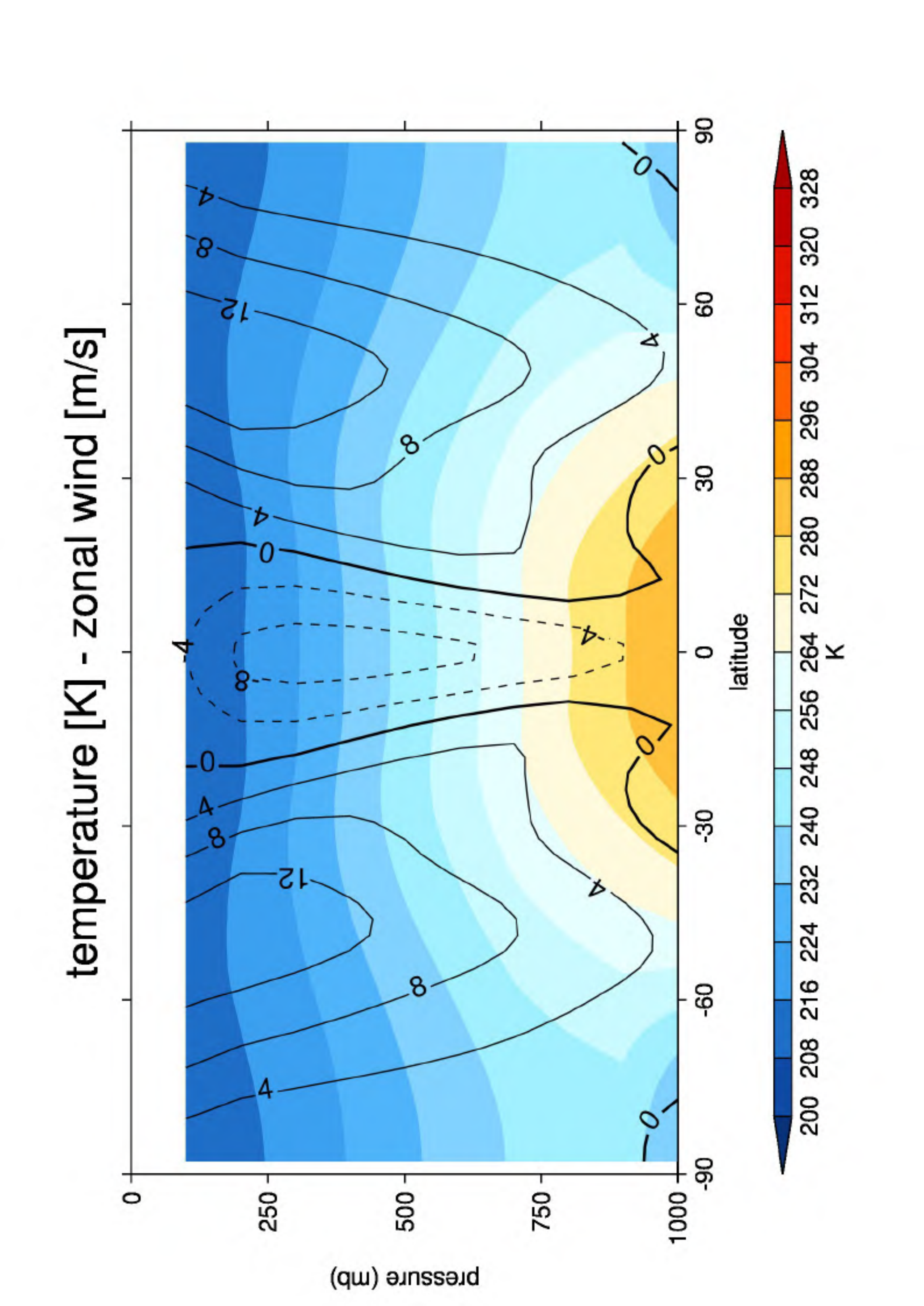}
    \label{circ1_t42_e}
   }
   \subfigure[$\mathcal{F}_f=4\times 10^{5}$, $\mathcal{R}o=0.08$]{
    \includegraphics[angle=-0, width=0.3\textwidth]{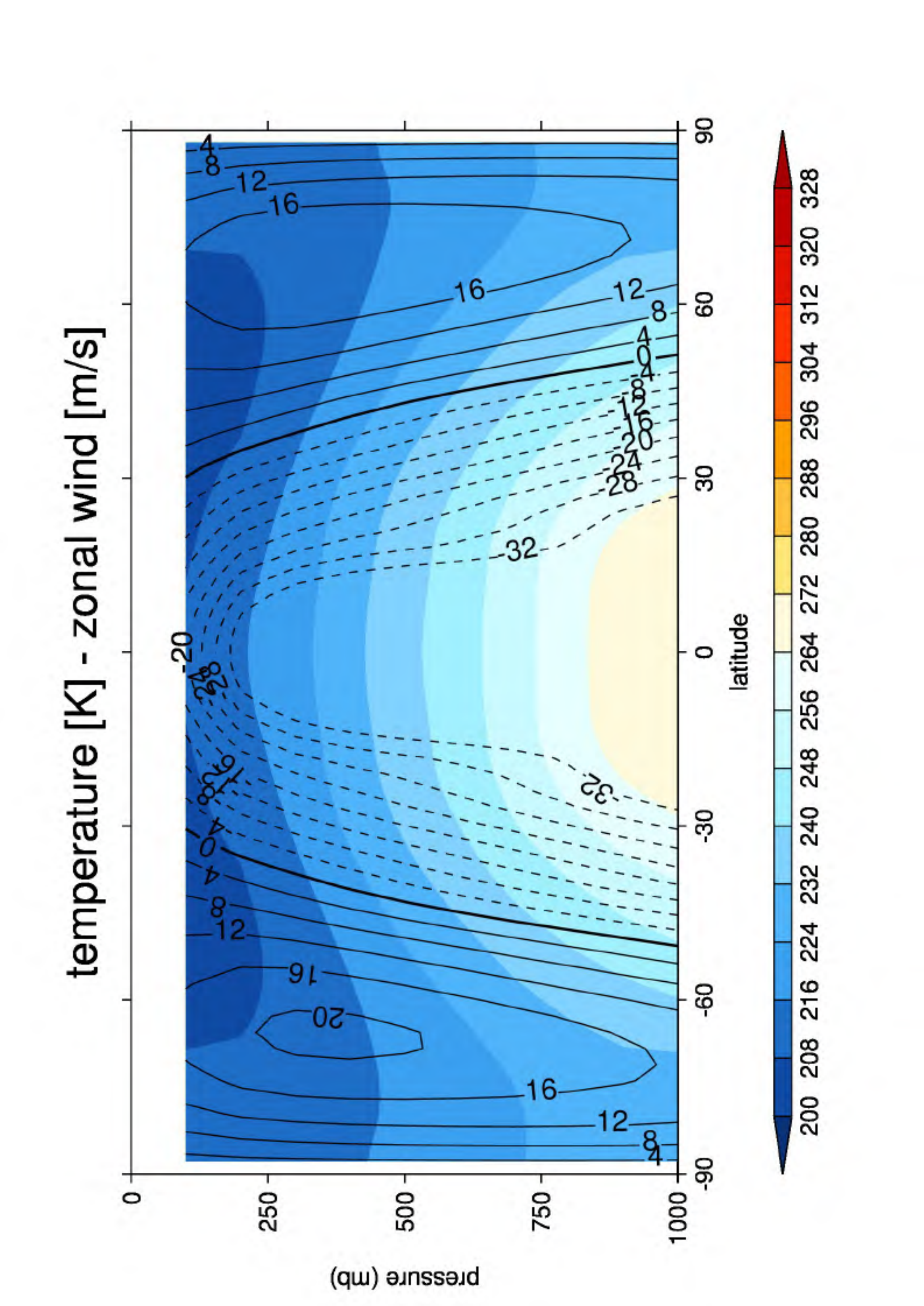}
     \label{circ1_t42_f}
      } 
 \subfigure[$\mathcal{F}_f=10$, $\mathcal{R}o=10^{-3}$]{
     \includegraphics[angle=-0, width=0.3\textwidth]{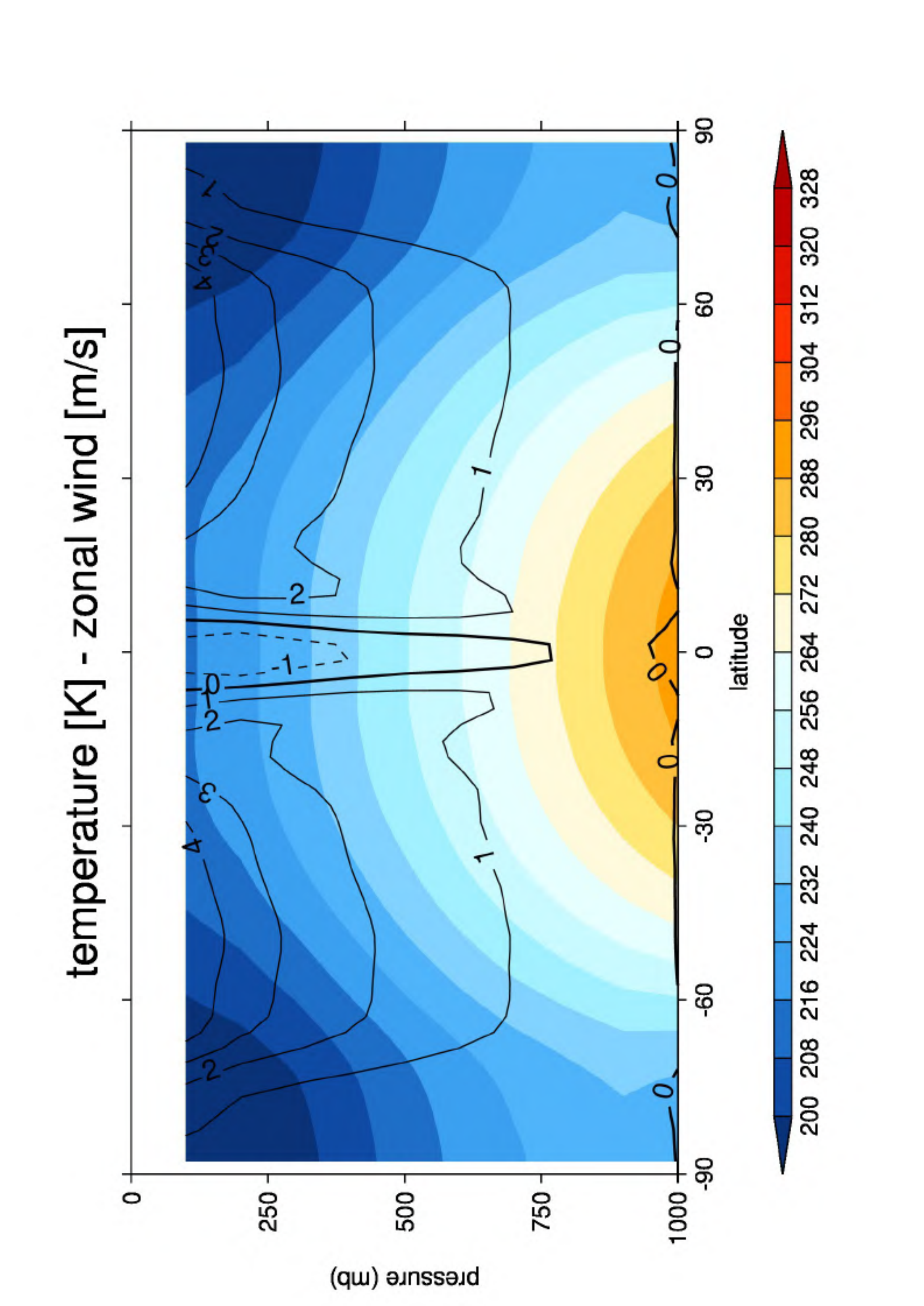}
    \label{circ1_t42_g}
   }
   \subfigure[$\mathcal{F}_f= 10^{4}$, $\mathcal{R}o=10^{-3}$]{
    \includegraphics[angle=-0, width=0.3\textwidth]{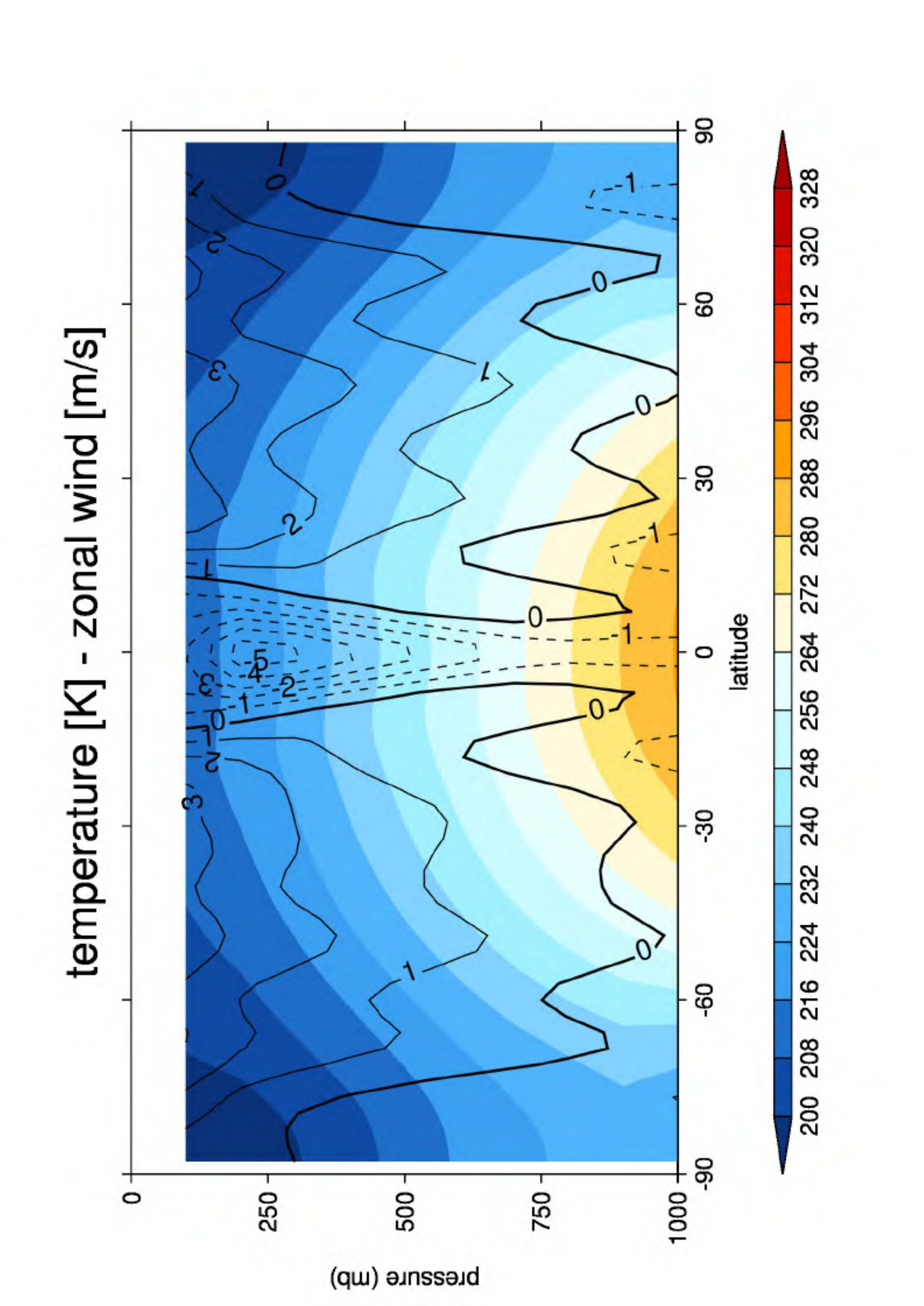}
    \label{circ1_t42_h}
     } 
   \subfigure[$\mathcal{F}_f= 10^{9}$, $\mathcal{R}o=10^{-3}$]{      
     \includegraphics[angle=-0, width=0.3\textwidth]{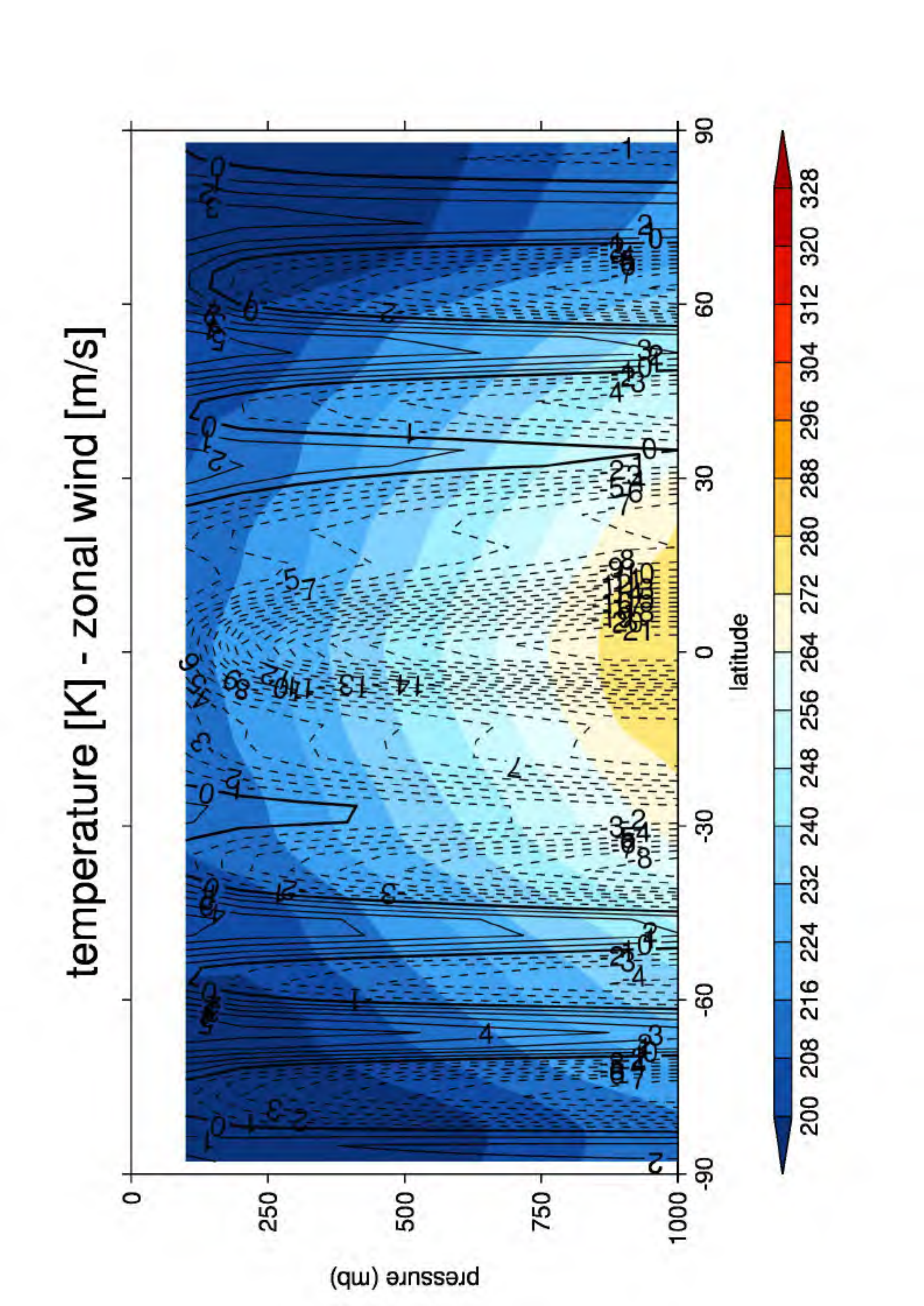}
    \label{circ1_t42_i}
   }     
\caption{    \label{circ1_t42}}
\end{figure}

\begin{figure}
 \centering
\subfigure[$\mathcal{F}_f=1.5\times 10^{-3}$, $ \mathcal{R}o=8 $]{
     \includegraphics[angle=0, width=0.3\textwidth]{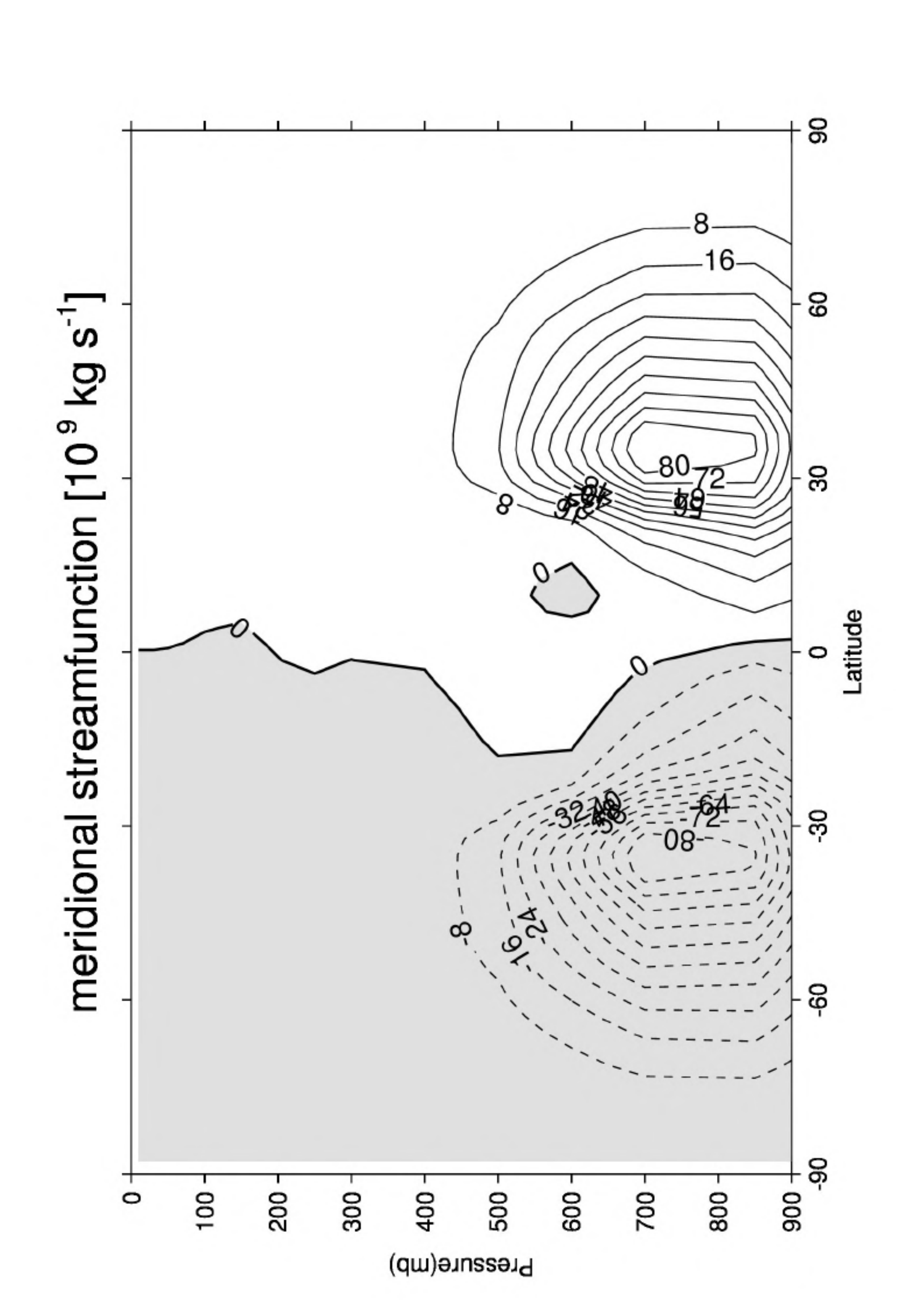}
    \label{circ2_t42_a}
   }
   \subfigure[$\mathcal{F}_f=1$, $\mathcal{R}o=8$]{
    \includegraphics[angle=0, width=0.3\textwidth]{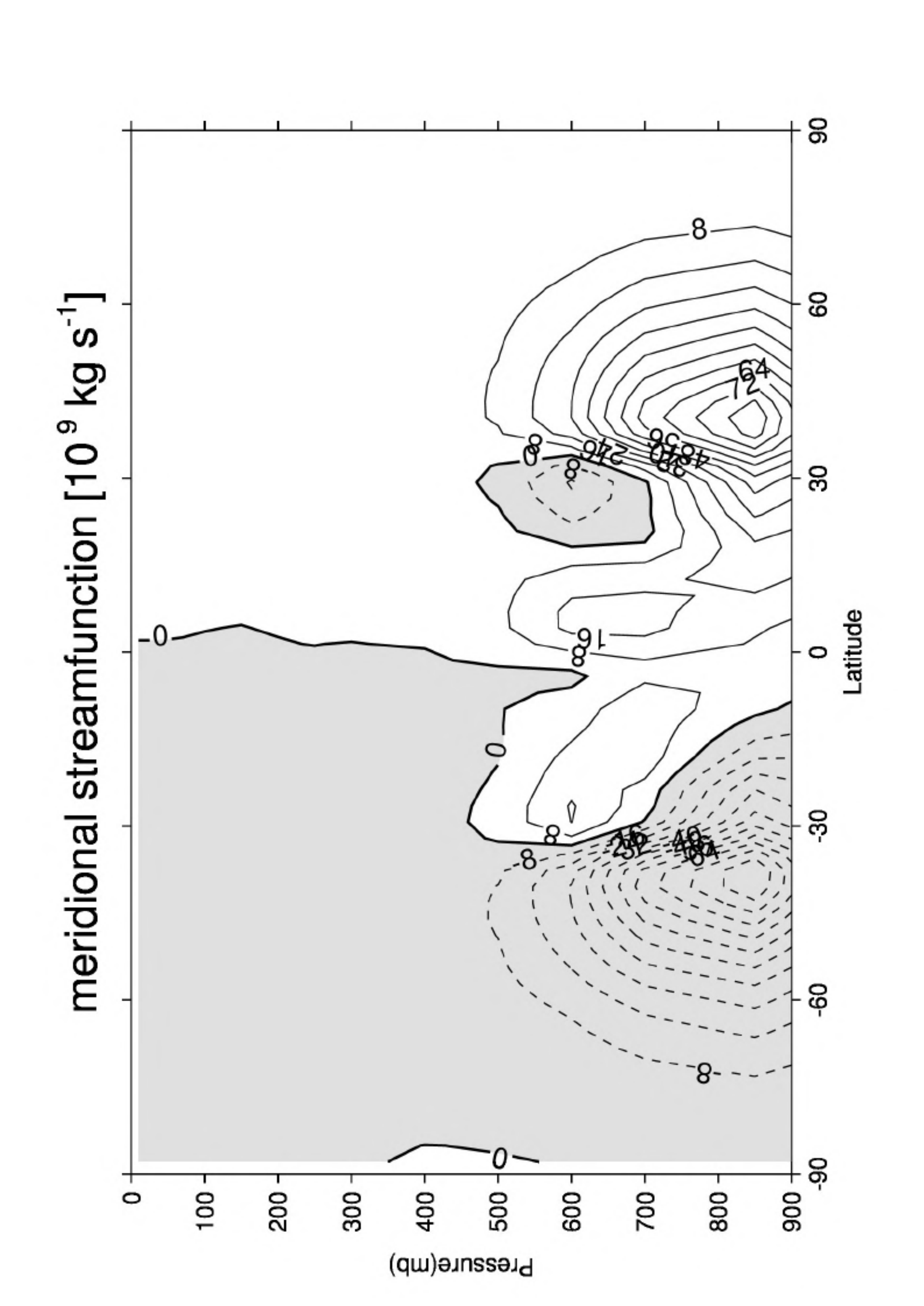}
    \label{circ2_t42_b}
     } 
   \subfigure[$\mathcal{F}_f=4\times 10^{5}$, $\mathcal{R}o=8$]{      
     \includegraphics[angle=0, width=0.3\textwidth]{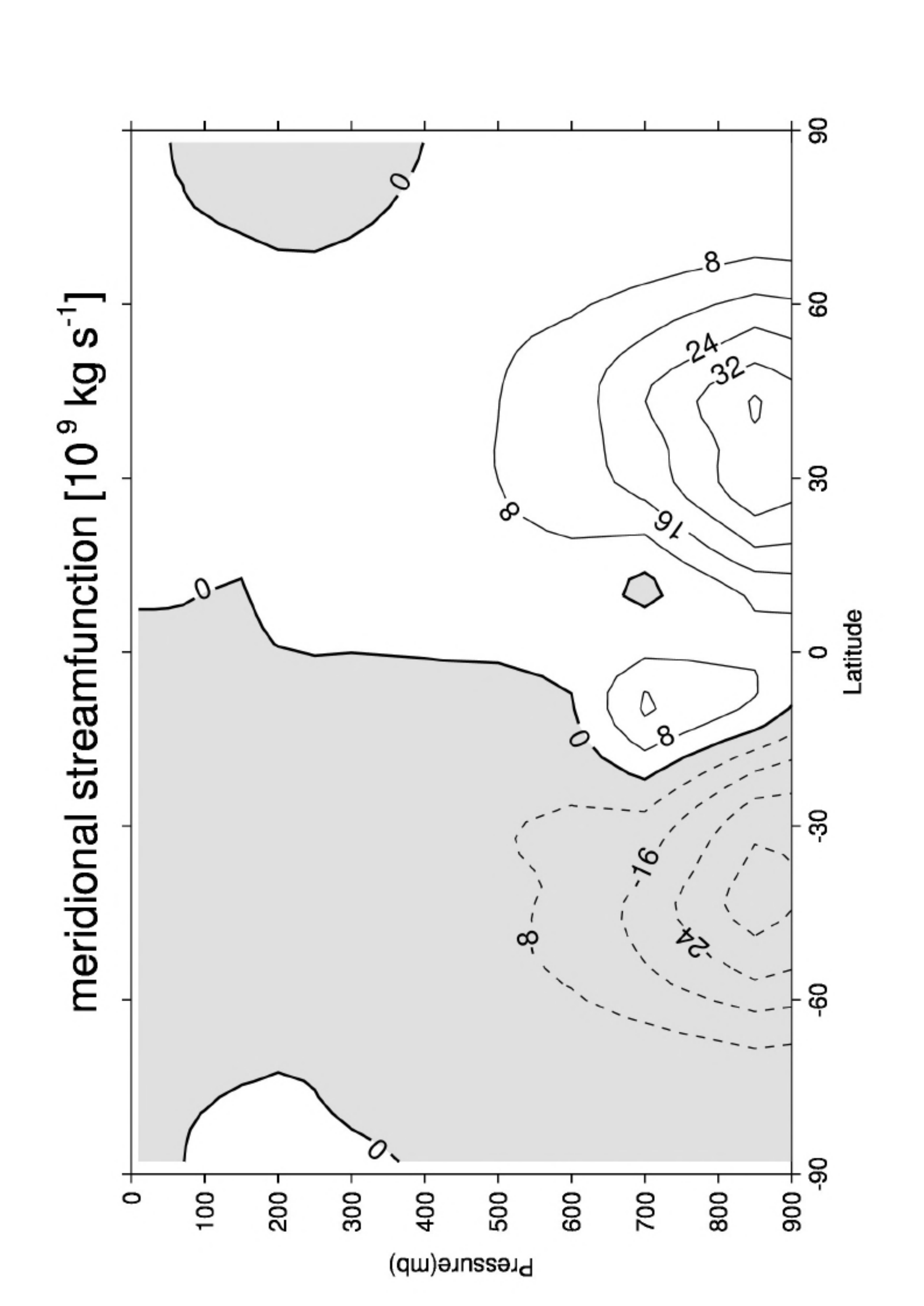}
    \label{circ2_t42_c}
   }
   \subfigure[$\mathcal{F}_f= 10^{-1}$, $\mathcal{R}o=0.08$]{
    \includegraphics[angle=0, width=0.3\textwidth]{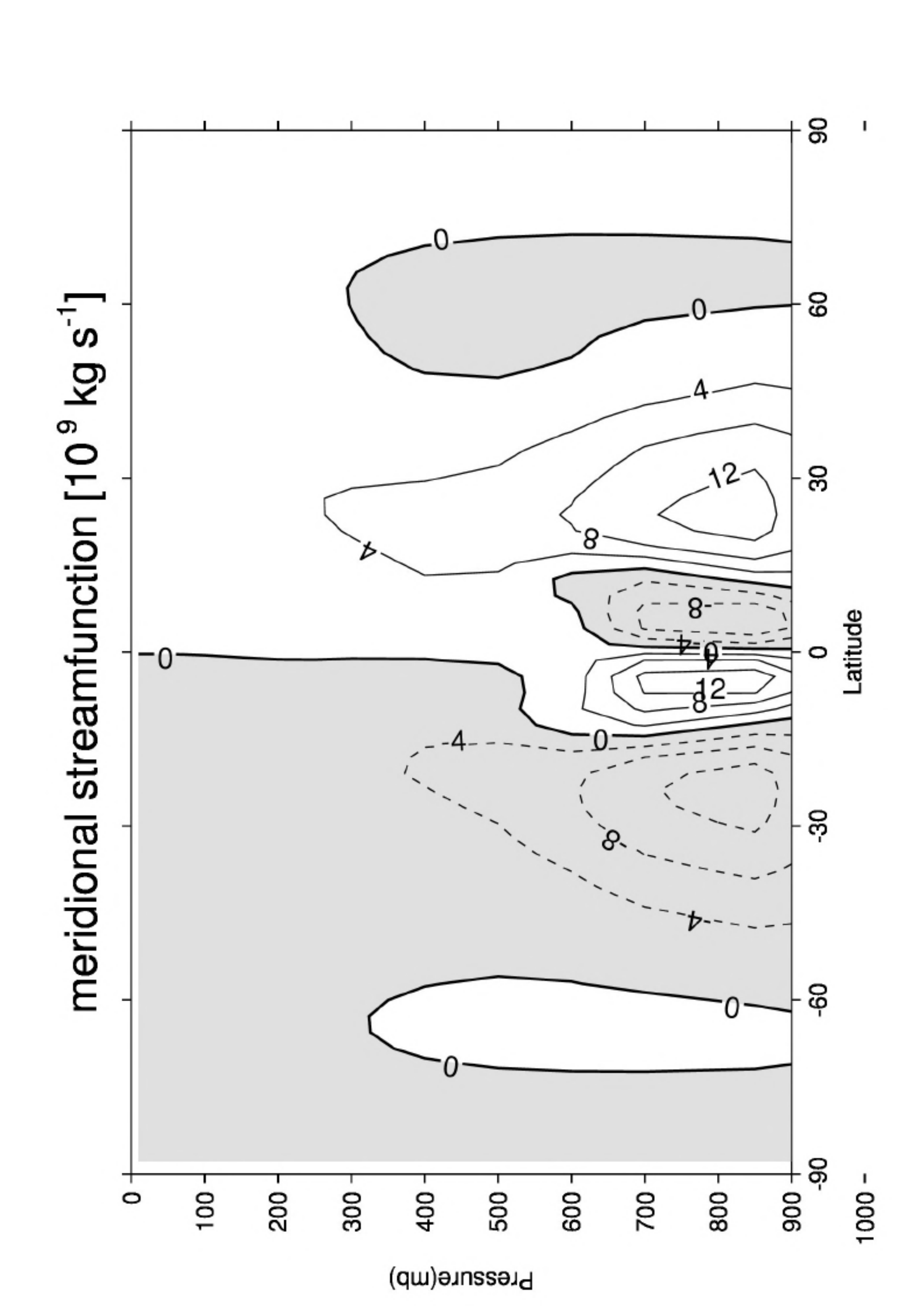}
     \label{circ2_t42_d}
      } 
 \subfigure[$\mathcal{F}_f=10^{2}$, $\mathcal{R}o=0.08$]{
    \includegraphics[angle=0, width=0.3\textwidth]{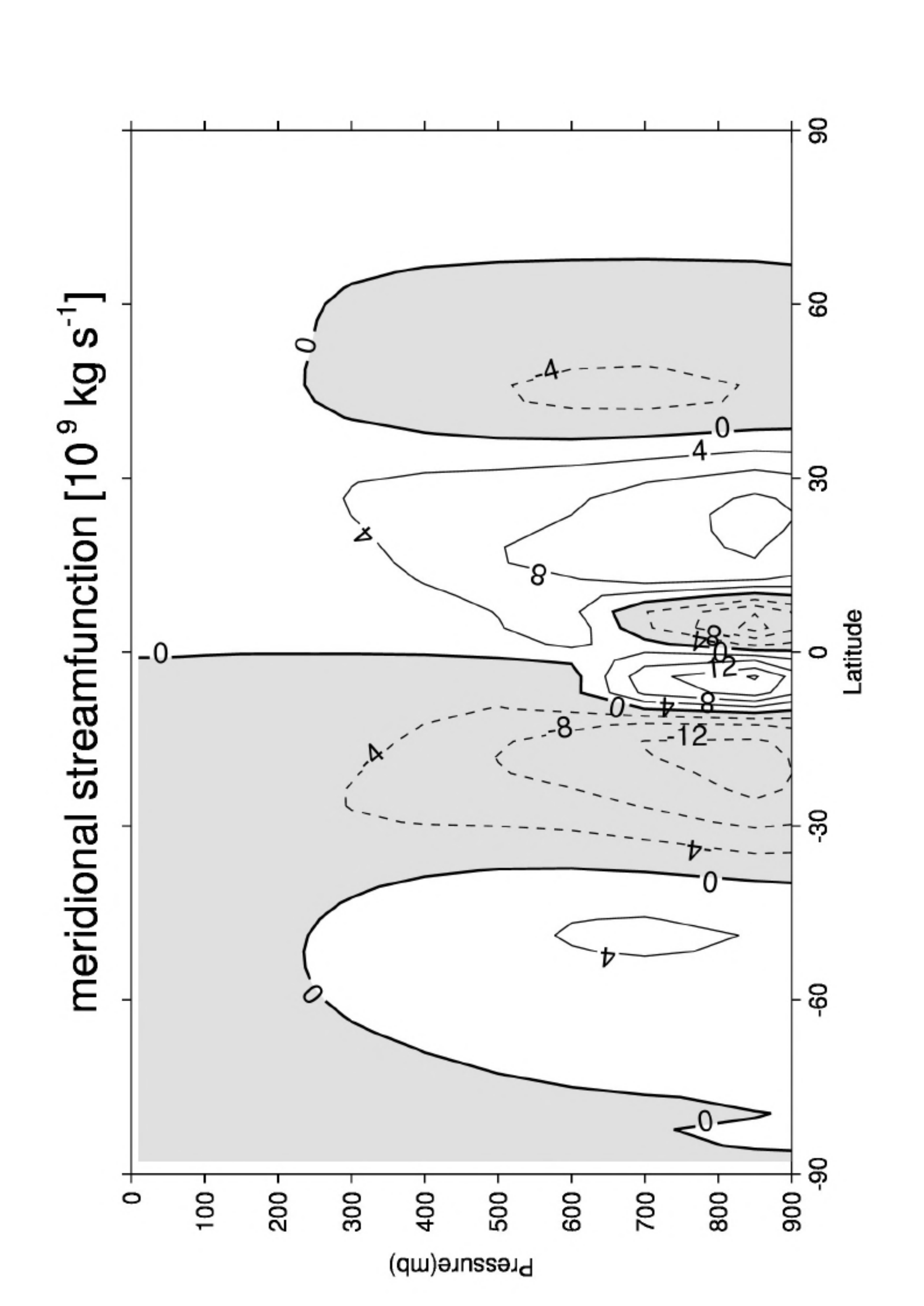}
    \label{circ2_t42_e}
   }
   \subfigure[$\mathcal{F}_f=4\times 10^{5}$, $\mathcal{R}o=0.08$]{
    \includegraphics[angle=0, width=0.3\textwidth]{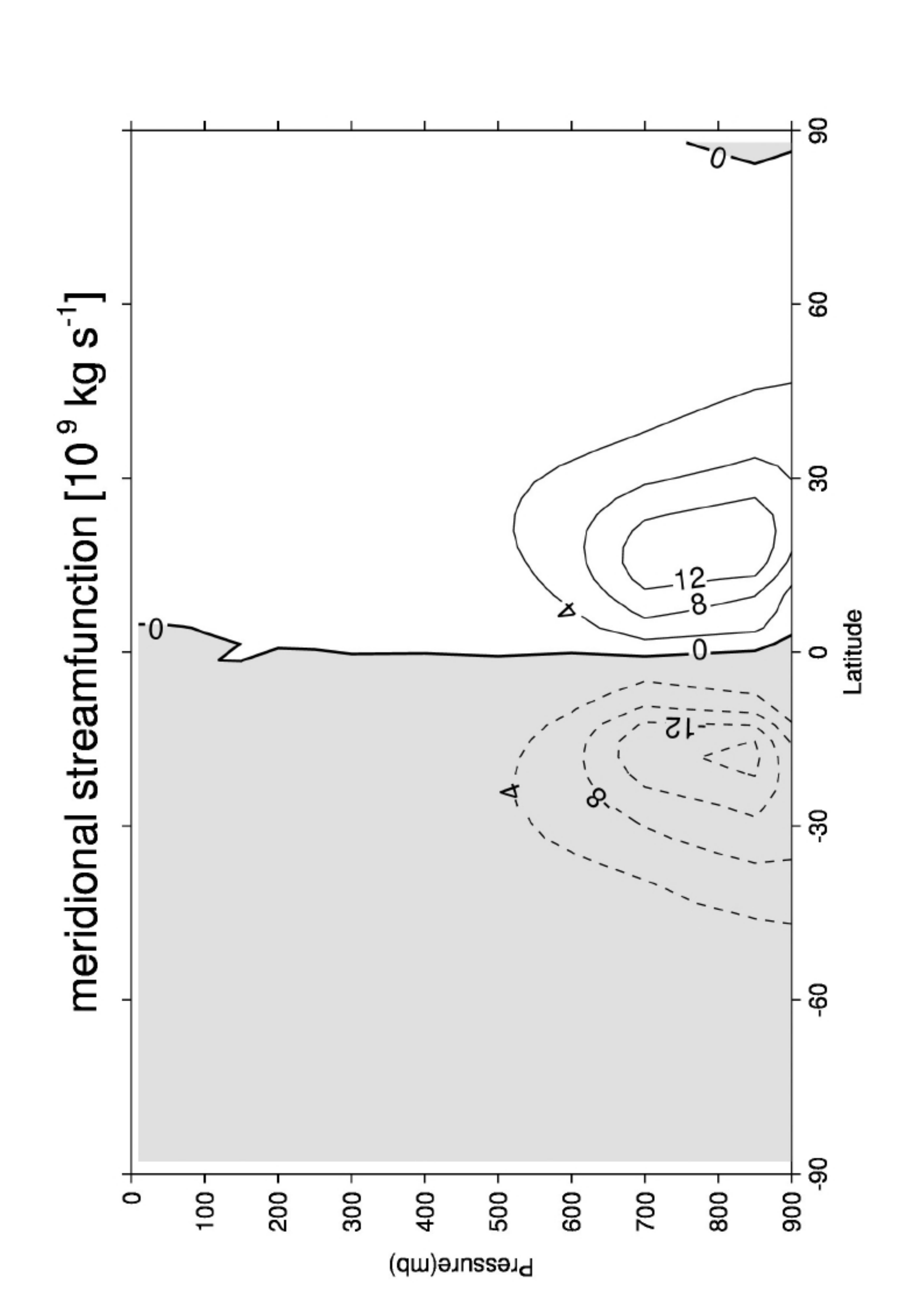}
     \label{circ2_t42_f}
      } 
 \subfigure[$\mathcal{F}_f=10$, $\mathcal{R}o=10^{-3}$]{
     \includegraphics[angle=0, width=0.3\textwidth]{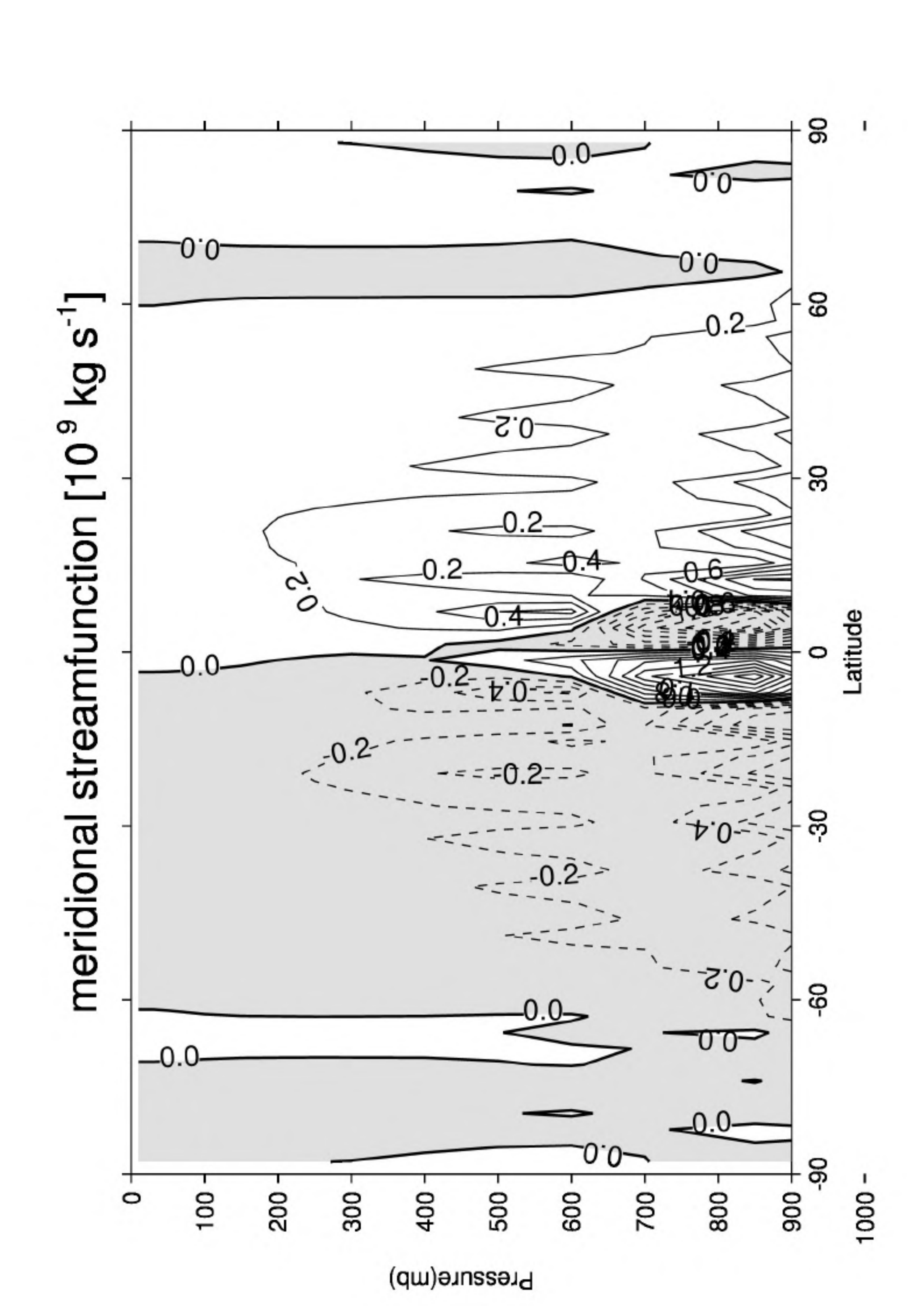}
    \label{circ2_t42_g}
   }
   \subfigure[$\mathcal{F}_f= 10^{4}$, $\mathcal{R}o=10^{-3}$]{
    \includegraphics[angle=0, width=0.3\textwidth]{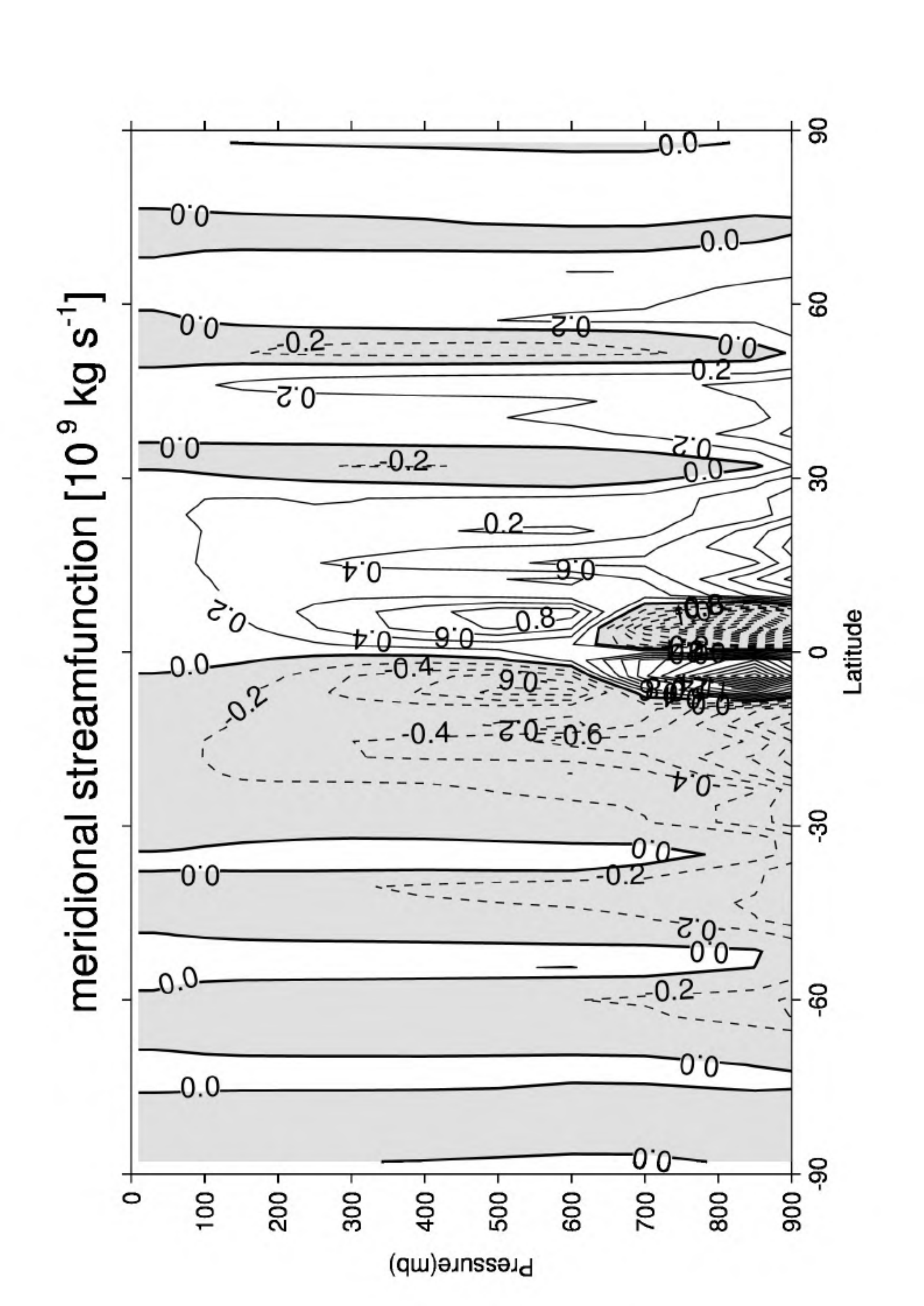}
    \label{circ2_t42_h}
     } 
   \subfigure[$\mathcal{F}_f= 10^{9}$, $\mathcal{R}o=10^{-3}$]{      
     \includegraphics[angle=0, width=0.3\textwidth]{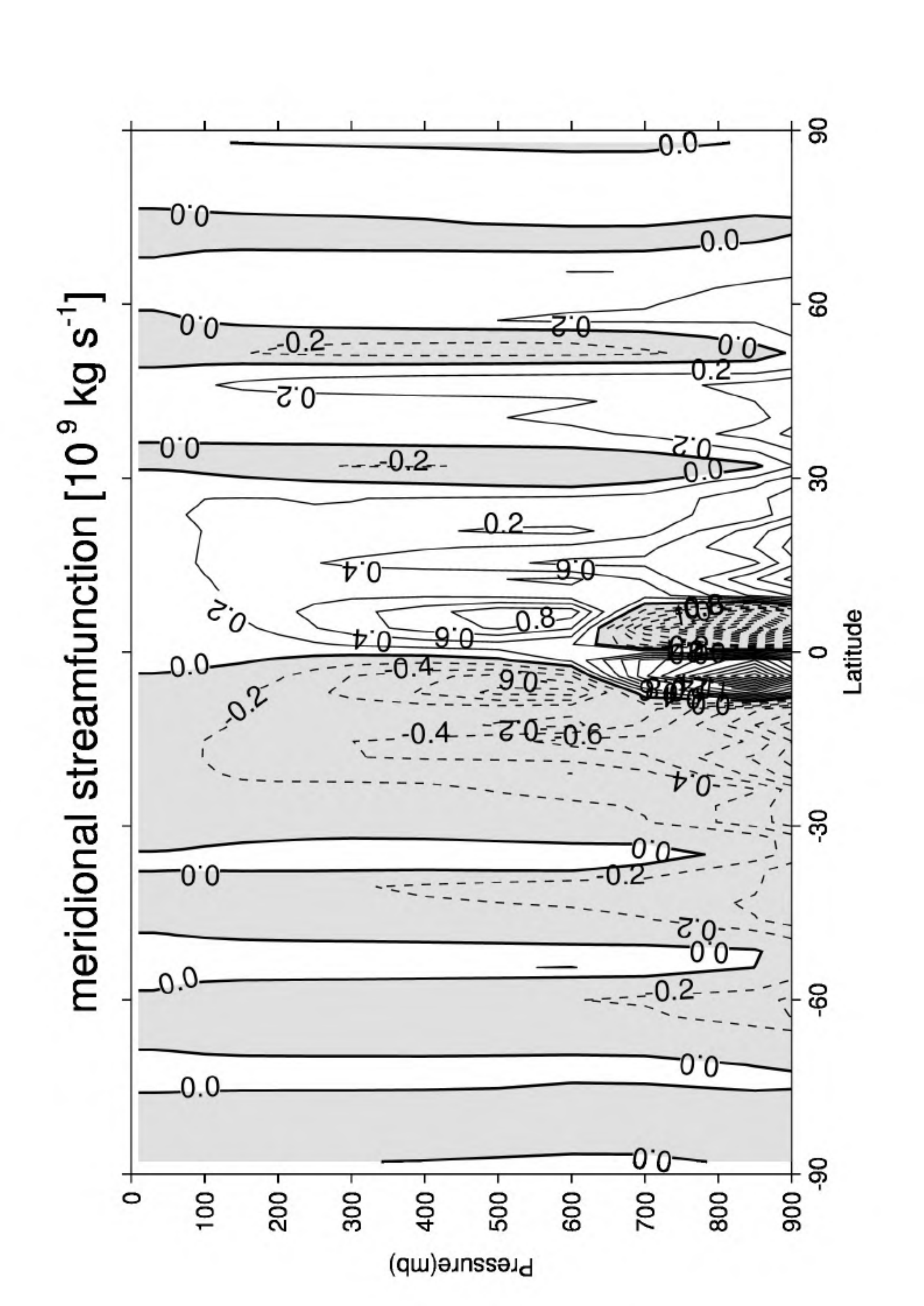}
    \label{circ2_t42_i}
   }     
\caption{   \label{circ2_t42}}
\end{figure}

\begin{itemize}

\item Figure \ref{diags} \\ \ref{diag2} Schematic diagram of the ($\mathcal{F}_f$, $\mathcal{R}o$) parametric space spanned in this study. Overplotted are the values of $\Omega^*$ (dashed-dotted) and $\tau$ (dotted). We have schematically scketched the boundaries between different circulation regimes found for dry PlaSim on the base of the circulations  (AS, axisymmetric;  BC, baroclinic; BT, barotropic; ZN, zonostrophic; SR, super-rotation). Circles, pentagons and triangles represent the simulations performed with $\Omega^*=0.1,1,8$ respectively (see Fig.\ref{circ2_t42} and $\ref{circ1_t42}$).  \ref{diag1} The same  regime diagram  is summarizing schematically the properties of  kinetic energy dissipation (continuos line, high and low D), meridional energy transport (dotted-dashed line, high  MHT), thermal material entropy production (dotted line, high and low $\dot{S}_{sense}$), efficiency (dashed line, high and low $\eta$). 

\item Figure \ref{circ1_t42}\\ Zonal winds and temperature for $\Omega^*=1/10$
 ($\tau=2700 s$ (a), $1$ day (b), $500$ days (c)), $\Omega^*=1$ ($\tau=2700 s$ (d), $1$ days (e), $500$ days (f)), 
 $\Omega^*=8$ ($\tau=2700 s$ (g), $1$ days (h), $500$ days (i)).

\item  Figure \ref{circ2_t42} \\As in Fig.\ref{circ2_t42} but for the meridional mass streamfunction (units 10$^9$ Kg\,s$^{-1}$).
 
% \item Figure     \ref{mars}\\  (a) Zonal mean  of temperature and zonal wind  and (b) meridional   mass transport stream function (10$^9$ Kg\,s$^{-1}$)   for $\mathcal{R}o\approx0.1$, $\mathcal{F}_f\approx 10^3$ show a good similarity with the mean state of Mars (e.g. see figure 2 of \cite{Read}), which in fact has a dry atmosphere and close dimensionless numbers ($\mathcal{R}o\approx0.2$, $\mathcal{F}_f\approx 50$).

\item Figure \ref{kediss} \\Total kinetic energy dissipation; overplotted (as in all the following plots) are the  values of $\log_{10}\mathcal{R}o$ (dashed) and $\log_{10}\mathcal{F}_f$ (dotted).

\item Figure \ref{kediss_bl} \\Contribution to the total kinetic energy dissipation due to 
parametrizations representing boundary layer 
stresses and gravity wave drag, $D_{phys}$. 

\item Figure \ref{dtmer} \\  Atmospheric  meridional energy transport   index $MHT$.

\item Figure \ref{carnot} \\ Carnot efficiency $\eta$. 

\item Figure \ref{entropybudget}  \\ Entropy production associated with surface sensible 
heat flux. Units in $10^{-3}$ W\,m$^{-2}$\,K$^{-1}$.

\item Figure \ref{entropybudget2}\\Material entropy production associated with  dissipation of kinetic energy. Units in $10^{-3}$ W\,m$^{-2}$\,K$^{-1}$.

\item Figure \ref{entropybudget3}\\  Total material entropy  production. Units in $10^{-3}$ W\,m$^{-2}$\,K$^{-1}$.

\item Figure \ref{Bej} \\ Irreversibility parameter $\alpha$.

\item Figure \ref{mepp} \\ $\dot{S}_{mat}$  (\ref{mep1}) and $\dot{S}_{kediss}$ (\ref{mep2}) for the control runs  BLS (continuous line), for $\tau_{max}(\Omega^*)$ maximizing $\dot{S}_{mat}$ (dashed) and for $\tau=0.1\,\tau_{max}$ (dotted) and $\tau= 10\, \tau_{max}$ (dotted-dashed) days.  \ref{mep3}-\ref{mep4} Same as in Fig. \ref{mep1} and \ref{mep2} but for $\tilde{\tau}_{max}$ maximising $\dot{S}_{kediss}$.  

\item Figure \ref{circ1_contr}\\  Zonal winds and temperature for $\Omega^*=1/10$ (a),  $\Omega^*=1$ (b) and  $\Omega^*=8$ for the BLS simulations.

\item Figure \ref{circ2_contr}\\  Meridional streamfunction  for $\Omega^*=1/10$ (a),  $\Omega^*=1$ (b) and  $\Omega^*=8$ for the BLS simulations.
 
\end{itemize}

\clearpage

\clearpage

\begin{figure}
 \centering
   \subfigure{      
     \includegraphics[angle=-0, width=0.9\textwidth]{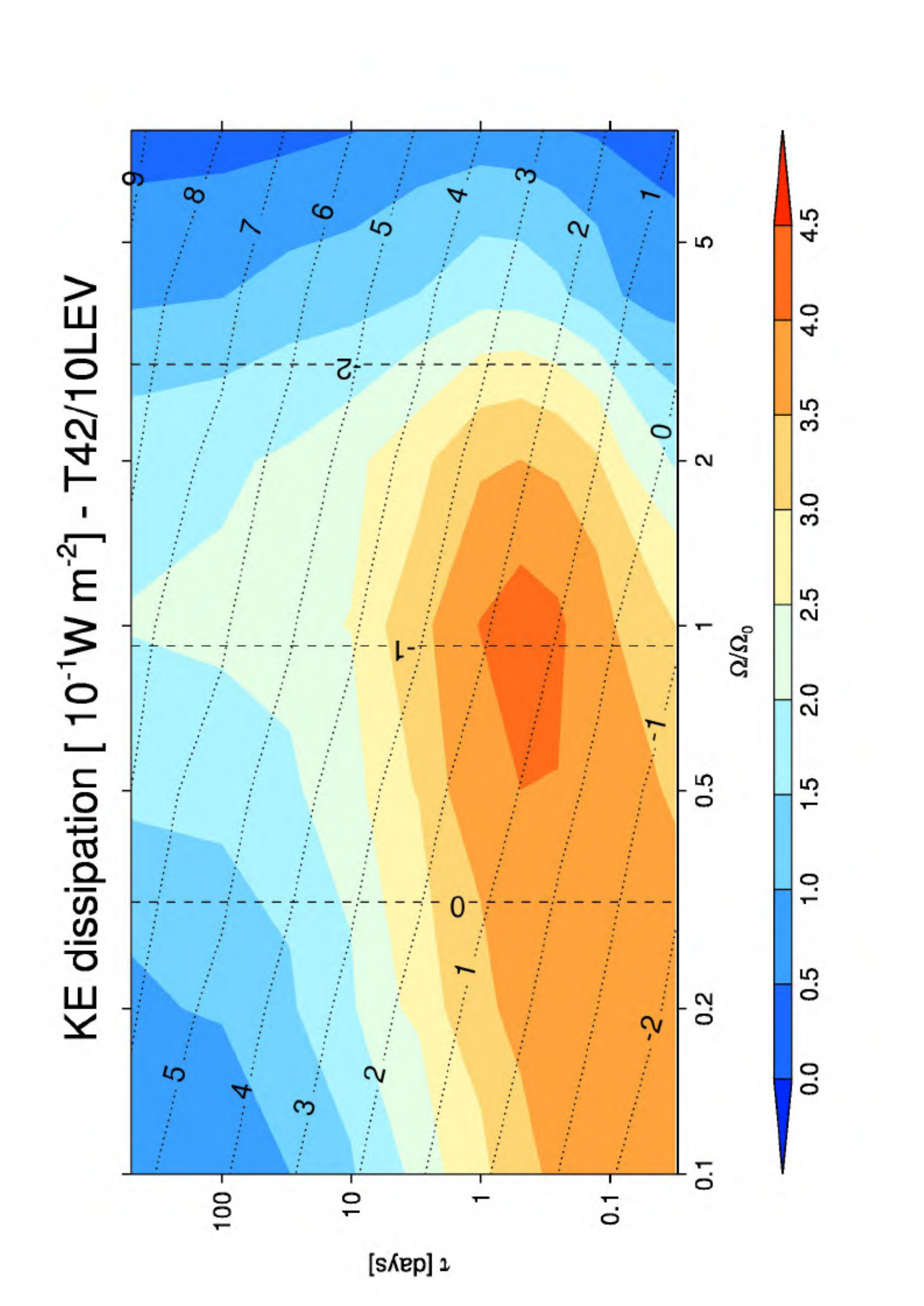}
    \label{kediss_oc}
   }
%   \subfigure[]{
%    \includegraphics[angle=-90, width=0.9\textwidth]{kediss_rock2.pdf}
%     \label{kediss_oc_phys}
%      } 
\caption{  \label{kediss}}
\end{figure}      
      
 \begin{figure}
 \centering     
 \subfigure{
    \includegraphics[angle=-0, width=0.9\textwidth]{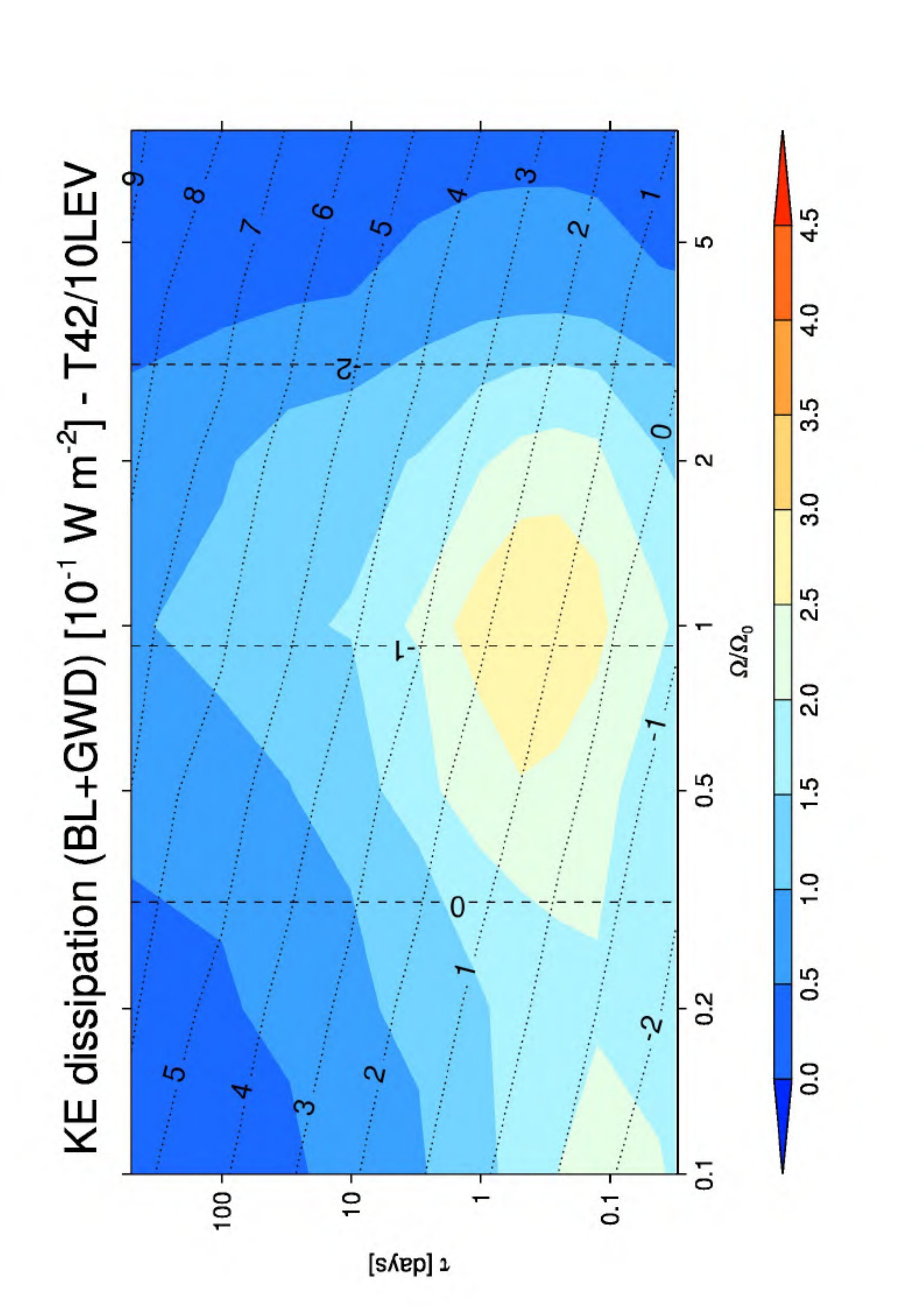}
    \label{kediss_rock}
   }
%   \subfigure[]{
%    \includegraphics[angle=-90, width=0.9\textwidth]{kediss_bl_rock2.pdf}
%     \label{kediss_rock_phys}
 %     } 
\caption{  \label{kediss_bl}}
\end{figure}

\begin{figure}
 \centering
 \subfigure{
    \includegraphics[angle=-0, width=0.9\textwidth]{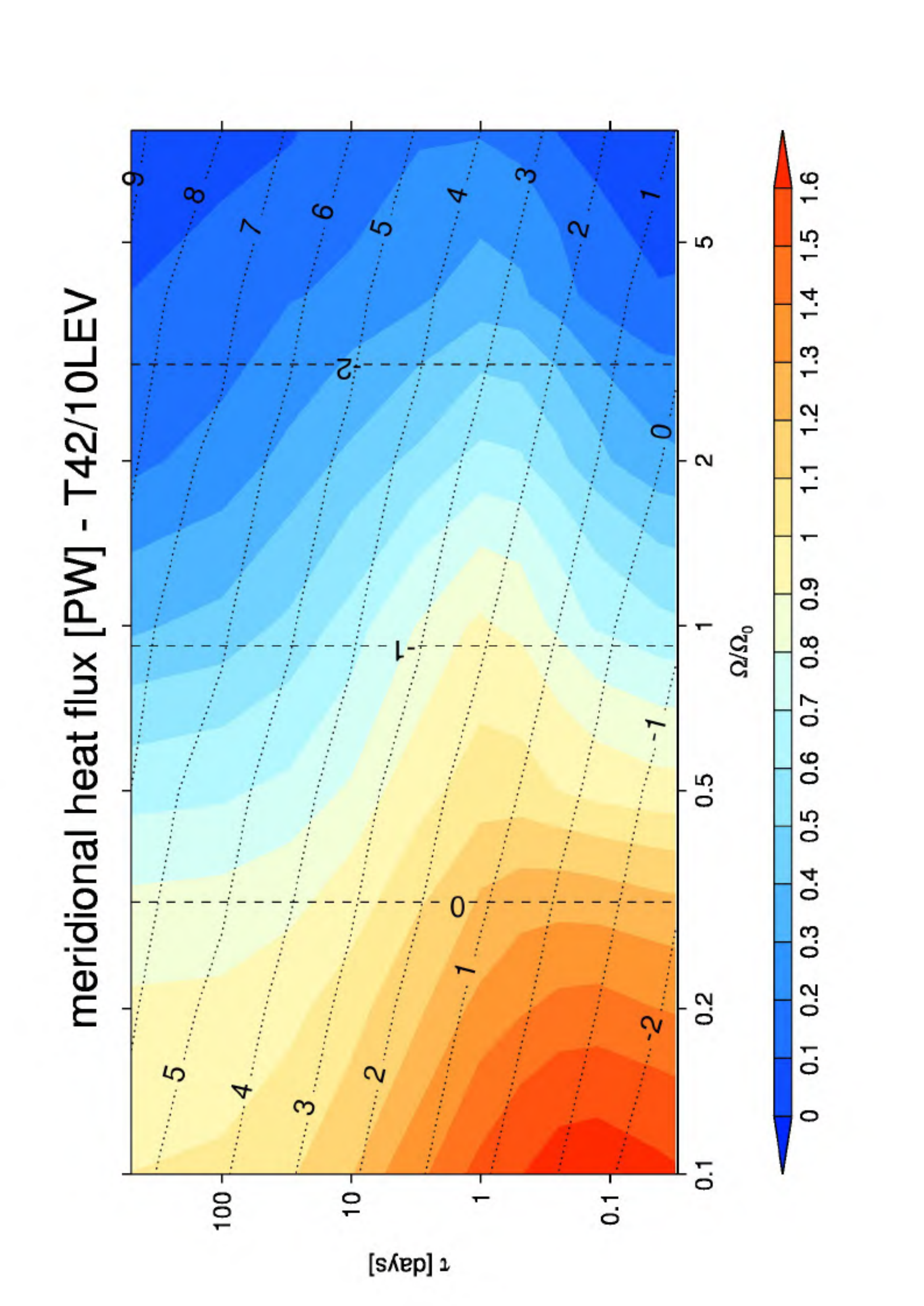}
    \label{dtmer_oc}
   }
%   \subfigure[]{
 %   \includegraphics[angle=-90, width=0.8\textwidth]{meridional_rock2.pdf}
 %    \label{meridional_oc}
%      } 
 \caption{ \label{dtmer}}
\end{figure}     
      
% \begin{figure}
% \centering     
%   \subfigure{      
 %   \includegraphics[angle=-0,width=0.9\textwidth]{DTeff_mer_rock2_t42.pdf}
 %        }
%   \subfigure[]{
%    \includegraphics[angle=-90, width=0.8\textwidth]{DTeff_mer_rock2.pdf}
 %    \label{meridional_rock}
%      } 
% \caption{ \label{dteff}}
% \end{figure}      
      
\begin{figure}
 \centering    
   \subfigure{
    \includegraphics[angle=-0, width=0.9\textwidth]{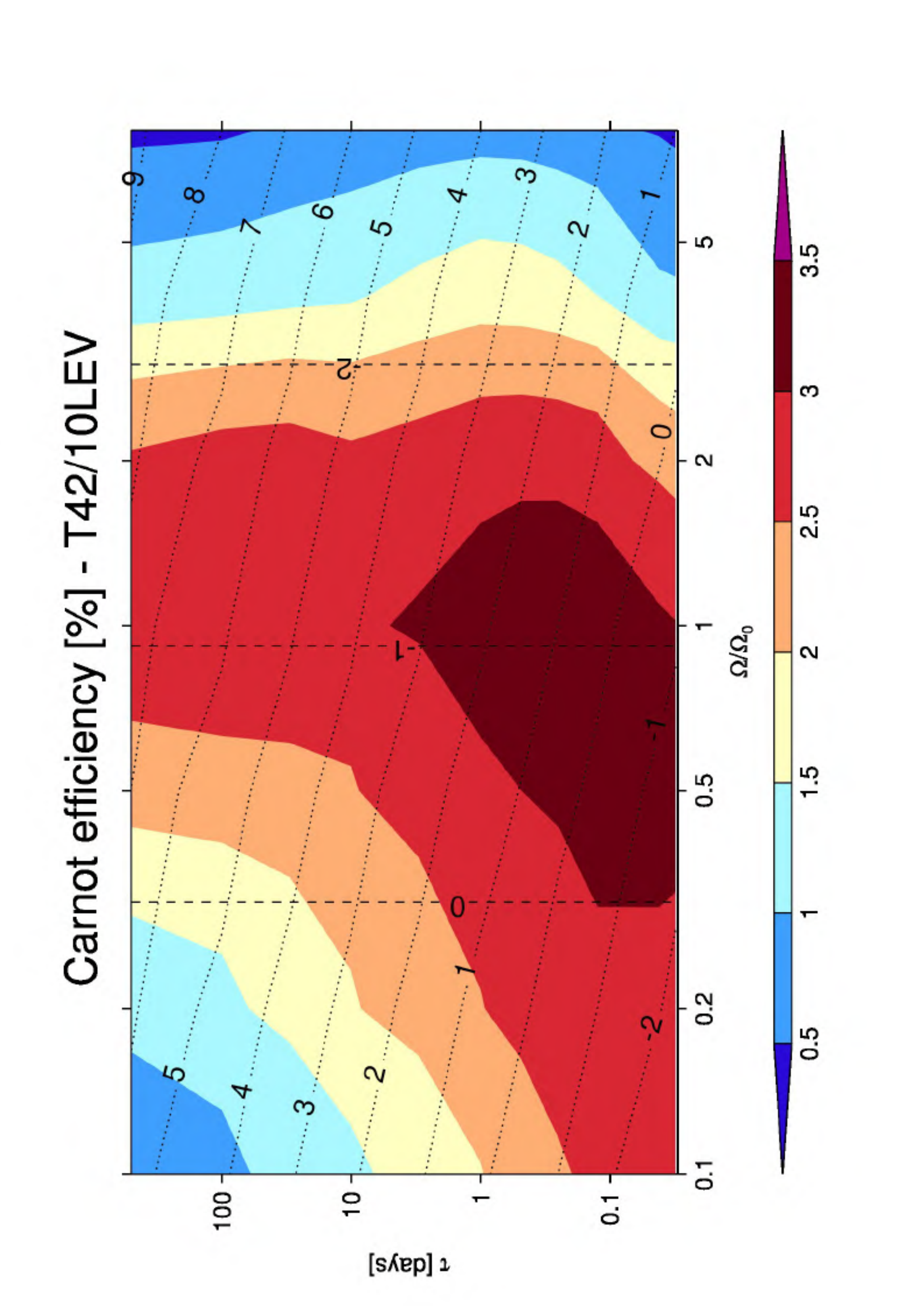}
    \label{carnot_t42}
   }
%   \subfigure[]{
%    \includegraphics[angle=-90, width=0.8\textwidth]{carnot_rock2_t21.pdf}
%     \label{carnot_t21}
%      } 
\caption{ \label{carnot}}
\end{figure}

%\begin{figure}
% \centering    
%   \subfigure[]{
%    \includegraphics[angle=90, width=0.8\textwidth]{phip_t42.pdf}
%    \label{phip_t42}
%   }
%    \subfigure[]{
%    \includegraphics[angle=90, width=0.8\textwidth]{phiplus_t21.pdf}
%     \label{phip_t21}
 %     }   
%\caption{ \label{varie3}}
%\end{figure}

\begin{figure}
 \centering
 \subfigure{
    \includegraphics[angle=-0, width=0.9\textwidth]{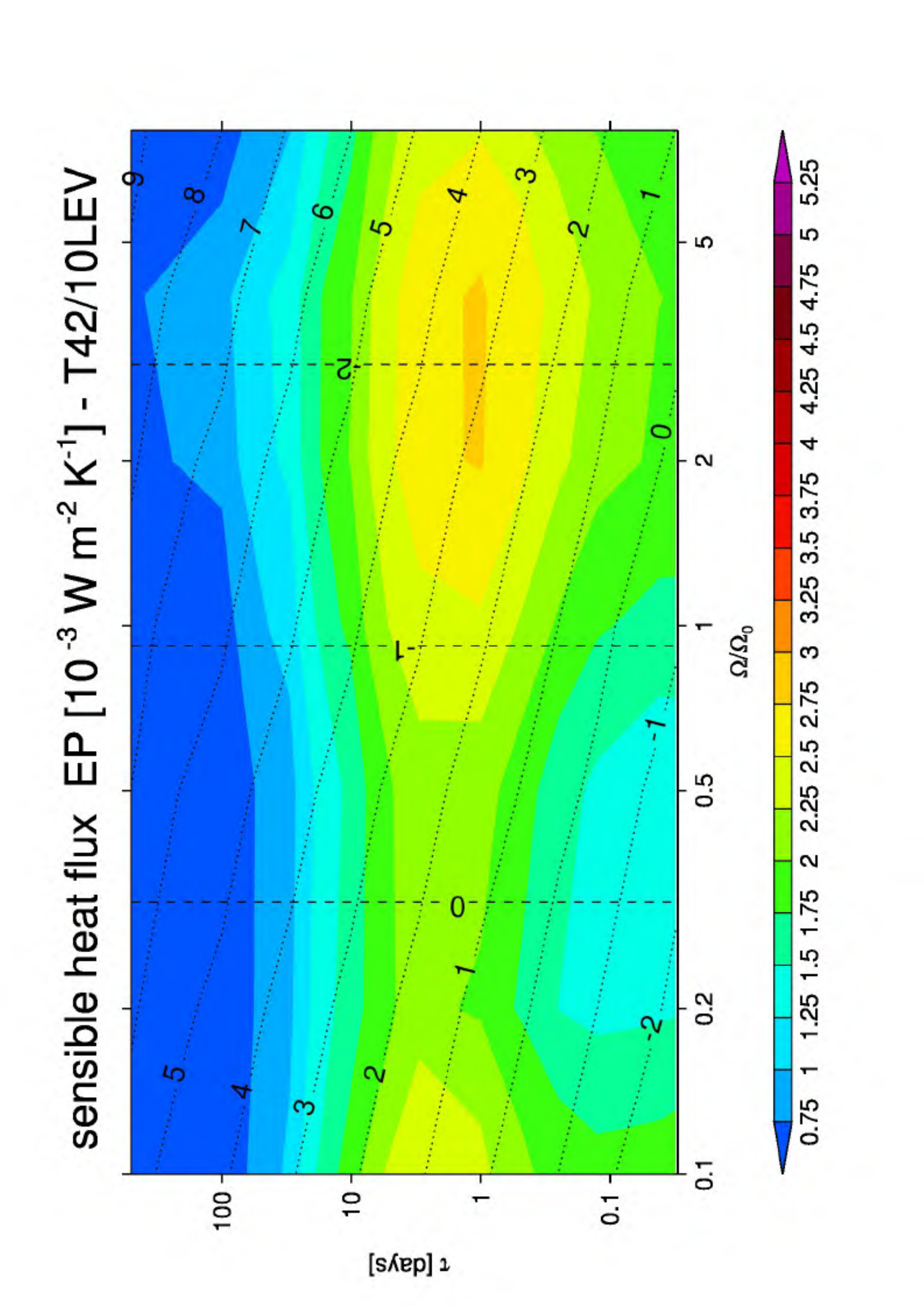}
    \label{SHentropy_t42}
   }
%   \subfigure[]{
 %   \includegraphics[angle=-90, width=0.8\textwidth]{SHentropy_rock2.pdf}
%     \label{SHentropy_t21}
%      } 
\caption{\label{entropybudget}}
\end{figure}      

\begin{figure}
 \centering      
   \subfigure{
    \includegraphics[angle=-0, width=0.9\textwidth]{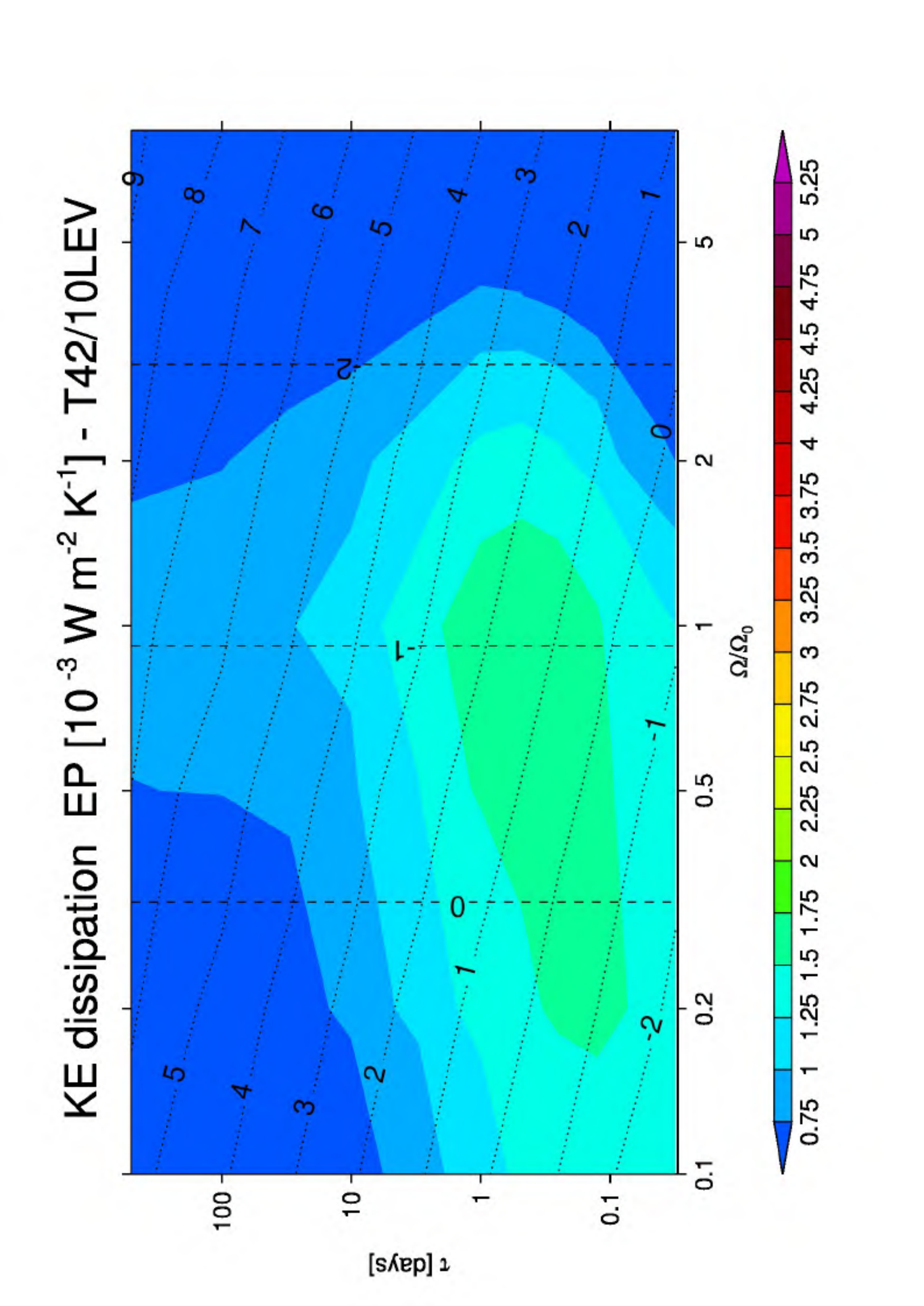}
    \label{kedissentr_oc}
   }
%   \subfigure[]{
%    \includegraphics[angle=-90, width=0.8\textwidth]{kedissentr_rock2.pdf}
%     \label{kedissentr_rock}
%      }     
\caption{\label{entropybudget2}}
\end{figure}

 \begin{figure}
 \centering
       \subfigure{
    \includegraphics[angle=-0, width=0.9\textwidth]{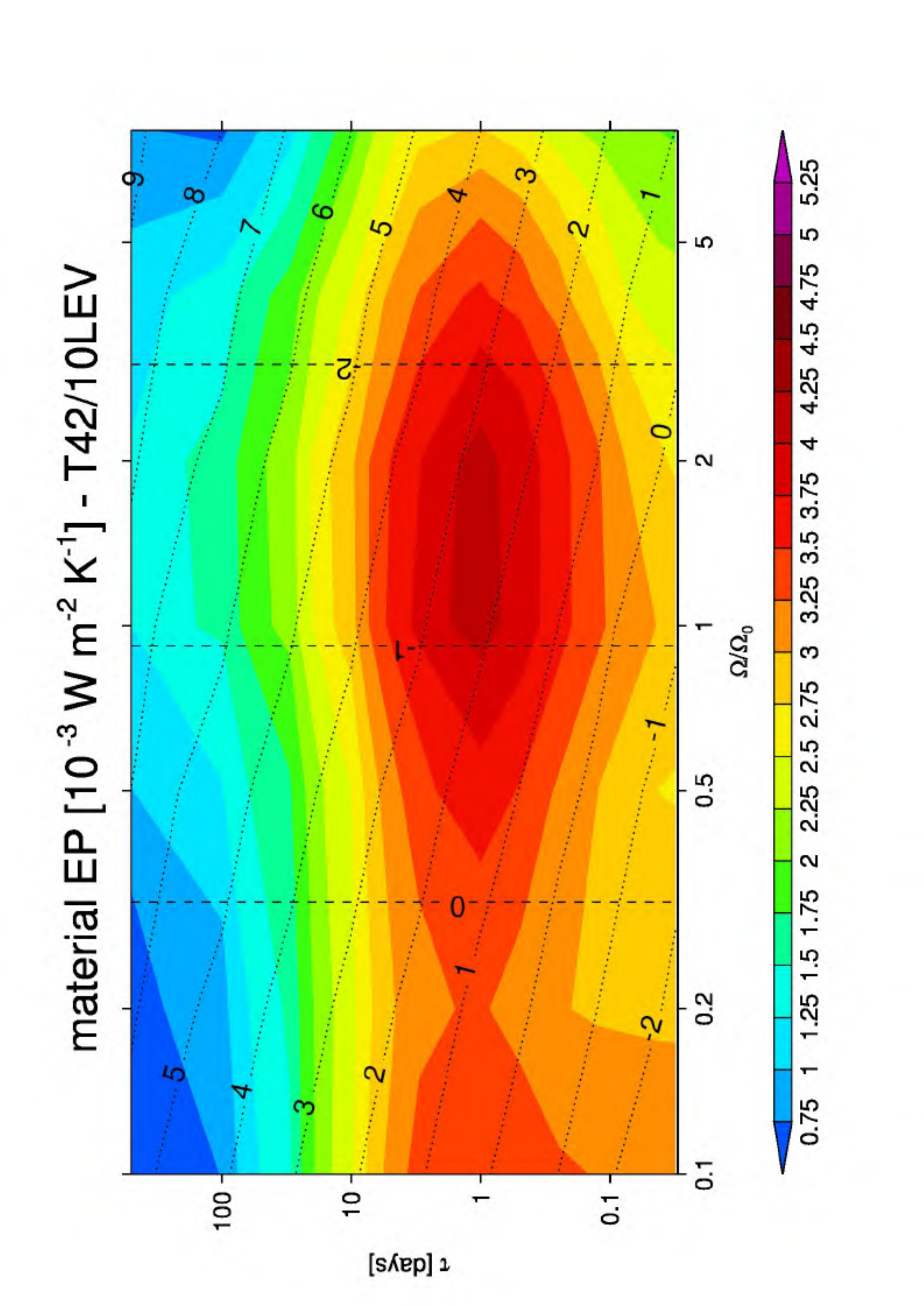}
    \label{mat_tauom_oc}
   }
%   \subfigure[]{
 %   \includegraphics[angle=-90, width=0.8\textwidth]{mat_tauom_rock2.pdf}
 %    \label{mat_tauom_rock}
 %     } 
\caption{\label{entropybudget3}}
\end{figure}

\begin{figure}
 \centering
 \subfigure{
    \includegraphics[angle=-0, width=0.9\textwidth]{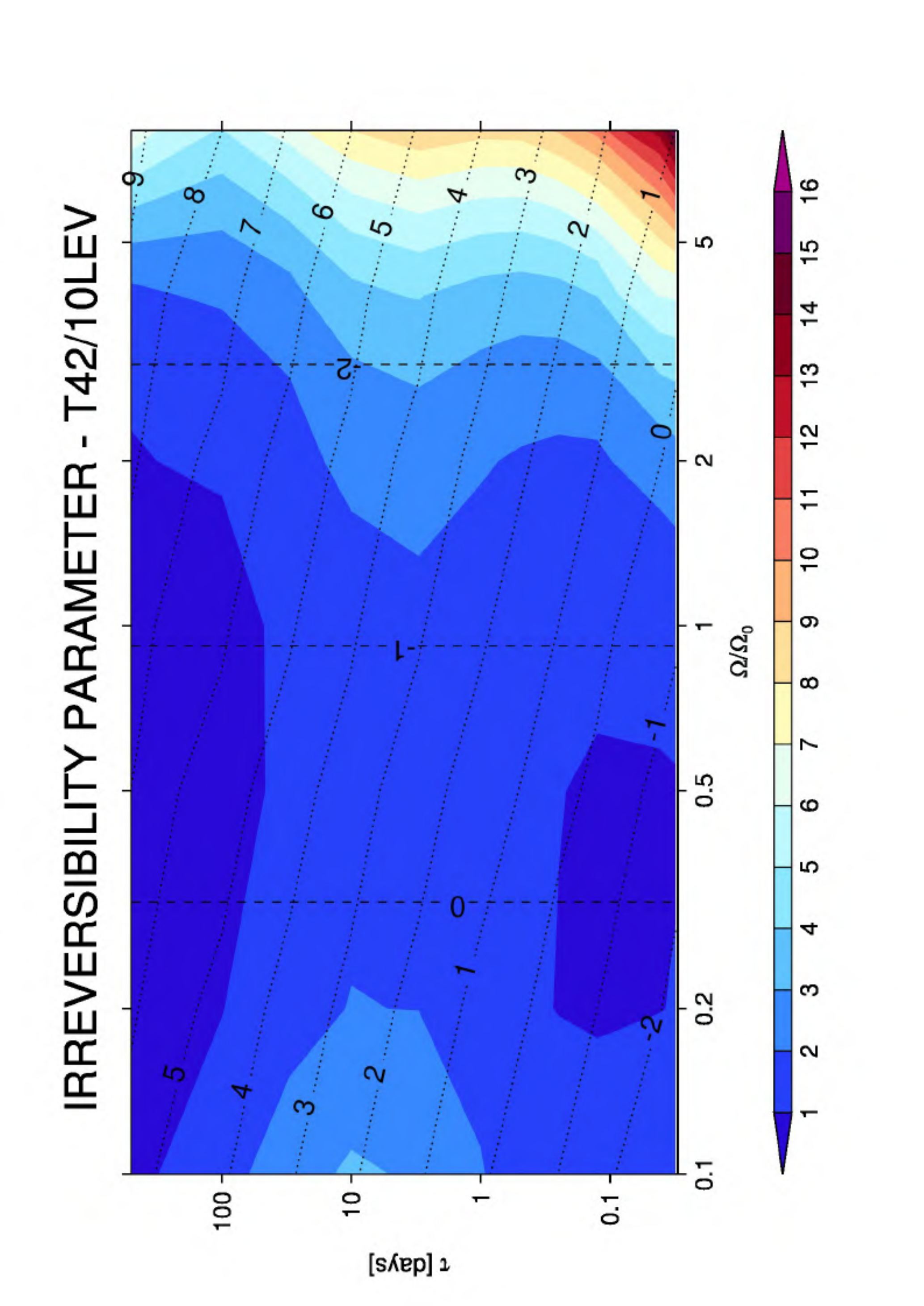}
    \label{Bej_oc}
   }
%   \subfigure[]{
 %   \includegraphics[angle=-90, width=0.8\textwidth]{Bejan_rock2.pdf}
 %    \label{Bej_rock}
 %     } 
  \caption{  \label{Bej}}
\end{figure}

\begin{figure}
 \centering
 \subfigure[]{
    \includegraphics[angle=-0, width=0.45\textwidth]{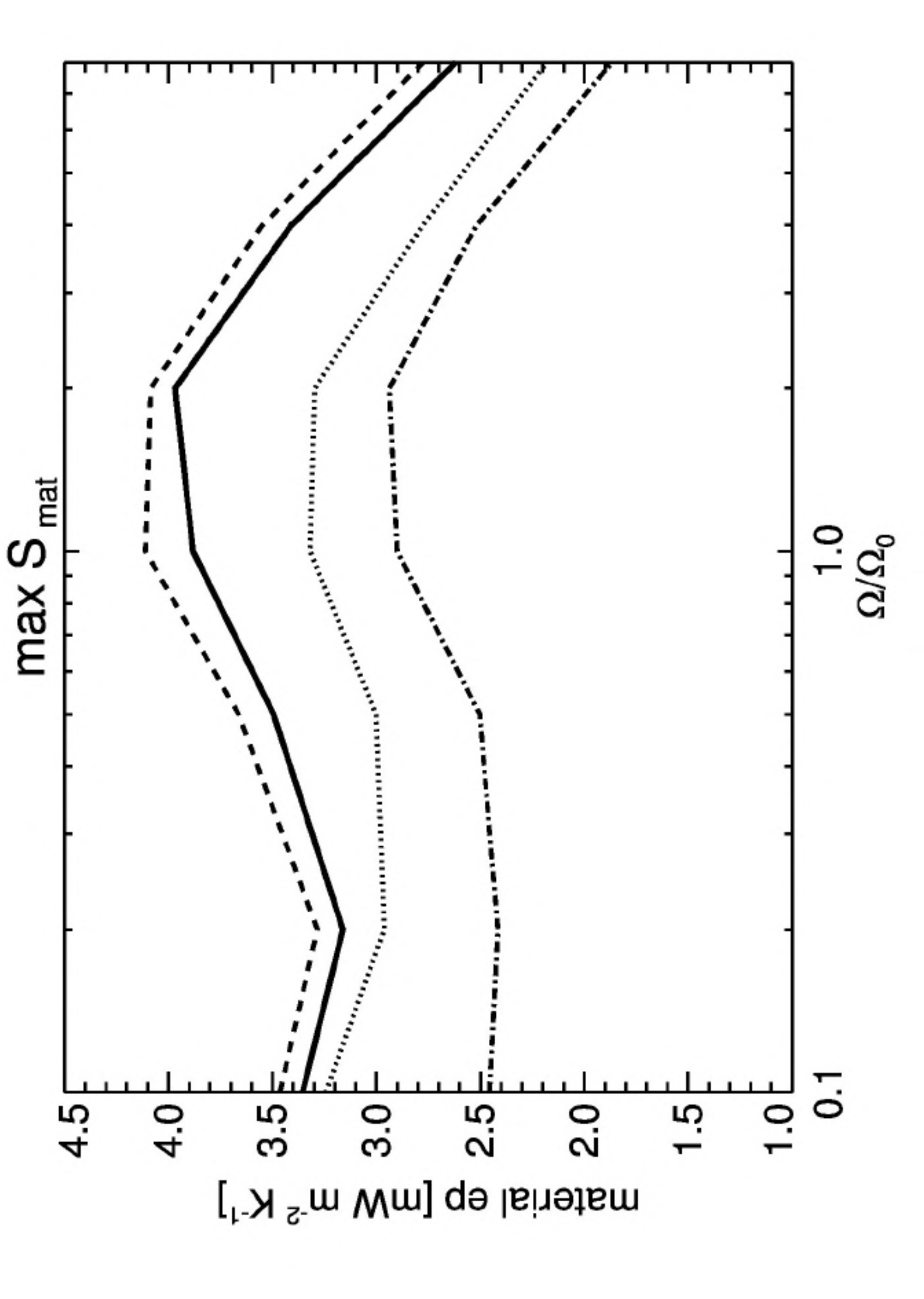}
    \label{mep1}
   }
      \subfigure[]{
    \includegraphics[angle=-0, width=0.45\textwidth]{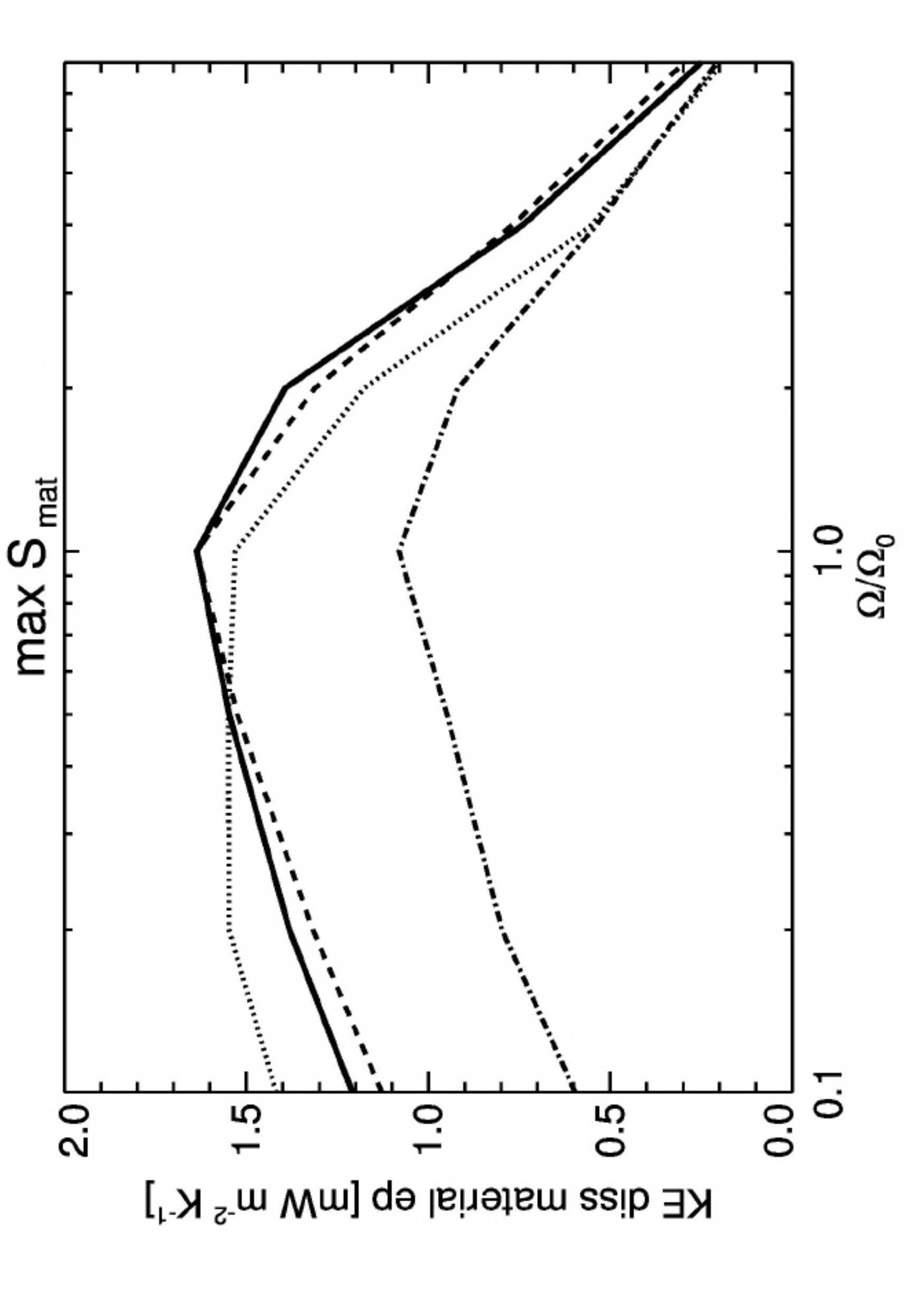}
     \label{mep2}
      } 
%  \caption{  \label{max_mat_entr}}
%\end{figure}
%\begin{figure}
% \centering
 \subfigure[]{
    \includegraphics[angle=-0, width=0.45\textwidth]{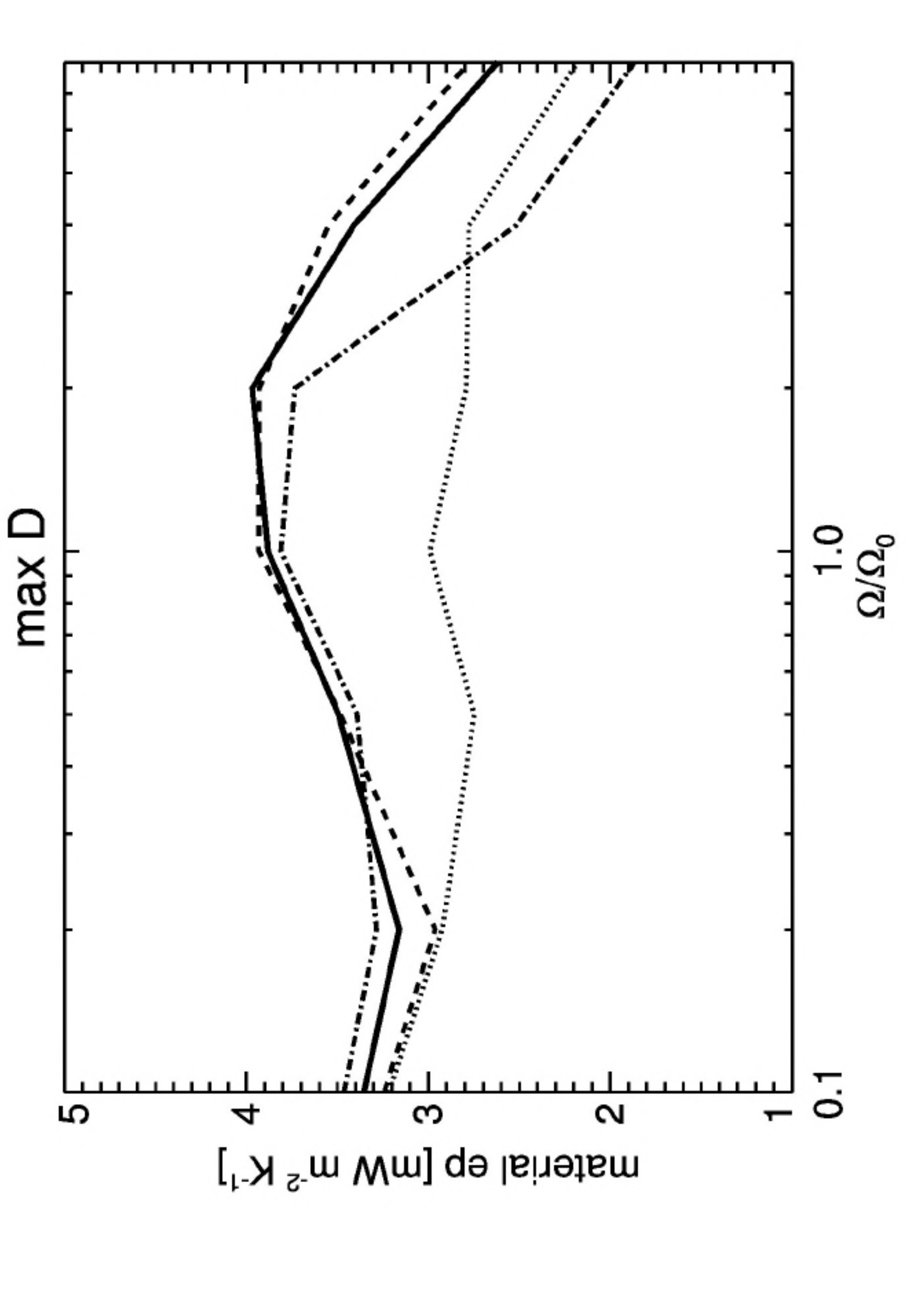}
    \label{mep3}
   }
      \subfigure[]{
    \includegraphics[angle=-0, width=0.45\textwidth]{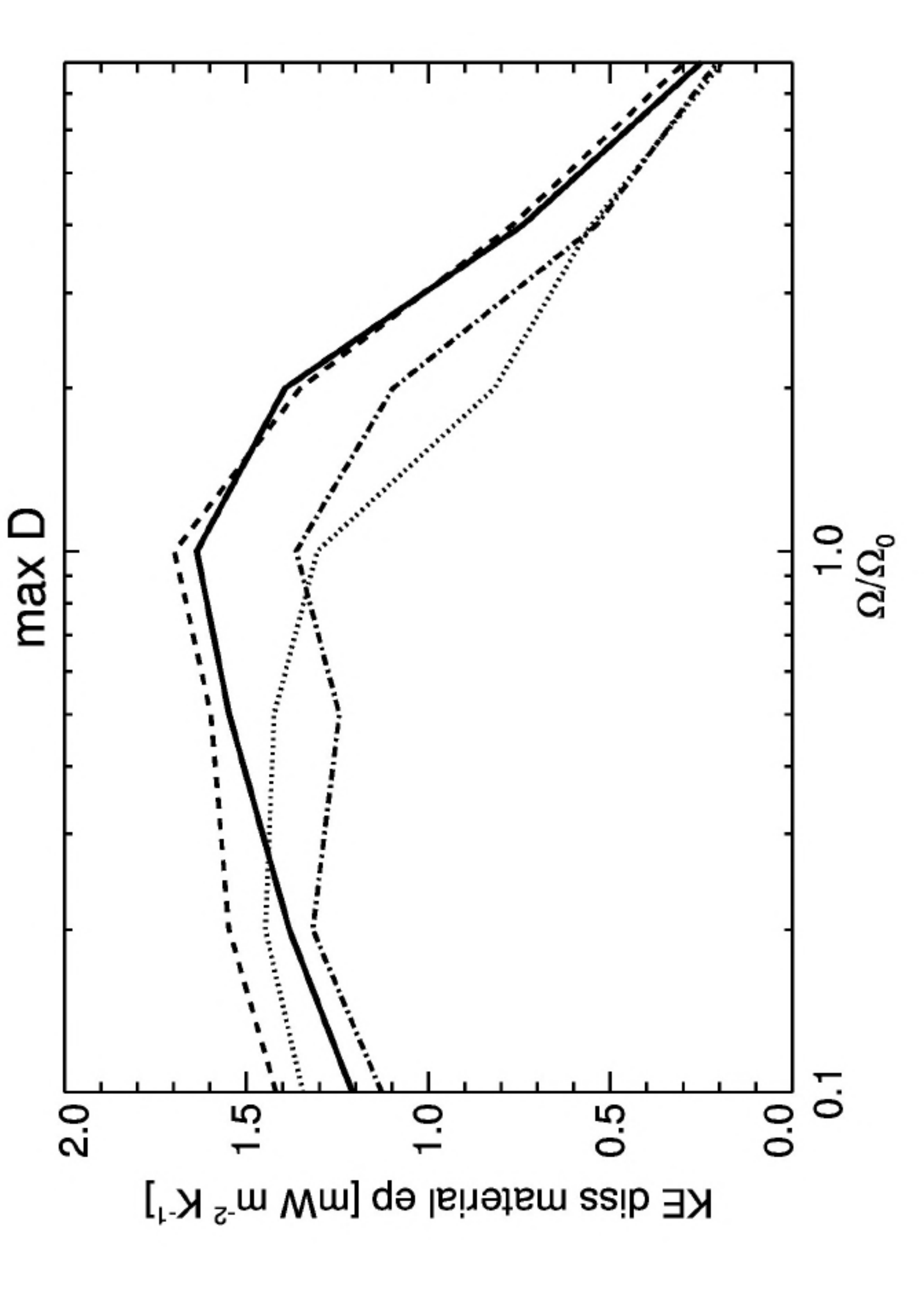}
     \label{mep4}
      } 
  \caption{  \label{mepp}}
\end{figure}

\begin{figure}
 \centering
\subfigure[]{
     \includegraphics[angle=-0, width=0.3\textwidth]{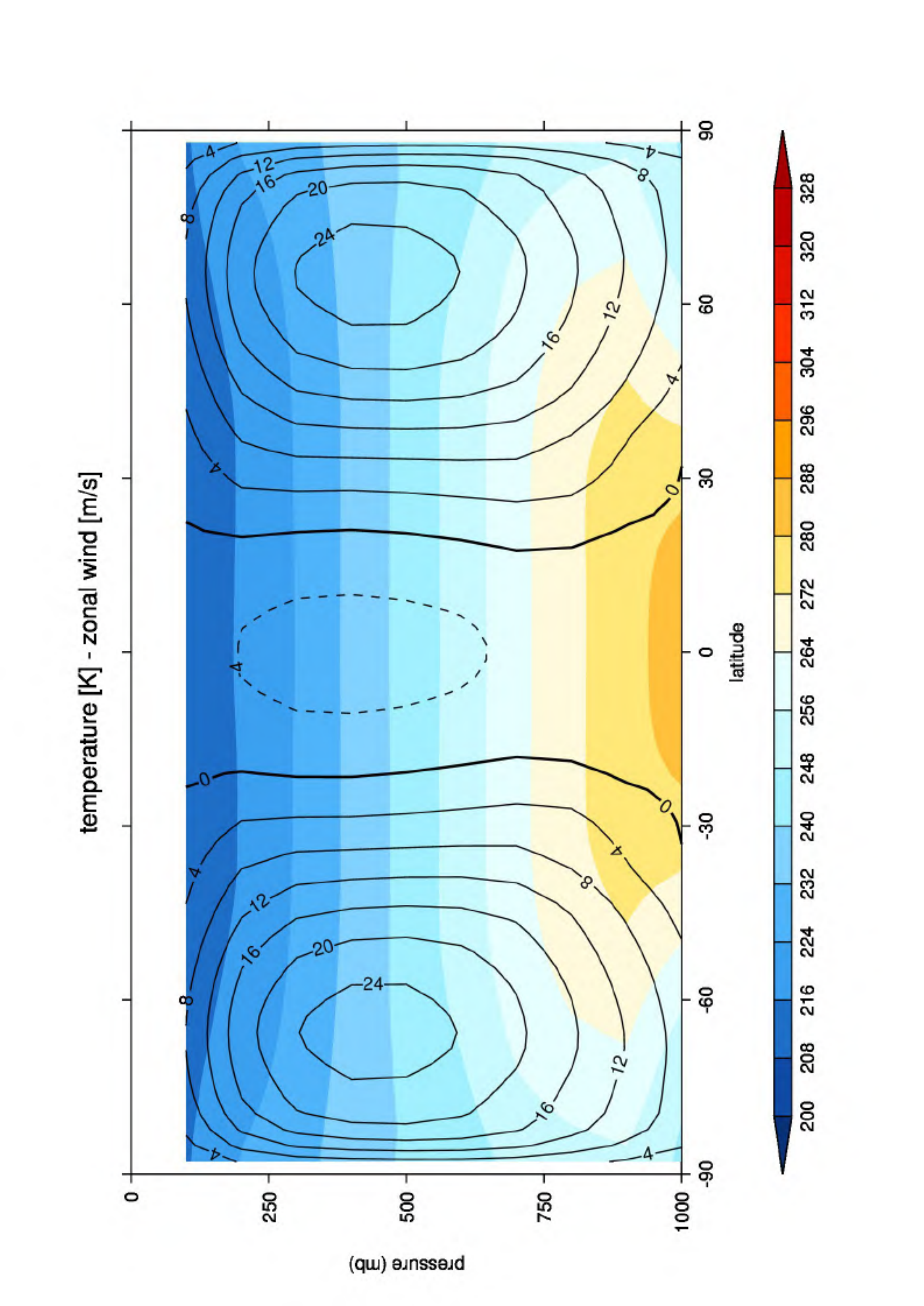}
    \label{circ1_contr_a}
   }
   \subfigure[]{
    \includegraphics[angle=-0, width=0.3\textwidth]{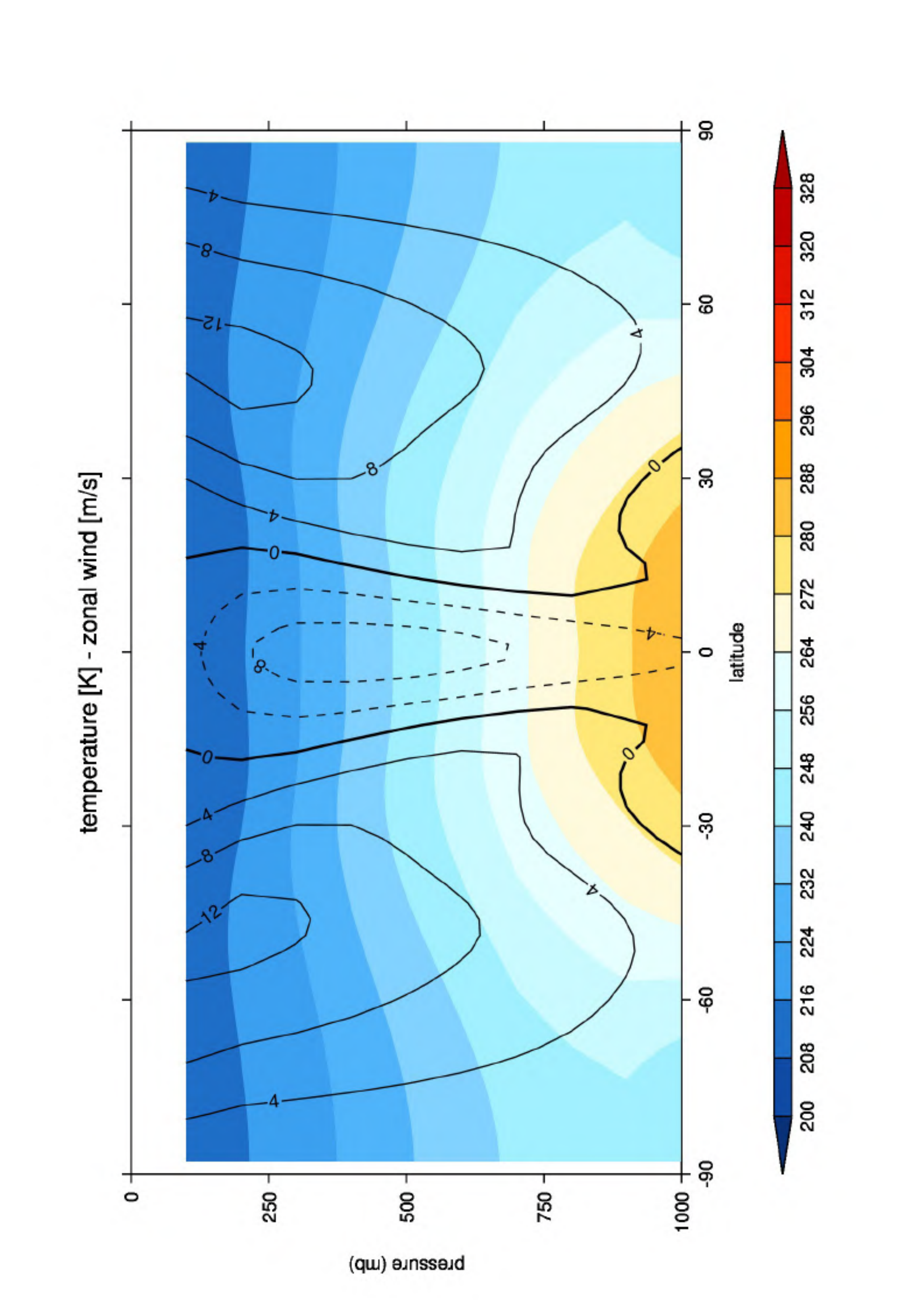}
    \label{circ1_contr_b}
     } 
   \subfigure[]{      
     \includegraphics[angle=-0, width=0.3\textwidth]{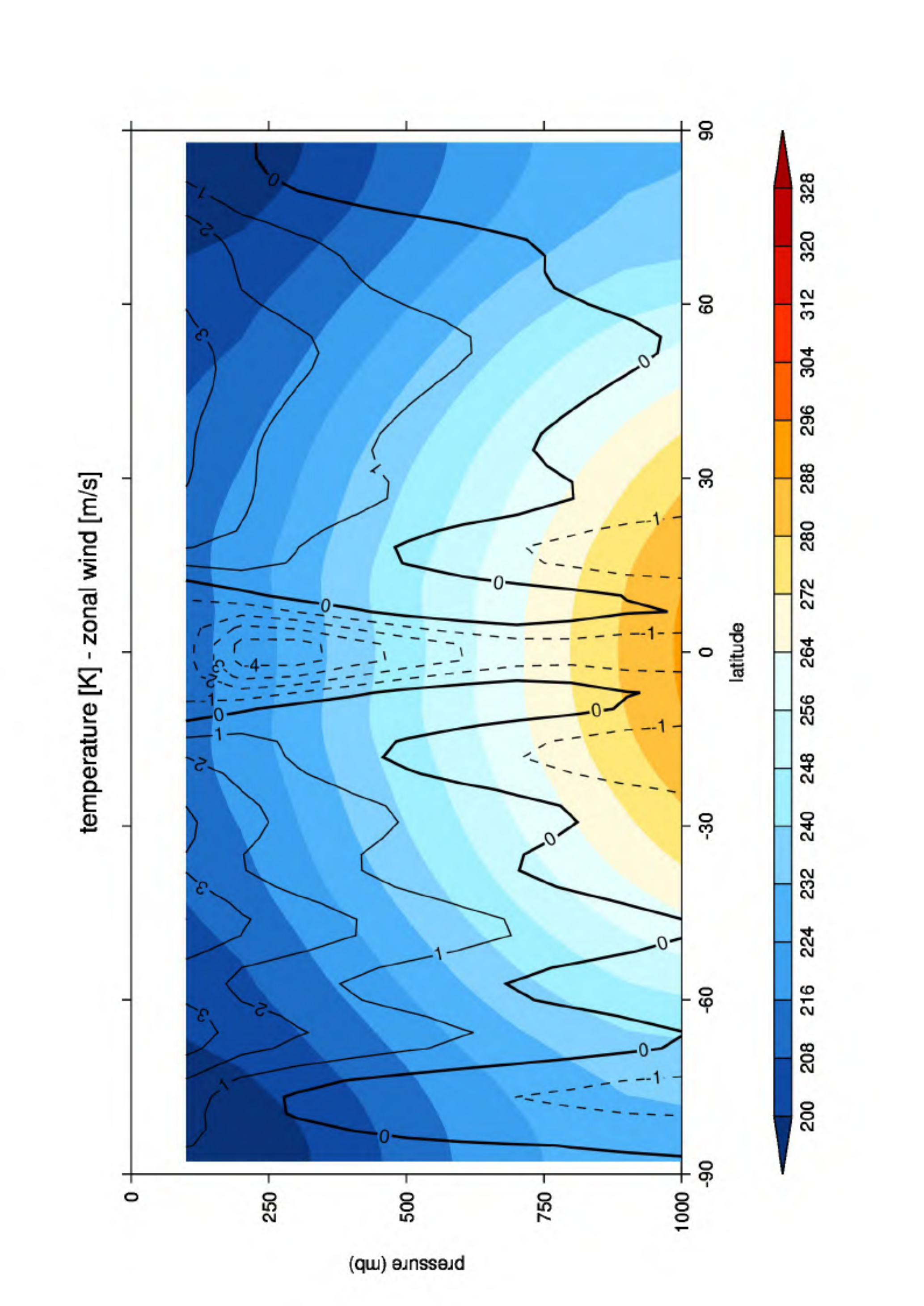}
    \label{circ1_contr_c}
   }
\caption{   \label{circ1_contr}}
\end{figure}

\begin{figure}
 \centering
\subfigure[]{
     \includegraphics[angle=0, width=0.3\textwidth]{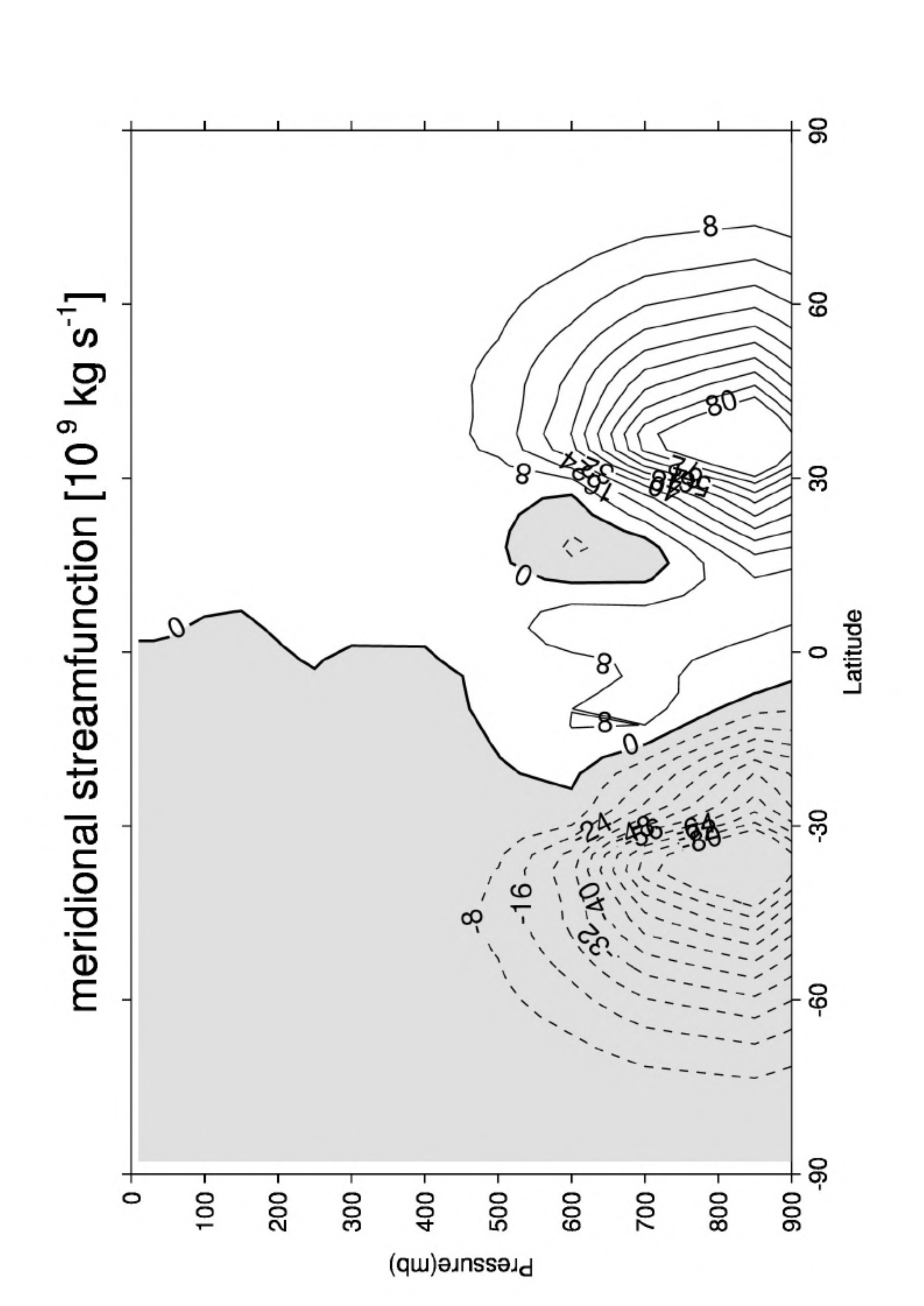}
    \label{circ2_contr_a}
   }
   \subfigure[]{
    \includegraphics[angle=0, width=0.3\textwidth]{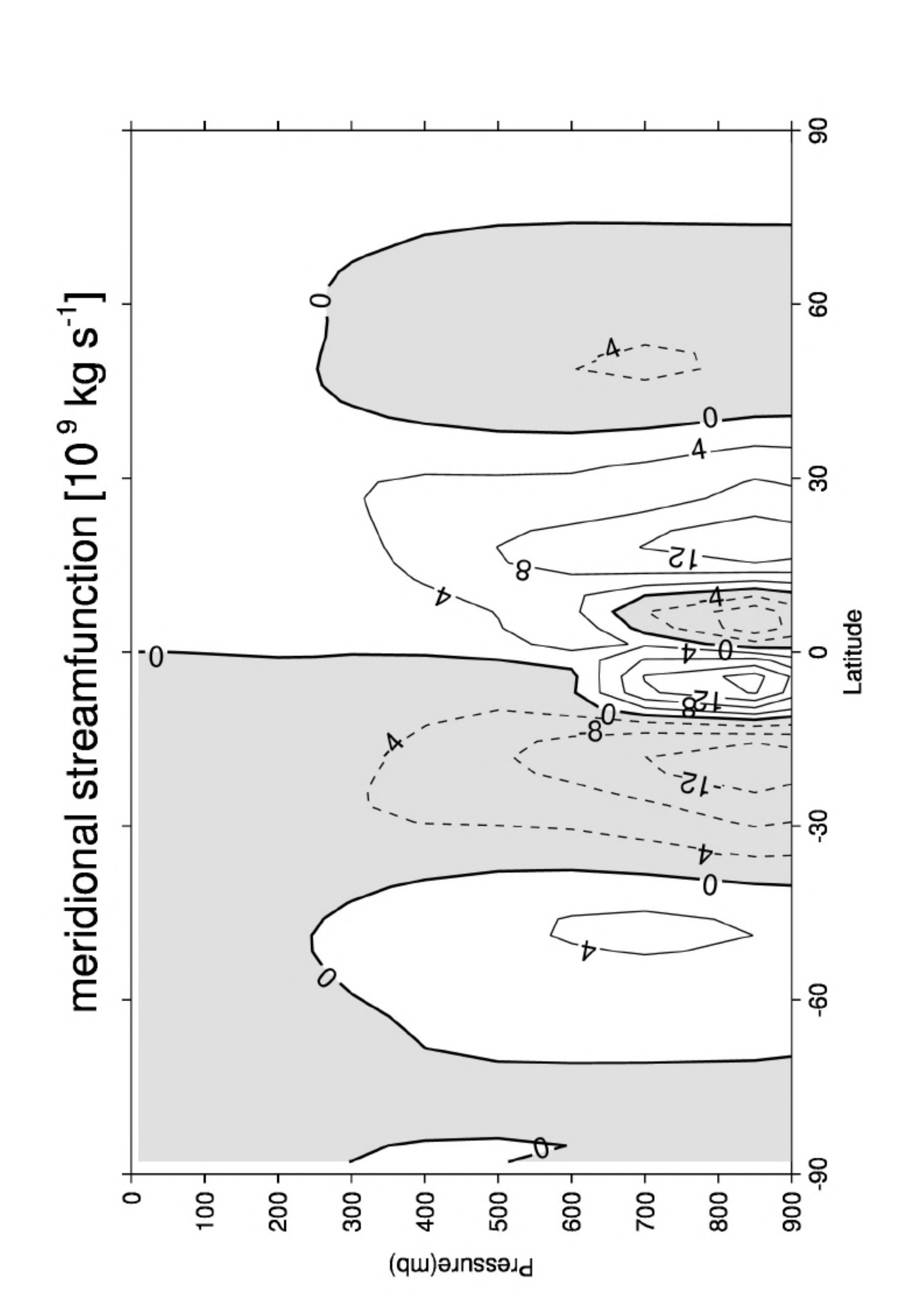}
    \label{circ2_contr_b}
     } 
   \subfigure[]{      
     \includegraphics[angle=0, width=0.3\textwidth]{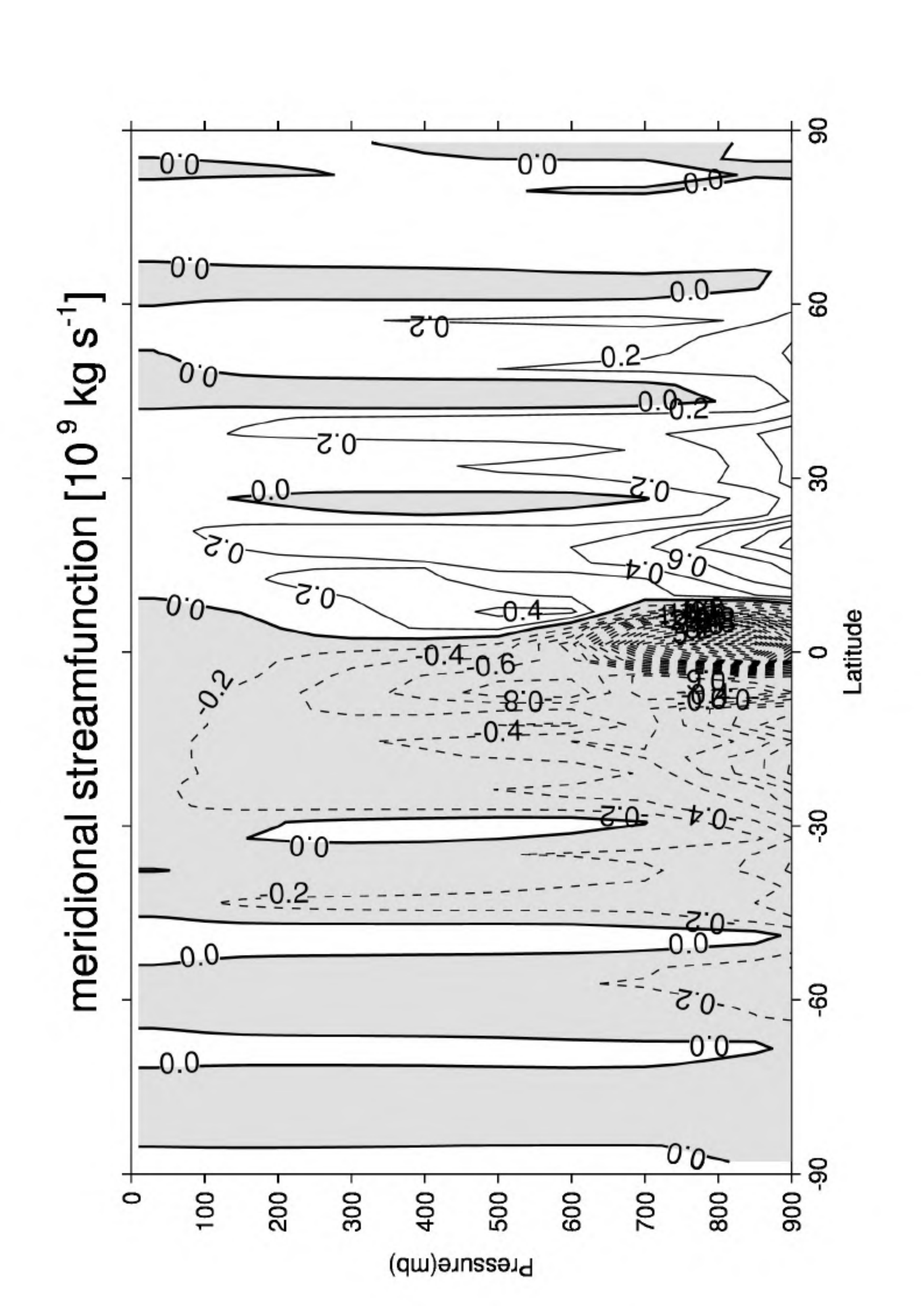}
    \label{circ2_contr_c}
   }
\caption{   \label{circ2_contr}}
\end{figure}

\end{document}